\documentclass[12pt]{iopart}
\usepackage{xcolor}
\usepackage{booktabs}

\input{definitions}

\let\Sectionmark\sectionmark
\def\sectionmark#1{\def\Sectionname{\uppercase{#1}}\Sectionmark{#1}}
\let\Subsectionmark\subsectionmark
\def\subsectionmark#1{\def\Subsectionname{#1}\Subsectionmark{#1}}

\DeclareAcronym{adv}{
  short=AdV,
  long=Advanced Virgo,
}

\DeclareAcronym{asd}{
  short=ASD,
  long=amplitude spectral density,
}

\DeclareAcronym{bh}{
  short=BH,
  long=black hole,
}

\DeclareAcronym{bns}{
  short=BNS,
  long=binary neutron star,
}

\DeclareAcronym{bristol}{
  short=\texttt{BRiSTOL},
  long=Band-limited RMS Stationarity Test Tool,
}

\DeclareAcronym{brms}{
  short=BRMS,
  long=band-limited RMS,
}

\DeclareAcronym{bruco}{
  short=\texttt{BruCo},
  long=brute-force coherence tool,
}

\DeclareAcronym{bs}{
  short=BS,
  long=beam splitter,
}

\DeclareAcronym{carm}{
  short=CARM,
  long=common (i.e. average) length of the two arm cavities,
}

\DeclareAcronym{ceb}{
  short=CEB,
  long=central building,
}

\DeclareAcronym{cw}{
  short=CW,
  long=continuous gravitational waves,
}

\DeclareAcronym{daq}{
  short=DAQ,
  long=data acquisition system,
}

\DeclareAcronym{darm}{
  short=DARM,
  long=difference of the two arm cavity lengths,
}

\DeclareAcronym{dms}{
  short=\texttt{DMS},
  long=Detector Monitoring System,
}

\DeclareAcronym{dof}{
  short=DOF,
  long=degree of freedom,
}

\DeclareAcronym{dq}{
  short=DQ,
  long=data quality,
}

\DeclareAcronym{dqr}{
  short=\texttt{DQR},
  long=data quality report,
}

\DeclareAcronym{dqsegdb}{
  short=\texttt{DQSEGDB},
  long=Data Quality Segment Database,
}

\DeclareAcronym{eom}{
  short=EOM,
  long=electro-optical modulator,
}

\DeclareAcronym{fft}{
  short=FFT,
  long=fast Fourier transform,
}

\DeclareAcronym{gracedb}{
  short=GraceDB,
  long=GRAvitational-wave Candidate Event Database,
}

\DeclareAcronym{gw}{
  short=GW,
  long=gravitational wave,
}

\DeclareAcronym{gwosc}{
  short=GWOSC,
  long=Gravitational Wave Open Science Center,
}

\DeclareAcronym{imc}{
  short=IMC,
  long=input mode-cleaner,
}

\DeclareAcronym{lvalert}{
  short=\texttt{LVAlert},
  long=LIGO-Virgo Alert System,
}

\DeclareAcronym{mich}{
  short=MICH,
  long=length difference between the Virgo Michelson interferometer short arms
}

\DeclareAcronym{monet}{
  short=\texttt{MONET},
  long=Modulated NoisE Tool
}

\DeclareAcronym{ne}{
  short=NE,
  long=north end,
}

\DeclareAcronym{neb}{
  short=NEB,
  long=north-end building,
}

\DeclareAcronym{ni}{
  short=NI,
  long=north input,
}

\DeclareAcronym{noemi}{
  short=\texttt{NoEMi},
  long=Noise Frequency Event Miner,
}

\DeclareAcronym{ns}{
  short=NS,
  long=neutron star,
}

\DeclareAcronym{omc}{
  short=OMC,
  long=output mode-cleaner,
}

\DeclareAcronym{pr}{
  short=PR,
  long=power recycling,
}

\DeclareAcronym{prcl}{
  short=PRCL,
  long=power recycling cavity length,
}

\DeclareAcronym{psd}{
  short=PSD,
  long=power spectral density,
}

\DeclareAcronym{rrt}{
  short=RRT,
  long=rapid-response team,
}

\DeclareAcronym{sgwb}{
  short=SGWB,
  long=stochastic gravitational-wave background,
}

\DeclareAcronym{sneb}{
  short=SNEB,
  long=suspended north-end bench,
}

\DeclareAcronym{snr}{
  short=SNR,
  long=signal-to-noise ratio,
}

\DeclareAcronym{sr}{
  short=SR,
  long=signal recycling,
}

\DeclareAcronym{ssfs}{
  short=SSFS,
  long=second-stage frequency stabilization system,
}

\DeclareAcronym{sweb}{
  short=SWEB,
  long=suspended west-end bench,
}

\DeclareAcronym{upv}{
  short=\texttt{UPV},
  long=use-percentage veto,
}

\DeclareAcronym{vim}{
  short=\texttt{VIM},
  long=Virgo Interferometer Monitor,
}

\DeclareAcronym{vpm}{
  short=VPM,
  long=Virgo Process Monitoring
}

\DeclareAcronym{we}{
  short=WE,
  long=west end,
}

\DeclareAcronym{web}{
  short=WEB,
  long=west-end building,
}

\DeclareAcronym{wi}{
  short=WI,
  long=west input,
}

\begin{document}
\leftline{Dated: \today}

\title{Virgo Detector Characterization and Data Quality during the O3 run}

\author{%
F~Acernese$^{1,2}$, 
M~Agathos$^{3}$, 
A~Ain$^{4}$, 
S~Albanesi$^{5,6}$, 
A~Allocca\orcidlink{0000-0002-5288-1351}$^{7,2}$, 
A~Amato\orcidlink{0000-0001-9557-651X}$^{8}$, 
T~Andrade$^{9}$, 
N~Andres\orcidlink{0000-0002-5360-943X}$^{10}$, 
M~Andr\'es-Carcasona\orcidlink{0000-0002-8738-1672}$^{11}$, 
T~Andri\'c\orcidlink{0000-0002-9277-9773}$^{12}$, 
S~Ansoldi$^{13,14}$, 
S~Antier\orcidlink{0000-0002-7686-3334}$^{15,16}$, 
T~Apostolatos$^{17}$, 
E~Z~Appavuravther$^{18,19}$, %
M~Ar\`ene$^{20}$, %
N~Arnaud\orcidlink{0000-0001-6589-8673}$^{21,22}$, 
M~Assiduo$^{23,24}$, 
S~Assis~de~Souza~Melo$^{22}$, 
P~Astone\orcidlink{0000-0003-4981-4120}$^{25}$, 
F~Aubin\orcidlink{0000-0003-1613-3142}$^{24}$, 
S~Babak\orcidlink{0000-0001-7469-4250}$^{20}$, 
F~Badaracco\orcidlink{0000-0001-8553-7904}$^{26}$, 
M~K~M~Bader$^{27}$, %
S~Bagnasco\orcidlink{0000-0001-6062-6505}$^{6}$, 
J~Baird$^{20}$, %
T~Baka$^{28}$, 
G~Ballardin$^{22}$, 
G~Baltus\orcidlink{0000-0002-0304-8152}$^{29}$, 
B~Banerjee\orcidlink{0000-0002-8008-2485}$^{12}$, 
C~Barbieri$^{30,31,32}$, %
P~Barneo\orcidlink{0000-0002-8883-7280}$^{9}$, 
F~Barone\orcidlink{0000-0002-8069-8490}$^{33,2}$, 
M~Barsuglia\orcidlink{0000-0002-1180-4050}$^{20}$, 
D~Barta\orcidlink{0000-0001-6841-550X}$^{34}$, %
A~Basti$^{35,4}$, 
M~Bawaj\orcidlink{0000-0003-3611-3042}$^{18,36}$, 
M~Bazzan$^{37,38}$, 
F~Beirnaert\orcidlink{0000-0002-4003-7233}$^{39}$, 
M~Bejger\orcidlink{0000-0002-4991-8213}$^{40}$, 
I~Belahcene$^{21}$, %
V~Benedetto$^{41}$, %
M~Berbel\orcidlink{0000-0001-6345-1798}$^{42}$, 
S~Bernuzzi\orcidlink{0000-0002-2334-0935}$^{3}$, 
D~Bersanetti\orcidlink{0000-0002-7377-415X}$^{43}$, 
A~Bertolini$^{27}$, 
U~Bhardwaj\orcidlink{0000-0003-1233-4174}$^{16,27}$, 
A~Bianchi$^{27,44}$, 
S~Bini$^{45,46}$, 
M~Bischi$^{23,24}$, 
M~Bitossi$^{22,4}$, 
M-A~Bizouard\orcidlink{0000-0002-4618-1674}$^{15}$, 
F~Bobba$^{47,48}$, 
M~Bo\"{e}r$^{15}$, 
G~Bogaert$^{15}$, 
M~Boldrini$^{49,25}$, 
L~D~Bonavena$^{37}$, 
F~Bondu$^{50}$, 
R~Bonnand\orcidlink{0000-0001-5013-5913}$^{10}$, 
B~A~Boom$^{27}$, %
V~Boschi\orcidlink{0000-0001-8665-2293}$^{4}$, 
V~Boudart\orcidlink{0000-0001-9923-4154}$^{29}$, 
Y~Bouffanais$^{37,38}$, 
A~Bozzi$^{22}$, 
C~Bradaschia$^{4}$, 
M~Branchesi\orcidlink{0000-0003-1643-0526}$^{12,51}$, 
M~Breschi\orcidlink{0000-0002-3327-3676}$^{3}$, 
T~Briant\orcidlink{0000-0002-6013-1729}$^{52}$, 
A~Brillet$^{15}$, 
J~Brooks$^{22}$, %
G~Bruno$^{26}$, 
F~Bucci$^{24}$, 
T~Bulik$^{53}$, 
H~J~Bulten$^{27}$, 
D~Buskulic$^{10}$, 
C~Buy\orcidlink{0000-0003-2872-8186}$^{54}$, 
G~S~Cabourn~Davies\orcidlink{0000-0002-4289-3439}$^{55}$, %
G~Cabras\orcidlink{0000-0002-6852-6856}$^{13,14}$, 
R~Cabrita\orcidlink{0000-0003-0133-1306}$^{26}$, 
G~Cagnoli\orcidlink{0000-0002-7086-6550}$^{8}$, 
E~Calloni$^{7,2}$, 
M~Canepa$^{56,43}$, 
S~Canevarolo$^{28}$, 
M~Cannavacciuolo$^{47}$, %
E~Capocasa\orcidlink{0000-0003-3762-6958}$^{20}$, 
G~Carapella$^{47,48}$, 
F~Carbognani$^{22}$, 
M~Carpinelli$^{57,58,22}$, 
G~Carullo\orcidlink{0000-0001-9090-1862}$^{35,4}$, 
J~Casanueva~Diaz$^{22}$, 
C~Casentini$^{59,60}$, 
S~Caudill$^{27,28}$, 
F~Cavalier\orcidlink{0000-0002-3658-7240}$^{21}$, %
R~Cavalieri\orcidlink{0000-0001-6064-0569}$^{22}$, 
G~Cella\orcidlink{0000-0002-0752-0338}$^{4}$, 
P~Cerd\'a-Dur\'an$^{61}$, 
E~Cesarini\orcidlink{0000-0001-9127-3167}$^{60}$, 
W~Chaibi$^{15}$, 
P~Chanial\orcidlink{0000-0003-1753-524X}$^{22}$, %
E~Chassande-Mottin\orcidlink{0000-0003-3768-9908}$^{20}$, 
S~Chaty\orcidlink{0000-0002-5769-8601}$^{20}$, 
F~Chiadini\orcidlink{0000-0002-9339-8622}$^{62,48}$, 
G~Chiarini$^{38}$, 
R~Chierici$^{63}$, 
A~Chincarini\orcidlink{0000-0003-4094-9942}$^{43}$, 
M~L~Chiofalo$^{35,4}$, 
A~Chiummo\orcidlink{0000-0003-2165-2967}$^{22}$, 
S~Choudhary\orcidlink{0000-0003-0949-7298}$^{64}$, %
N~Christensen\orcidlink{0000-0002-6870-4202}$^{15}$, 
G~Ciani\orcidlink{0000-0003-4258-9338}$^{37,38}$, 
P~Ciecielag$^{40}$, 
M~Cie\'slar\orcidlink{0000-0001-8912-5587}$^{40}$, 
M~Cifaldi$^{59,60}$, 
R~Ciolfi\orcidlink{0000-0003-3140-8933}$^{65,38}$, 
F~Cipriano$^{15}$, %
S~Clesse$^{66}$, 
F~Cleva$^{15}$, 
E~Coccia$^{12,51}$, 
E~Codazzo\orcidlink{0000-0001-7170-8733}$^{12}$, 
P-F~Cohadon\orcidlink{0000-0003-3452-9415}$^{52}$, 
D~E~Cohen\orcidlink{0000-0002-0583-9919}$^{21}$, %
A~Colombo\orcidlink{0000-0002-7439-4773}$^{30,31}$, 
M~Colpi$^{30,31}$, 
L~Conti\orcidlink{0000-0003-2731-2656}$^{38}$, 
I~Cordero-Carri\'on\orcidlink{0000-0002-1985-1361}$^{67}$, 
S~Corezzi$^{36,18}$, 
D~Corre$^{21}$, %
S~Cortese\orcidlink{0000-0002-6504-0973}$^{22}$, 
J-P~Coulon$^{15}$, 
M~Croquette\orcidlink{0000-0002-8581-5393}$^{52}$, %
J~R~Cudell\orcidlink{0000-0002-2003-4238}$^{29}$, 
E~Cuoco$^{22,68,4}$, 
M~Cury{\l}o$^{53}$, 
P~Dabadie$^{8}$, %
T~Dal~Canton\orcidlink{0000-0001-5078-9044}$^{21}$, 
S~Dall'Osso\orcidlink{0000-0003-4366-8265}$^{12}$, %
G~D\'alya\orcidlink{0000-0003-3258-5763}$^{39}$, 
B~D'Angelo\orcidlink{0000-0001-9143-8427}$^{56,43}$, 
S~Danilishin\orcidlink{0000-0001-7758-7493}$^{69,27}$, 
S~D'Antonio$^{60}$, 
V~Dattilo$^{22}$, 
M~Davier$^{21}$, 
D~Davis\orcidlink{0000-0001-5620-6751}$^{70}$, %
J~Degallaix\orcidlink{0000-0002-1019-6911}$^{71}$, 
M~De~Laurentis$^{7,2}$, 
S~Del\'eglise\orcidlink{0000-0002-8680-5170}$^{52}$, 
F~De~Lillo\orcidlink{0000-0003-4977-0789}$^{26}$, 
D~Dell'Aquila\orcidlink{0000-0001-5895-0664}$^{57}$, 
W~Del~Pozzo$^{35,4}$, 
F~De~Matteis$^{59,60}$, %
A~Depasse\orcidlink{0000-0003-1014-8394}$^{26}$, 
R~De~Pietri\orcidlink{0000-0003-1556-8304}$^{72,73}$, 
R~De~Rosa\orcidlink{0000-0002-4004-947X}$^{7,2}$, 
C~De~Rossi$^{22}$, 
R~De~Simone$^{62}$, %
L~Di~Fiore$^{2}$, 
C~Di~Giorgio\orcidlink{0000-0003-2127-3991}$^{47,48}$, %
F~Di~Giovanni\orcidlink{0000-0001-8568-9334}$^{61}$, 
M~Di~Giovanni$^{12}$, 
T~Di~Girolamo\orcidlink{0000-0003-2339-4471}$^{7,2}$, 
A~Di~Lieto\orcidlink{0000-0002-4787-0754}$^{35,4}$, 
A~Di~Michele\orcidlink{0000-0002-0357-2608}$^{36}$, %
S~Di~Pace\orcidlink{0000-0001-6759-5676}$^{49,25}$, 
I~Di~Palma\orcidlink{0000-0003-1544-8943}$^{49,25}$, 
F~Di~Renzo\orcidlink{0000-0002-5447-3810}$^{35,4}$, 
L~D'Onofrio\orcidlink{0000-0001-9546-5959}$^{7,2}$, %
M~Drago\orcidlink{0000-0002-3738-2431}$^{49,25}$, 
J-G~Ducoin$^{21}$, 
U~Dupletsa$^{12}$, 
O~Durante$^{47,48}$, 
D~D'Urso\orcidlink{0000-0002-8215-4542}$^{57,58}$, 
P-A~Duverne$^{21}$, 
M~Eisenmann$^{10}$, %
L~Errico$^{7,2}$, 
D~Estevez\orcidlink{0000-0002-3021-5964}$^{74}$, 
F~Fabrizi\orcidlink{0000-0002-3809-065X}$^{23,24}$, 
F~Faedi$^{24}$, 
V~Fafone\orcidlink{0000-0003-1314-1622}$^{59,60,12}$, 
S~Farinon$^{43}$, %
G~Favaro\orcidlink{0000-0002-0351-6833}$^{37}$, 
M~Fays\orcidlink{0000-0002-4390-9746}$^{29}$, 
E~Fenyvesi\orcidlink{0000-0003-2777-3719}$^{34,75}$, %
I~Ferrante\orcidlink{0000-0002-0083-7228}$^{35,4}$, 
F~Fidecaro\orcidlink{0000-0002-6189-3311}$^{35,4}$, 
P~Figura\orcidlink{0000-0002-8925-0393}$^{53}$, 
A~Fiori\orcidlink{0000-0003-3174-0688}$^{4,35}$, 
I~Fiori\orcidlink{0000-0002-0210-516X}$^{22}$, 
R~Fittipaldi$^{76,48}$, %
V~Fiumara$^{77,48}$, %
R~Flaminio$^{10,78}$, 
J~A~Font\orcidlink{0000-0001-6650-2634}$^{61,79}$, 
S~Frasca$^{49,25}$, 
F~Frasconi\orcidlink{0000-0003-4204-6587}$^{4}$, 
A~Freise\orcidlink{0000-0001-6586-9901}$^{27,44}$, 
O~Freitas$^{80}$, 
G~G~Fronz\'e\orcidlink{0000-0003-0966-4279}$^{6}$, 
B~U~Gadre\orcidlink{0000-0002-1534-9761}$^{81,28}$, %
R~Gamba$^{3}$, 
B~Garaventa\orcidlink{0000-0003-2490-404X}$^{43,56}$, 
F~Garufi\orcidlink{0000-0003-1391-6168}$^{7,2}$, 
G~Gemme\orcidlink{0000-0002-1127-7406}$^{43}$, 
A~Gennai\orcidlink{0000-0003-0149-2089}$^{4}$, 
Archisman~Ghosh\orcidlink{0000-0003-0423-3533}$^{39}$, 
B~Giacomazzo\orcidlink{0000-0002-6947-4023}$^{30,31,32}$, 
L~Giacoppo$^{49,25}$, %
P~Giri\orcidlink{0000-0002-4628-2432}$^{4,35}$, 
F~Gissi$^{41}$, %
S~Gkaitatzis\orcidlink{0000-0001-9420-7499}$^{4,35}$, %
B~Goncharov\orcidlink{0000-0003-3189-5807}$^{12}$, 
M~Gosselin$^{22}$, 
R~Gouaty$^{10}$, 
A~Grado\orcidlink{0000-0002-0501-8256}$^{82,2}$, 
M~Granata\orcidlink{0000-0003-3275-1186}$^{71}$, 
V~Granata$^{47}$, 
G~Greco$^{18}$, 
G~Grignani$^{36,18}$, 
A~Grimaldi\orcidlink{0000-0002-6956-4301}$^{45,46}$, 
S~J~Grimm$^{12,51}$, %
P~Gruning$^{21}$, %
D~Guerra\orcidlink{0000-0003-0029-5390}$^{61}$, 
G~M~Guidi\orcidlink{0000-0002-3061-9870}$^{23,24}$, 
G~Guix\'e$^{9}$, 
Y~Guo$^{27}$, 
P~Gupta$^{27,28}$, 
L~Haegel\orcidlink{0000-0002-3680-5519}$^{20}$, 
O~Halim\orcidlink{0000-0003-1326-5481}$^{14}$, 
O~Hannuksela$^{28,27}$, 
T~Harder$^{15}$, 
K~Haris$^{27,28}$, 
J~Harms\orcidlink{0000-0002-7332-9806}$^{12,51}$, 
B~Haskell$^{40}$, 
A~Heidmann\orcidlink{0000-0002-0784-5175}$^{52}$, 
H~Heitmann\orcidlink{0000-0003-0625-5461}$^{15}$, 
P~Hello$^{21}$, 
G~Hemming\orcidlink{0000-0001-5268-4465}$^{22}$, 
E~Hennes\orcidlink{0000-0002-2246-5496}$^{27}$, 
S~Hild$^{69,27}$, 
D~Hofman$^{71}$, %
V~Hui\orcidlink{0000-0002-0233-2346}$^{10}$, 
B~Idzkowski\orcidlink{0000-0001-5869-2714}$^{53}$, 
A~Iess$^{59,60}$, %
P~Iosif\orcidlink{0000-0003-1621-7709}$^{83}$, 
T~Jacqmin\orcidlink{0000-0002-0693-4838}$^{52}$, 
P-E~Jacquet\orcidlink{0000-0001-9552-0057}$^{52}$, %
S~P~Jadhav$^{64}$, %
J~Janquart$^{28,27}$, 
K~Janssens\orcidlink{0000-0001-8760-4429}$^{84,15}$, 
P~Jaranowski\orcidlink{0000-0001-8085-3414}$^{85}$, 
V~Juste$^{74}$, 
C~Kalaghatgi$^{28,27,86}$, 
C~Karathanasis\orcidlink{0000-0002-0642-5507}$^{11}$, 
S~Katsanevas\orcidlink{0000-0003-0324-0758}$^{22}$, 
F~K\'ef\'elian$^{15}$, 
N~Khetan$^{12,51}$, %
G~Koekoek$^{27,69}$, 
S~Koley\orcidlink{0000-0002-5793-6665}$^{12}$, 
M~Kolstein\orcidlink{0000-0002-5482-6743}$^{11}$, 
A~Kr\'olak\orcidlink{0000-0003-4514-7690}$^{87,88}$, 
P~Kuijer\orcidlink{0000-0002-6987-2048}$^{27}$, 
P~Lagabbe$^{10}$, 
D~Laghi\orcidlink{0000-0001-7462-3794}$^{54}$, 
M~Lalleman$^{84}$, 
A~Lamberts$^{15,89}$, 
I~La~Rosa$^{10}$, 
A~Lartaux-Vollard$^{21}$, %
C~Lazzaro$^{37,38}$, 
P~Leaci\orcidlink{0000-0002-3997-5046}$^{49,25}$, 
A~Lema{\^i}tre$^{90}$, 
M~Lenti\orcidlink{0000-0002-2765-3955}$^{24,91}$, 
E~Leonova$^{16}$, %
N~Leroy\orcidlink{0000-0002-2321-1017}$^{21}$, 
N~Letendre$^{10}$, 
K~Leyde$^{20}$, 
F~Linde$^{86,27}$, 
L~London$^{16}$, 
A~Longo\orcidlink{0000-0003-4254-8579}$^{92}$, 
M~Lopez~Portilla$^{28}$, %
M~Lorenzini\orcidlink{0000-0002-2765-7905}$^{59,60}$, 
V~Loriette$^{93}$, 
G~Losurdo\orcidlink{0000-0003-0452-746X}$^{4}$, 
D~Lumaca\orcidlink{0000-0002-3628-1591}$^{59,60}$, 
A~Macquet$^{15}$, 
C~Magazz\`u\orcidlink{0000-0002-9913-381X}$^{4}$, %
M~Magnozzi\orcidlink{0000-0003-4512-8430}$^{43,56}$, 
E~Majorana$^{49,25}$, 
I~Maksimovic$^{93}$, %
N~Man$^{15}$, 
V~Mangano\orcidlink{0000-0001-7902-8505}$^{49,25}$, 
M~Mantovani\orcidlink{0000-0002-4424-5726}$^{22}$, 
M~Mapelli\orcidlink{0000-0001-8799-2548}$^{37,38}$, 
F~Marchesoni$^{19,18,94}$, 
D~Mar\'{\i}n~Pina\orcidlink{0000-0001-6482-1842}$^{9}$, 
F~Marion$^{10}$, 
A~Marquina$^{67}$, 
S~Marsat\orcidlink{0000-0001-9449-1071}$^{20}$, 
F~Martelli$^{23,24}$, 
M~Martinez$^{11}$, 
V~Martinez$^{8}$, 
A~Masserot$^{10}$, 
S~Mastrogiovanni\orcidlink{0000-0003-1606-4183}$^{20}$, 
Q~Meijer$^{28}$, 
A~Menendez-Vazquez$^{11}$, %
L~Mereni$^{71}$, 
M~Merzougui$^{15}$, %
A~Miani\orcidlink{0000-0001-7737-3129}$^{45,46}$, 
C~Michel\orcidlink{0000-0003-0606-725X}$^{71}$, 
L~Milano$^{7}$\footnote{Deceased, April 2021.}, %
A~Miller$^{26}$, 
B~Miller$^{16,27}$, %
E~Milotti$^{95,14}$, 
Y~Minenkov$^{60}$, 
Ll~M~Mir$^{11}$, 
M~Miravet-Ten\'es\orcidlink{0000-0002-8766-1156}$^{61}$, 
M~Montani$^{23,24}$, 
F~Morawski$^{40}$, 
B~Mours\orcidlink{0000-0002-6444-6402}$^{74}$, 
C~M~Mow-Lowry\orcidlink{0000-0002-0351-4555}$^{27,44}$
% ...
}%
\address{$^{1}$Dipartimento di Farmacia, Universit\`a di Salerno, I-84084 Fisciano, Salerno, Italy}
\address{$^{2}$INFN, Sezione di Napoli, Complesso Universitario di Monte S. Angelo, I-80126 Napoli, Italy}
\address{$^{3}$Theoretisch-Physikalisches Institut, Friedrich-Schiller-Universit\"at Jena, D-07743 Jena, Germany}
\address{$^{4}$INFN, Sezione di Pisa, I-56127 Pisa, Italy}
\address{$^{5}$Dipartimento di Fisica, Universit\`a degli Studi di Torino, I-10125 Torino, Italy}
\address{$^{6}$INFN Sezione di Torino, I-10125 Torino, Italy}
\address{$^{7}$Universit\`a di Napoli ``Federico II'', Complesso Universitario di Monte S. Angelo, I-80126 Napoli, Italy}
\address{$^{8}$Universit\'e de Lyon, Universit\'e Claude Bernard Lyon 1, CNRS, Institut Lumi\`ere Mati\`ere, F-69622 Villeurbanne, France}
\address{$^{9}$Institut de Ci\`encies del Cosmos (ICCUB), Universitat de Barcelona, C/ Mart\'{\i} i Franqu\`es 1, Barcelona, 08028, Spain}
\address{$^{10}$Univ. Savoie Mont Blanc, CNRS, Laboratoire d'Annecy de Physique des Particules - IN2P3, F-74000 Annecy, France}
\address{$^{11}$Institut de F\'{\i}sica d'Altes Energies (IFAE), Barcelona Institute of Science and Technology, and  ICREA, E-08193 Barcelona, Spain}
\address{$^{12}$Gran Sasso Science Institute (GSSI), I-67100 L'Aquila, Italy}
\address{$^{13}$Dipartimento di Scienze Matematiche, Informatiche e Fisiche, Universit\`a di Udine, I-33100 Udine, Italy}
\address{$^{14}$INFN, Sezione di Trieste, I-34127 Trieste, Italy}
\address{$^{15}$Artemis, Universit\'e C\^ote d'Azur, Observatoire de la C\^ote d'Azur, CNRS, F-06304 Nice, France}
\address{$^{16}$GRAPPA, Anton Pannekoek Institute for Astronomy and Institute for High-Energy Physics, University of Amsterdam, Science Park 904, 1098 XH Amsterdam, Netherlands}
\address{$^{17}$Department of Physics, National and Kapodistrian University of Athens, School of Science Building, 2nd floor, Panepistimiopolis, 15771 Ilissia, Greece}
\address{$^{18}$INFN, Sezione di Perugia, I-06123 Perugia, Italy}
\address{$^{19}$Universit\`a di Camerino, I-62032 Camerino, Italy}
\address{$^{20}$Universit\'e de Paris, CNRS, Astroparticule et Cosmologie, F-75006 Paris, France}
\address{$^{21}$Universit\'e Paris-Saclay, CNRS/IN2P3, IJCLab, 91405 Orsay, France}
\address{$^{22}$European Gravitational Observatory (EGO), I-56021 Cascina, Pisa, Italy}
\address{$^{23}$Universit\`a degli Studi di Urbino ``Carlo Bo'', I-61029 Urbino, Italy}
\address{$^{24}$INFN, Sezione di Firenze, I-50019 Sesto Fiorentino, Firenze, Italy}
\address{$^{25}$INFN, Sezione di Roma, I-00185 Roma, Italy}
\address{$^{26}$Universit\'e catholique de Louvain, B-1348 Louvain-la-Neuve, Belgium}
\address{$^{27}$Nikhef, Science Park 105, 1098 XG Amsterdam, Netherlands}
\address{$^{28}$Institute for Gravitational and Subatomic Physics (GRASP), Utrecht University, Princetonplein 1, 3584 CC Utrecht, Netherlands}
\address{$^{29}$Universit\'e de Li\`ege, B-4000 Li\`ege, Belgium}
\address{$^{30}$Universit\`a degli Studi di Milano-Bicocca, I-20126 Milano, Italy}
\address{$^{31}$INFN, Sezione di Milano-Bicocca, I-20126 Milano, Italy}
\address{$^{32}$INAF, Osservatorio Astronomico di Brera sede di Merate, I-23807 Merate, Lecco, Italy}
\address{$^{33}$Dipartimento di Medicina, Chirurgia e Odontoiatria ``Scuola Medica Salernitana'', Universit\`a di Salerno, I-84081 Baronissi, Salerno, Italy}
\address{$^{34}$Wigner RCP, RMKI, H-1121 Budapest, Konkoly Thege Mikl\'os \'ut 29-33, Hungary}
\address{$^{35}$Universit\`a di Pisa, I-56127 Pisa, Italy}
\address{$^{36}$Universit\`a di Perugia, I-06123 Perugia, Italy}
\address{$^{37}$Universit\`a di Padova, Dipartimento di Fisica e Astronomia, I-35131 Padova, Italy}
\address{$^{38}$INFN, Sezione di Padova, I-35131 Padova, Italy}
\address{$^{39}$Universiteit Gent, B-9000 Gent, Belgium}
\address{$^{40}$Nicolaus Copernicus Astronomical Center, Polish Academy of Sciences, 00-716, Warsaw, Poland}
\address{$^{41}$Dipartimento di Ingegneria, Universit\`a del Sannio, I-82100 Benevento, Italy}
\address{$^{42}$Departamento de Matem\'aticas, Universitat Aut\`onoma de Barcelona, Edificio C Facultad de Ciencias 08193 Bellaterra (Barcelona), Spain}
\address{$^{43}$INFN, Sezione di Genova, I-16146 Genova, Italy}
\address{$^{44}$Vrije Universiteit Amsterdam, 1081 HV Amsterdam, Netherlands}
\address{$^{45}$Universit\`a di Trento, Dipartimento di Fisica, I-38123 Povo, Trento, Italy}
\address{$^{46}$INFN, Trento Institute for Fundamental Physics and Applications, I-38123 Povo, Trento, Italy}
\address{$^{47}$Dipartimento di Fisica ``E.R. Caianiello'', Universit\`a di Salerno, I-84084 Fisciano, Salerno, Italy}
\address{$^{48}$INFN, Sezione di Napoli, Gruppo Collegato di Salerno, Complesso Universitario di Monte S. Angelo, I-80126 Napoli, Italy}
\address{$^{49}$Universit\`a di Roma ``La Sapienza'', I-00185 Roma, Italy}
\address{$^{50}$Univ Rennes, CNRS, Institut FOTON - UMR6082, F-3500 Rennes, France}
\address{$^{51}$INFN, Laboratori Nazionali del Gran Sasso, I-67100 Assergi, Italy}
\address{$^{52}$Laboratoire Kastler Brossel, Sorbonne Universit\'e, CNRS, ENS-Universit\'e PSL, Coll\`ege de France, F-75005 Paris, France}
\address{$^{53}$Astronomical Observatory Warsaw University, 00-478 Warsaw, Poland}
\address{$^{54}$L2IT, Laboratoire des 2 Infinis - Toulouse, Universit\'e de Toulouse, CNRS/IN2P3, UPS, F-31062 Toulouse Cedex 9, France}
\address{$^{55}$University of Portsmouth, Portsmouth, PO1 3FX, United Kingdom}
\address{$^{56}$Dipartimento di Fisica, Universit\`a degli Studi di Genova, I-16146 Genova, Italy}
\address{$^{57}$Universit\`a degli Studi di Sassari, I-07100 Sassari, Italy}
\address{$^{58}$INFN, Laboratori Nazionali del Sud, I-95125 Catania, Italy}
\address{$^{59}$Universit\`a di Roma Tor Vergata, I-00133 Roma, Italy}
\address{$^{60}$INFN, Sezione di Roma Tor Vergata, I-00133 Roma, Italy}
\address{$^{61}$Departamento de Astronom\'{\i}a y Astrof\'{\i}sica, Universitat de Val\`encia, E-46100 Burjassot, Val\`encia, Spain}
\address{$^{62}$Dipartimento di Ingegneria Industriale (DIIN), Universit\`a di Salerno, I-84084 Fisciano, Salerno, Italy}
\address{$^{63}$Universit\'e Lyon, Universit\'e Claude Bernard Lyon 1, CNRS, IP2I Lyon / IN2P3, UMR 5822, F-69622 Villeurbanne, France}
\address{$^{64}$Inter-University Centre for Astronomy and Astrophysics, Post Bag 4, Ganeshkhind, Pune 411 007, India}
\address{$^{65}$INAF, Osservatorio Astronomico di Padova, I-35122 Padova, Italy}
\address{$^{66}$Universit\'e libre de Bruxelles, Avenue Franklin Roosevelt 50 - 1050 Bruxelles, Belgium}
\address{$^{67}$Departamento de Matem\'aticas, Universitat de Val\`encia, E-46100 Burjassot, Val\`encia, Spain}
\address{$^{68}$Scuola Normale Superiore, Piazza dei Cavalieri, 7 - 56126 Pisa, Italy}
\address{$^{69}$Maastricht University, P.O. Box 616, 6200 MD Maastricht, Netherlands}
\address{$^{70}$LIGO Laboratory, California Institute of Technology, Pasadena, CA 91125, USA}
\address{$^{71}$Universit\'e Lyon, Universit\'e Claude Bernard Lyon 1, CNRS, Laboratoire des Mat\'eriaux Avanc\'es (LMA), IP2I Lyon / IN2P3, UMR 5822, F-69622 Villeurbanne, France}
\address{$^{72}$Dipartimento di Scienze Matematiche, Fisiche e Informatiche, Universit\`a di Parma, I-43124 Parma, Italy}
\address{$^{73}$INFN, Sezione di Milano Bicocca, Gruppo Collegato di Parma, I-43124 Parma, Italy}
\address{$^{74}$Universit\'e de Strasbourg, CNRS, IPHC UMR 7178, F-67000 Strasbourg, France}
\address{$^{75}$Institute for Nuclear Research, Bem t'er 18/c, H-4026 Debrecen, Hungary}
\address{$^{76}$CNR-SPIN, c/o Universit\`a di Salerno, I-84084 Fisciano, Salerno, Italy}
\address{$^{77}$Scuola di Ingegneria, Universit\`a della Basilicata, I-85100 Potenza, Italy}
\address{$^{78}$Gravitational Wave Science Project, National Astronomical Observatory of Japan (NAOJ), Mitaka City, Tokyo 181-8588, Japan}
\address{$^{79}$Observatori Astron\`omic, Universitat de Val\`encia, E-46980 Paterna, Val\`encia, Spain}
\address{$^{80}$Centro de F\'{\i}sica das Universidades do Minho e do Porto, Universidade do Minho, Campus de Gualtar, PT-4710 - 057 Braga, Portugal}
\address{$^{81}$Max Planck Institute for Gravitational Physics (Albert Einstein Institute), D-14476 Potsdam, Germany}
\address{$^{82}$INAF, Osservatorio Astronomico di Capodimonte, I-80131 Napoli, Italy}
\address{$^{83}$Department of Physics, Aristotle University of Thessaloniki, University Campus, 54124 Thessaloniki, Greece}
\address{$^{84}$Universiteit Antwerpen, Prinsstraat 13, 2000 Antwerpen, Belgium}
\address{$^{85}$University of Bia{\l}ystok, 15-424 Bia{\l}ystok, Poland}
\address{$^{86}$Institute for High-Energy Physics, University of Amsterdam, Science Park 904, 1098 XH Amsterdam, Netherlands}
\address{$^{87}$Institute of Mathematics, Polish Academy of Sciences, 00656 Warsaw, Poland}
\address{$^{88}$National Center for Nuclear Research, 05-400 {\' S}wierk-Otwock, Poland}
\address{$^{89}$Laboratoire Lagrange, Universit\'e C\^ote d'Azur, Observatoire C\^ote d'Azur, CNRS, F-06304 Nice, France}
\address{$^{90}$NAVIER, \'{E}cole des Ponts, Univ Gustave Eiffel, CNRS, Marne-la-Vall\'{e}e, France}
\address{$^{91}$Universit\`a di Firenze, Sesto Fiorentino I-50019, Italy}
\address{$^{92}$INFN, Sezione di Roma Tre, I-00146 Roma, Italy}
\address{$^{93}$ESPCI, CNRS, F-75005 Paris, France}
\address{$^{94}$School of Physics Science and Engineering, Tongji University, Shanghai 200092, China}
\address{$^{95}$Dipartimento di Fisica, Universit\`a di Trieste, I-34127 Trieste, Italy}
% ...

\begin{abstract}

The Advanced Virgo detector has contributed with its data to the rapid growth
of the number of detected gravitational-wave signals in the past few years,
alongside the two LIGO instruments.
First, during the last month of the Observation Run 2 (O2) in August 2017 (with,
most notably, the compact binary mergers GW170814 and GW170817) and then during
the full Observation Run 3 (O3): an 11 months data taking period, between April
2019 and March 2020, that led to the addition of about 80 events to the catalog
of transient gravitational-wave sources maintained by LIGO, Virgo and KAGRA.
These discoveries and the manifold exploitation of the detected waveforms
require an accurate characterization of the quality of the data, such as
continuous study and monitoring of the detector noise.
These activities, collectively named {\em detector characterization} or
{\em DetChar}, span the whole workflow of the Virgo data, from the instrument
front-end to the final analysis.
They are described in details in the following article, with a focus on the
associated tools, the results achieved by the Virgo DetChar group during the O3
run and the main prospects for future data-taking periods with an improved
detector.

\end{abstract}

\maketitle

\tableofcontents

\clearpage

\setlength{\parindent}{0pt}
\setlength{\parskip}{\medskipamount}

\section{Introduction}
\markboth{\thesection. \Sectionname}{}

A century after being predicted by Albert Einstein in the framework of general relativity, \acp{gw} have been detected by a global network of ground-based interferometric detectors~\cite{Abbott:2016blz}.
The LIGO~\cite{TheLIGOScientific:2014jea} and Virgo~\cite{TheVirgo:2014hva} collaborations, now joined by the KAGRA~\cite{10.1093/ptep/ptab018} collaboration, have observed in the past six years dozens of \ac{gw} signals coming from merging compact binary systems.
Compact binaries composed of two \acp{bh}, two \acp{ns}, or both kinds of compact object have all been observed so far.
GW150914~\cite{Abbott:2016blz}, the first \ac{gw} signal ever detected (at that time by the two Advanced LIGO detectors) was a binary \ac{bh} merger.
Two years later, shortly after the \ac{adv} detector had started operating, the 3-interferometer network detected the signal GW170817~\cite{TheLIGOScientific:2017qsa}, emitted by the fusion of two \acp{ns} and associated with counterparts in the entire electromagnetic spectrum, leading to the birth of multi-messenger astronomy with \acp{gw}.
More recently, LIGO, Virgo and KAGRA (``LVK'') have announced the first detections of \ac{ns}-\ac{bh} mergers in data taken in January 2020~\cite{NSBHDiscovery}.

All events add up in a \ac{gw} Transient Catalog whose first four versions--- GWTC-1~\cite{LIGOScientific:2018mvr}, GWTC-2~\cite{Abbott:2020niy}, GWTC-2.1~\cite{GWTC2p1} and GWTC-3~\cite{GWTC3}--- have been released.
Such a catalog allows scientists to go beyond the detections themselves and probe the populations of compact stars, estimate the merger rate of binary systems, test general relativity in the strong-field regime, and perform searches for counterparts using archival data from other observatories.
The reconstructed \ac{gw} strain data---the so-called $h(t)$ streams---are regularly released in chunks of several months on the \ac{gwosc} website~\cite{GWOSC}.

Producing these results requires a thorough characterization of the data quality, a large part of which involves studying and monitoring the noise of \ac{gw} detectors.
This activity, often referred to as {\em detector characterization} or {\em DetChar}, is an expertise which has been constructed over many years, starting with the initial detectors~\cite{Aasi:2012wd,LIGOScientific:2014qfs}.
Analysis methods and tools have been developed and implemented to characterize the Virgo data both with low latency and offline.
They cover a physics-driven chain that starts from the raw data recorded by the instrument and extends all the way to the final set of \ac{gw} events and the related analysis.
The results of DetChar analyses are used to improve the detector performance during commissioning periods and to maximize the sensitivity to \ac{gw} signals during an {\em Observing Run}, when science-quality data is recorded.

This paper reports the work of the Virgo DetChar group over the past few years, with an emphasis on the preparation for the third LIGO-Virgo observing run, O3 (April 2019--- March 2020), on the activities during O3, as well as on the final results achieved after the run and the experience accumulated in view of the future runs of the LVK network.
These achievements stem from the developments made before and during the O2 run (for Virgo: 25 days of data taking in August 2017) and those will also be described here when appropriate.

Section~\ref{section:AdV} provides an overview of the \ac{adv} detector configuration during the O3 run, preceded by a short summary of the path that led to this data taking period.
The same section also introduces notions and concepts that will be extensively used in the rest of the article, and defines a few related abbreviations.
Then, Section~\ref{section:O3} summarizes the O3 run from a Virgo perspective: how the data taking was organized, what the performance of the detector and the final O3 dataset were.
Section~\ref{section:tools} describes the main DetChar tools, classified by categories: monitoring tools; generic multi-purpose tools; tools to study noise transients; tools to investigate the noise spectrum; and finally, common LIGO-Virgo tools.
The description of each tool has been kept short in the main text; additional information regarding those that are not referenced in the literature is provided in \ref{section:appendix}.
Section~\ref{sec:onlinedq} presents the Virgo online data quality framework built for the O3 run.
Section~\ref{section:public_alerts} deals with the framework developed to vet signals candidate to be released as public alerts to the astronomical community.
Section~\ref{section:dq_studies} presents the main DetChar analyses done on the O3 dataset to study noise transients; their impact on \ac{gw} searches; the noise spectrum; and the final validation of events.
Finally, Section~\ref{section:outlook} provides some information about the ongoing preparation of the future O4 run, that is scheduled to start at the end of 2022.

A list of the main abbreviations used throughout the article is provided as well, for reference.

\section {The Advanced Virgo Detector}
\label{section:AdV}
\markboth{\thesection. \Sectionname}{}

This section focuses on the Virgo detector during the O3 run. First, we briefly review the main steps of the \ac{adv} project up to the beginning of O3. In particular, we emphasize its participation to the last four weeks of the O2 run in August 2017 that were rich of discoveries. Then, we summarize the activities during the 1.5 year-long shutdown between O2 and O3 that allowed the Virgo Collaboration to improve the instrument significantly. Finally, we describe the detector configuration during O3 and  present the main features of the data it has collected.

\subsection{The path to the O3 run}
Virgo~\cite{2012JInst...7.3012A} is an interferometric detector of \acp{gw} located at the European Gravitational Observatory (EGO) in Cascina, Italy. The \ac{adv} project~\cite{TheVirgo:2014hva} allowed to upgrade the original instrument to a second-generation detector, similarly to what LIGO has done with its two interferometers~\cite{TheLIGOScientific:2014jea}, located in Hanford (WA, USA) and Livingston (LA, USA). The funding of \ac{adv} was approved in December 2009 by CNRS (France) and INFN (Italy), with an in-kind contribution from Nikhef (The Netherlands). The decommissioning of the first-generation Virgo detector started in Fall 2011, after the completion of the science run VSR4~\cite{PhysRevD.91.022004}, pursued together with the GEO600 detector~\cite{0264-9381-33-7-075009} (the Advanced LIGO upgrade project had already started).
The installation of the Avanced Virgo equipment started mid-2012 and was completed in 2016. The upgraded interferometer was robustly controlled in March 2017 and the next few months were dedicated to commissioning activities: noise hunting and sensitivity improvement. At the end of July, the detector was very stable and had a sensitivity corresponding to a \ac{bns} range\footnote{The \ac{bns} range is the average distance up to which the merger of a \ac{bns} system can be detected. The average is taken over the source location in the sky and the \ac{bns} system orientation, while a detection is defined as a signal-to-noise ratio of 8 or above} of ${\sim} 30$~Mpc, that is more than a factor two above the performance of the Virgo+ detector during the VSR4 run.

Therefore, \ac{adv} started taking data on August 1 2017, joining the second Observing Run, O2, which had started on November 30, 2016 for the two LIGO interferometers~\cite{0264-9381-32-7-074001}. 
On August 14, 2017, the \ac{adv} detector made its first detection of a \ac{gw}. That event, labelled GW170814, was also recorded by the two LIGO interferometers. It was the first ever triple detection of a binary black hole coalescence, allowing an unprecedented accuracy in the localization of the source in the sky~\cite{2017PhRvL.119n1101A}. A few days later, on August 17, the three interferometers jointly detected, for the first time, a \ac{gw} emitted by the coalescence of two neutron stars~\cite{2017PhRvL.119p1101A}. This event, known as GW170817, was accompanied by the almost simultaneous detection of a gamma-ray burst by the Fermi Gamma-ray and Integral space telescopes~\cite{GW-GRB}. The accuracy in the localization of the \ac{gw} source (approx. 30$\deg^2$) allowed to identify the optical counterpart in the galaxy NGC4993~\cite{MMA}. The O2 run ended on August 25, 2017.

The LIGO-Virgo shutdown between O2 and the third Observing Run, O3, lasted 19 months.
On the Virgo side, it was divided into four periods.
\begin{itemize}
\item A post-O2 commissioning phase, until early December 2017. \\
The goal was twofold: to make a series of measurements on the O2 detector configuration that would have been too invasive during the run and to perform some tests to try to further improve the instrument.
\item Hardware upgrades until mid-March 2018. Four main projects were pursued:
\begin{itemize}
\item {\bf The re-installation of the monolithic suspensions and various vacuum upgrades.} \\
The steel wires, with which the \ac{adv} arm cavity mirrors had been suspended for the O2 run, were replaced with quartz fibers. Monolithic suspensions, successfully tested in the Virgo+ configuration~\cite{doi:10.1063/1.1392338,art4-MSProc}, had been foreseen in the \ac{adv} Technical Design Report~\cite{TDR}. Yet, multiple breakings of fused silica fibers when installed in vacuum were observed in Fall 2016, forcing the recourse of steel wires to preserve the participation of Virgo to the O2 run. The fiber breaking issue was eventually demonstrated to be caused by a spurious dust contamination generated by some vacuum pumps~\cite{Travasso_2018}. Therefore, the Virgo vacuum system was improved in order to avoid contamination with dust particles while the suspension fibers were shielded.
\item {\bf A higher laser power.} \\
The laser power injected into the interferometer was increased, in order to reduce the photon shot noise, limiting the high-frequency sensitivity: 10~W (19~W) were injected in Virgo during the O2 run (at the beginning of the O3 run).
\item {\bf The installation of a squeezed light source.} \\
This allows to further reduce the shot noise limit at high frequencies by modifying the quantum properties of the light coming out of the interferometer~\cite{PhysRevLett.123.231108}.
\item {\bf The test installation of an array of seismic sensors.} \\
An in-depth characterization of the seismic noise field at the test mass locations was performed in order to prepare for the subtraction of the Newtonian noise contribution that may limit the low-frequency sensitivity in the future~\cite{2015LRR....18....3H}.
\end{itemize}
\item A commissioning period, until Fall 2018, to improve the sensitivity and the duty cycle of the detector.
\item Finally, the transition phase to the O3 run, that officially started on April 1st, 2019 at 15:00 UTC.
\end{itemize}

\subsection{The O3 configuration}
The \ac{adv} detector has been designed to achieve a sensitivity about one order of magnitude better than that of the initial Virgo, corresponding to an increase in the detection rate by about three orders of magnitude.
The \ac{adv} design choices were made on the basis of the outcome of the different research and development activities carried out within the \ac{gw} community and the experience gained with Virgo, while also taking into account budget and schedule constraints.

\begin{figure}
  \centering
  \includegraphics[width=0.95\textwidth]{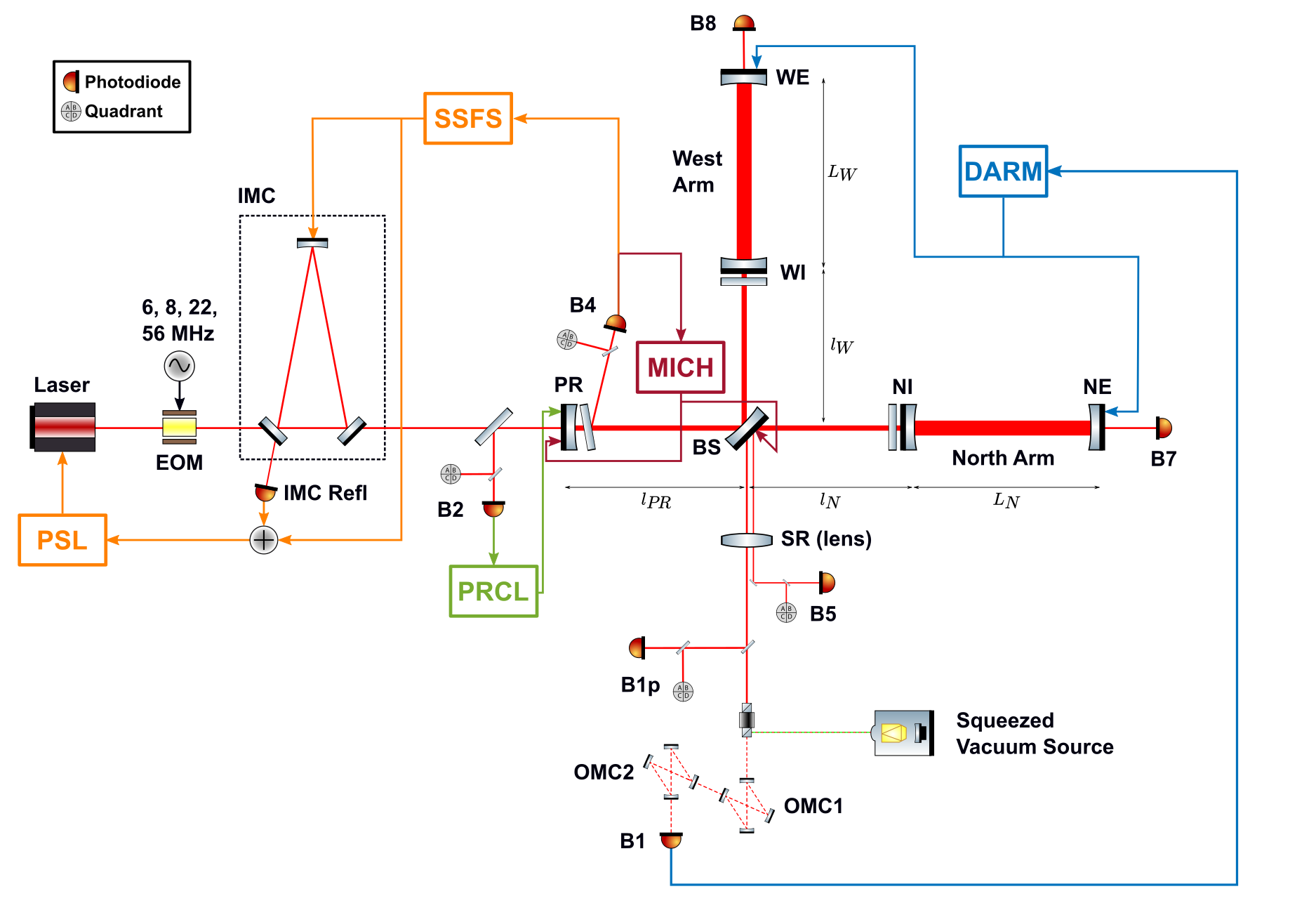}
  \caption{Schematics of the \ac{adv} configuration during the O3 run (not to scale), showing optics, photodiodes and quadrant photodiodes, such as the main components of the global feedback system used to steer the detector. The suspended optical benches introduced in the text are not represented here. This figure is taken from~\cite{galaxies8040085}.}
  \label{fig:AdVsetup}
\end{figure}

The simplified optical schematic of \ac{adv} during the O2 and O3 runs is shown in Figure~\ref{fig:AdVsetup}. In the following, we briefly outline the different parts of the detector layout and define the main abbreviations that are labelled on the schematic or used later in the article. Further information about the O3 configuration and control system of the Virgo detector  can be found in~\cite{galaxies8040085}.

The Virgo power-stabilized infrared laser beam (PSL, wavelength: 1.064~$\mu$m) is filtered at the interferometer input by a 144~m triangular cavity called the \ac{imc}; the two flat mirrors of the \ac{imc} are located on the first suspended injection bench (SIB1), that also hosts various optics for beam matching. Then, the beam goes through the partially reflective \ac{pr} mirror  before being split into two perpendicular beams at the \ac{bs} mirror. The two 3~km-long arms hosting Fabry-Perot cavities are called "North" and "West" as they are roughly oriented along these geographical directions. The cavity mirrors closest (furthest away) from the \ac{bs} are called "input" ("end") mirrors. So, following these conventions, the test masses (the four mirrors forming the two 3-km long Fabry-Perot cavities) are labelled \ac{ni}, \ac{ne}, \ac{wi} and \ac{we}. Both arms end with a suspended terminal bench---called \ac{sneb} or \ac{sweb}---hosting a photodiode (B7 or B8) receiving the cavity transmitted beam. After propagation and storage in the kilometric cavities, the arm beams recombine on the \ac{bs} and the beam resulting from this interference goes to the interferometer output port.
As indicated in Figure~\ref{fig:AdVsetup}, the location of the foreseen \ac{sr} mirror was occupied by the first lens of the detection system during the O3 run (and during O2 as well). The installation of that additional mirror only took place during the shutdown period that followed the end of O3.
Further downstream is the place where the beam from the frequency-indepedent squeezed light source enters the detector. Finally, prior to being detected on the dark fringe port B1 photodiode located on the suspended detection bench 2 (SDB2), the output port beam is filtered in sequence by two \ac{omc} cavities, \ac{omc}1 and \ac{omc}2, located on the suspended detection bench 1 (SDB1).

A complex active feedback system, made of several automated control feedback loops, is necessary to bring the detector to its global working point and maintain it there. In particular, it aims at controlling the four main longitudinal \acp{dof} of the \ac{adv} detector in its O2-O3 configuration that are defined below.
This global control relies on radio-frequency sidebands for the carrier beam that are generated by the \ac{eom} located in between the laser source and the IMC on Figure~\ref{fig:AdVsetup}. The 6, 8 and 56~Mhz sidebands are used to control the interferometer, while the 22~MHz one is used to control the injection system.

\begin{itemize}
\item The \ac{mich}, $l_N - l_W$, sets the destructive interference (`dark fringe') optimal condition.
\item The\ac{prcl}, $l_{PR} + (l_N + l_W) / 2$, that must be resonant.
\item The lengths of the kilometric Fabry-Perot cavities, $L_N$ and $L_W$, that must be resonant as well, or rather their sum and difference that are more physical.
\begin{itemize}
\item The \ac{carm}, $(L_N + L_W) / 2$, used as a length etalon by the \ac{ssfs} to stabilize further the frequency of the input laser.
\item The \ac{darm}, $L_N - L_W$, the quantity sensitive to a passing \ac{gw}.
\end{itemize}
\end{itemize}

\subsection{Virgo data and DetChar products}
\label{section:data_detchar}
The GW strain data stream reconstructed at the Virgo detector is dominated by noise with, up-to-now, rare and weak \ac{gw} signals. That noise results from several contributions that can be roughly classified into two main categories.

\begin{itemize}
\item Fundamental noises, that are inherent to the instrument and represent the ultimate limit of its sensitivity. Their combination is usually Gaussian and stationary, meaning that their properties do not change in time.
\item Various varying noise artifacts, whose origins are manifold (hardware components of the detector, feedback control loops, interaction with the external environment, etc.) and that represent potential issues, not only because they may impact the running of the instrument but also---and above all---because they show up in the background of searches for \acp{gw}, limiting thus their sensitivity. Noise transients, called {\em glitches}, can either look like real signals or overlap in time with one, either imparing its detection or confusing the inference of its source parameters. They are monitored and studied using time-frequency representations that are used to classify their numerous signatures into families and separate them from real \ac{gw} events. In addition, long-lasting noise excesses, also called {\it spectral noises}, are also seen around particular frequencies (power main frequency and its harmonics, suspension resonating modes, etc.): the narrow ones, (nearly) monochromatic, are called {\em lines} and the wider ones {\em bumps}.
Both can manifest themselves in several ``flavours''. For instance, lines can exist individually, but sometimes appear as {\it combs}, that is families of lines separated by a constant frequency interval. They are typically due to processes with a strict time periodicity, like electronic clock signals. Bumps may have some specific structure, depending on the source. Both lines and bumps can exhibit structures symmetric around their main frequency, called {\it sidebands}, that are due to non-linear interactions among different disturbances. Moreover, spectral noises can be persistent across a full run, be present only in a portion of it, or evolve in time. 

 Both the glitch rate in a particular frequency band and the properties (amplitude, peak frequency and bandwidth) of spectral noises can vary in time to reflect changes occuring at the level of the detector or its environment.
\end{itemize}

To allow investigating these variations, hundreds of {\em auxiliary channels} are acquired by the Virgo \ac{daq}, providing both a detailed status of the detector control systems and a complete monitoring of the local environment~\cite{env_hunt,o3virgoenv}. When characterizing the detector or studying the quality of some data, the Virgo DetChar group often singles out integer GPS ranges of interest, that are called {\em segments} in the following.

\subsection{Noise Budget}
\label{sec:tools:noise_budget}
The noise budget compares the measured detector sensitivity with the incoherent sum of all known noise contributions. Each noise projection depends on the noise level, as measured by external probes, and of its coupling to the strain channel $h(t)$, that is estimated by dedicated measurements called noise injections~\cite{EnvHuntVirgoO3}.

The \ac{adv} noise budget is based on the \texttt{SimulinkNb}
\cite{SimulinkNb_git} software package. It includes a complete model
of the four main longitudinal \acp{dof} of the interferometer
(\ac{darm}, \ac{carm}, \ac{mich}, \ac{prcl}), with the interferometer optical response
simulated using \texttt{Optickle}~\cite{optickle_git}, the mirror
suspension approximated with a double pendulum state space model of
the mirror and marionette
(the steel body to which the mirror is suspended, a component of the Virgo suspension's last stage, called payload~\cite{art2-PayloadProceeding}), 
and the feedback response measured from the
transfer function between the photodiode signal and the mirror and
marionette corrections. This approach allows to simply add different
noise sources at their physical entry into the interferometer control
loop, and also includes the expected cross couplings between the
longitudinal \acp{dof}.

This model has been verified to match the measured open loop transfer
functions of the four modeled \acp{dof}, and to reproduce the
interferometer strain data calibration with errors smaller than
10\%. In total more than 100 noise sources are taken into
account, and the total of those noises is summarized in
Figure~\ref{fig:noise_budget_O3b}. The noise is summed in log spaced
frequency bins, which allows resolving narrow lines at low
frequency and a precise representation of broadband noise at high
frequency. The noises taken into account are as follows:

\begin{figure}
  \center
\includegraphics[width=\textwidth]{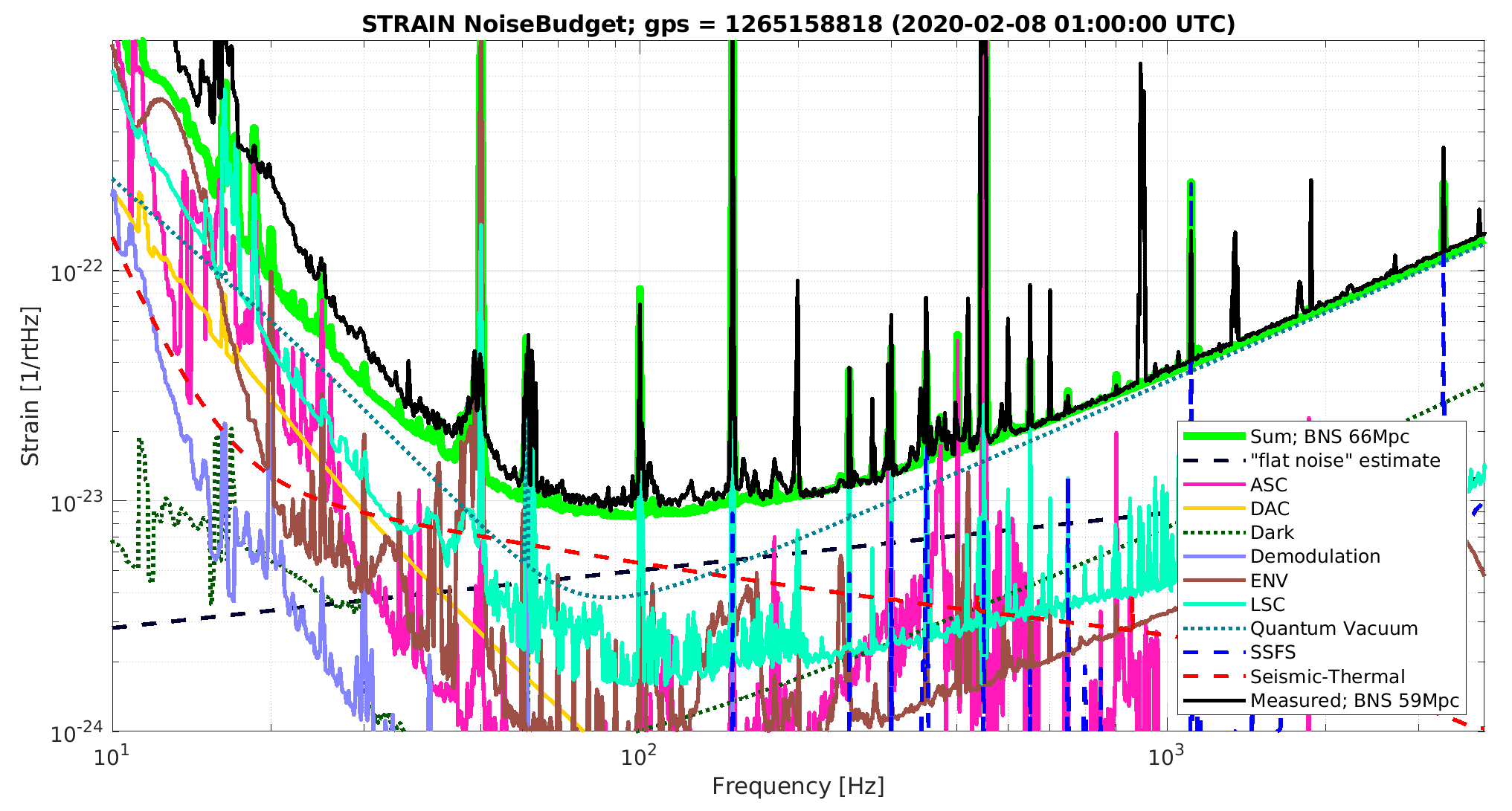}
\caption{Snapshot of the \ac{adv} O3 noise budget generated at a time
    of near best sensitivity of the detector (February 8th, 2020). The different noise
    sources shown are described in the text, the green line (\ac{bns} range: 66~Mpc) represents the
    sum of these noises and it can be compared to the measured total
    noise shown in black (\ac{bns} range: 59~Mpc).}
 \label{fig:noise_budget_O3b}  
\end{figure}

\begin{description}
\item[ASC] -- Angular Sensing and Control. This represents the control
  noise of 12 angular \acp{dof} of the interferometer (two per
  mirror) and four \acp{dof} of the beam injected into the
  interferometer. The coupling of these noises has been measured by
  injecting broadband noise into each \ac{dof}~\cite{thesisBader}.
\item[DAC] -- Digital Analog Converter. This is the electronic noise
  of the digital to analog converters used to drive the six main mirrors
  and marionettes of the interferometer. This electronic noise has
  been measured in the laboratory before installation, and the
  noise coupling is modeled using \texttt{SimulinkNb}.
\item[Dark.] This is the electronic and dark noise of the photodiodes used in the four
  longitudinal \acp{dof} control. The noise is measured by closing shutters
  of each photodiode, and the noise coupling is modeled.
\item[Demodulation.] This is the phase noise of the demodulation of radio
  frequency signal from photodiodes to control \ac{carm}, \ac{mich} and
  \ac{prcl}. That phase noise mixes the two demodulation quadratures. This
  bi-linear noise source is measured, and the noise coupling is
  modeled using \texttt{SimulinkNb}.
\item[ENV] -- Environment. This is the sum of three contributions:
  acoustic, magnetic and scattered light. The acoustic and magnetic
  noises are measured with four microphones and three 3-axis
  magnetometers, located in the experimental buildings near the interferometer
  components (see~\cite{env_hunt,o3virgoenv} for details).
  Their couplings are measured by broadband and sweeping
  sine noise injections. Scattered light is measured in two ways:
  i) using the signal from auxiliary photodiodes which have a linear
  coupling that is modeled; ii) using position sensors of suspended
  benches that couple in a non-linear way modeled with a measured
  scaling factor~\cite{Was2021}.
\item[LSC] -- Length Sensing and Control. This represents the control
  noise of four \acp{dof}: \ac{mich}, \ac{prcl}, \ac{omc} length, and residual intensity
  noise. The noise is measured in all cases, the coupling is measured
  for all except for the \ac{omc} length where it is modeled. Note that this
  results in double counting the dark and quantum noise of the sensors
  used for \ac{mich} and \ac{prcl} control, however these double counted
  contributions are negligible.
\item[Quantum.] Quantum noise of the detector and shot noise of the
  sensors used for \ac{mich}, \ac{prcl} and \ac{carm} control. The noise and the
  coupling are modeled using \texttt{SimulinkNb}.
\item[\ac{ssfs}.] This represents the
  control noise of the relative error between \ac{carm} and the laser
  wavelength. The noise is measured, the  frequency dependent coupling
  is modeled using \texttt{SimulinkNb} and a time dependent scaling factor is measured.  
\item[Seismic-Thermal.] This is the sum of the negligible seismic noise
  and three thermal noise contributions: suspension, mirror coatings and
  residual gas pressure in the arm vacuum tubes. The noise sources and
  the couplings are modeled using analytical functions in separate dedicated codes.
\item[``flat noise''.] It is a noise source of not yet understood physical origin. Its level has been measured proportional to the
  square root of the \ac{darm} offset used to obtain the interferometer DC
  readout~\cite{Hild2009, Fricke2011}. 
\end{description}

The sum of the noises described above correspond to a \ac{bns} range of
66~Mpc, while the actual \ac{bns} range in the corresponding
data was measured at 59~Mpc. Hence, about 10\% of the noise limiting \ac{bns} detections
is unaccounted for.

More in details, at frequencies above 1~kHz the sensitivity is mostly
limited by quantum shot noise. The measured level is about 5\% higher
than expected. This is due to a slow degradation of the frequency independent light squeezing during O3,
from 3~dB at the beginning of the run to about 2.5~dB at the end of it.

In the most sensitive frequency range, between 80~Hz and 200~Hz, there
are significant contributions from three sources: quantum shot noise,
mirror coating thermal noise and the ``flat noise'' of unknown physical
origin. Assuming that the ``flat noise'' estimate is correct,
removing completely this unknown noise source would have resulted in
10~Mpc improvement in the \ac{bns} range.

At low frequencies between 20~Hz and 50~Hz, the dominant noise
sources are quantum radiation pressure noise that is increased by the
frequency independent light squeezing and the laser intensity noise. However
30\% of the noise remains not understood in that frequency range, so
other significant noise sources are yet to be identified.

\section{The O3 run}
\label{section:O3}
\markboth{\thesection. \Sectionname}{}
The joint LIGO-Virgo Observing Run 3--- "O3"--- has been divided into two consecutive sub-data-taking periods, separated by a one-month commissioning break in October 2019.
\begin{itemize}
\item O3a: from April 1, 2019 at 15:00 UTC (GPS: 1238166018), to October 1, 2019 at 15:00 UTC (GPS: 1253977218).
\item O3b: from November 1, 2019 at 15:00 UTC (GPS: 1256655618), to March 27, 2020 at 17:00 UTC (GPS: 1269363618).
\end{itemize}
All three detectors have participated to the whole run. The O3b end date has been anticipated by about a month, due to the worldwide covid-19 pandemic.

This section presents the LIGO-Virgo O3 run, seen from a Virgo perspective. First, we describe the main activities into which the data taking was divided, before summarizing how the detector was steered from the EGO control room. Then, we focus on actions taken to maximize the amount of data collected and to ensure their good quality. In particular, we highlight the main DetChar activities during O3, explaining how they fit and complement each other, following the flow of data from the detector to the final analysis. Key to achieve this level of performance and to maintain it over almost a year, were the 24/7 on-call duty service and the rapid response team: both are briefly described as well.

Then, we review the performance of the Virgo detector during O3, mainly from the point of view of the duty cycle. A high duty cycle requires not only a stable and robust detector against external disturbances (see~\cite{o3virgoenv} for a comprehensive study of that topic) but also a quick and reliable procedure to bring the instrument to its working point (the {\em lock acquisition}), starting from an uncontrolled global state. The main statistics of the Virgo O3 global control acquisition are thus provided, before studying the actual duty cycle. We also present the evolution of the \ac{adv} detector sensitivity, from the O2 run to the end of O3.

This section ends with a brief overview of the final Virgo O3 dataset, describing how it was constructed offline, building upon the preliminary dataset estabilished by the live monitoring and data quality checks.

\subsection{Data taking}
\label{subsection:data_taking}
While data acquisition was the highest priority during the O3 run, a limited fraction of the time had to be dedicated to other activities. The two main recurring ones were:
\begin{itemize}
\item the maintenance periods, held every Tuesday morning, staggered with respect to the similar times in LIGO, in order to maximize the two-detector network coverage. Maintenance, limited to about 4 hours per week, was used to look after the detector components, to perform various cleaning activities, and to host noisy activities incompatible with data taking--- for instance the refilling of liquid nitrogen tanks located nearby the \ac{ceb}, \ac{neb} and \ac{web}, delivered by heavy trucks.
\item the calibration shifts, held almost every week on Wednesday afternoons or evenings. These campaigns allowed to check the accuracy of the reconstruction of the $h(t)$ strain stream~\cite{VIRGO:2021umk}, to monitor its stability over time and to test new, complimentary calibration methods, like the use of a Newtonian calibration~\cite{Estevez_2021_2} in addition to the usual photon calibrators~\cite{Estevez_2021_1}.
\end{itemize}

In addition, commissioning time was allocated irregularly to tune or optimize some aspects of the detector, depending on the needs and opportunities. Finally, some time was spent studying and fixing problems impacting the data taking.

\subsection{Detector steering}
\label{subsubsection:metatron}

The Virgo data taking is largely automated and usually only requires a single operator on duty in the control room. Operators are present 24/7 during a run and take shifts every 8 hours.

The \ac{adv} detector automation, called \texttt{Metatron}, relies on the Guardian~\cite{Rollins:2016hlk,guardian} framework, developed by LIGO and based on hierarchical finite state machines. The Virgo implementation links this framework to the \ac{daq}: automation nodes become \ac{daq} nodes that get data directly from shared memories and are synchronized with the one-second data availability period. A generic mechanism to read and write \ac{daq} channels has been introduced and can be used within user codes via dedicated functions.

\begin{figure}[htb!]
  \centering
  \includegraphics[width=\textwidth]{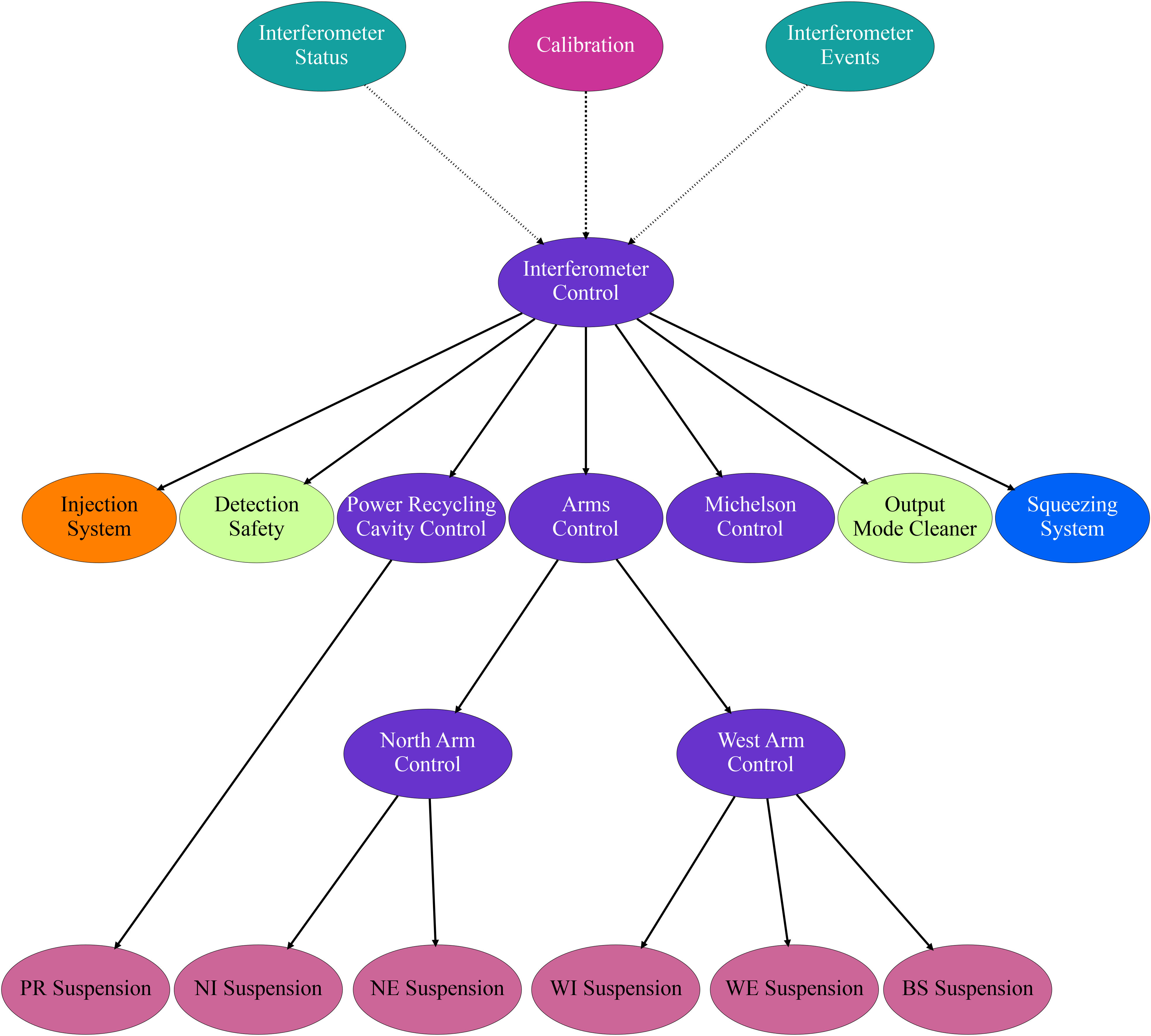}
  \caption{\texttt{Metatron} nodes hierarchy used during the O3 run.}
  \label{fig:metatron_hierarchy}
\end{figure}

The full Virgo control acquisition procedure has been implemented in \texttt{Metatron}, initially prior to the O2 run and then updated for the O3 configuration---the main difference being the addition of the frequency-independent squeezing~\cite{PhysRevLett.123.231108}. The scheme adopted, depicted in Figure~\ref{fig:metatron_hierarchy}, strictly follows a bottom-up approach, with the lower-level nodes being automatically managed by higher-level ones.

The suspension nodes (violet background in the graph) are tasked to align/misalign the Virgo optics; each of them is managed by the most appropriate control nodes (dark blue background), divided on the basis of the degrees of freedom to be controlled. The main node---\texttt{Interferometer Control}--- is usually the only one operated manually to steer the detector. It defines the control paths such as for instance the main global control procedure that allows reaching the Science mode (the nominal data taking state), plus other procedures to control various configurations of the optics or to perform automated calibrations, etc. It relies on the underlying managed nodes to perform these actions on the instrument. During the final steps of the control procedure, each single part of the interferometer is ultimately entangled with the others, and the interferometer is naturally treated as a single system; for these reasons, the last part of the procedure is directly managed by the upper level node, which sets the control parameters to the whole system, while the lower level nodes are only used as watchdogs for the correct functioning of their own sub-systems.

Additionally, the \texttt{Metatron} main node manages:
\begin{itemize}
\item the laser Injection System, from the laser source to the \ac{imc} (orange background);
\item the two Output Mode Cleaners that are controlled in sequence in the final steps of the nominal control acquisition procedure (pale green background);
\item the Detection System at the interferometer output port (pale green background); 
\item the frequency-independent Squeezing System (light blue background), whose control proceeds in parallel to the main detector control procedure. As Virgo can take valid Science data with or without this system being in its nominal state, the corresponding \texttt{Metatron} node is a bit apart from the others logic-wise.
\end{itemize}

Only during the calibration measurements, the \texttt{Interferometer Control} node is automatically managed by the \texttt{Calibration} node (magenta background). 

The \texttt{Metatron} framework also takes care of generating high-level flags that provide the overall status of the interferometer: this is done within the \texttt{Interferometer Status} node (dark green background). Finally, the \texttt{Interferometer Events} node (dark green background) records all state transitions of the detector. The Interferometer Status and Interferometer Events information is passed onto the Virgo live monitoring system, documented in Section~\ref{section:monitoring_tools}.

\subsection{DetChar organization}

\begin{figure}
  \center
  \includegraphics[width=\textwidth]{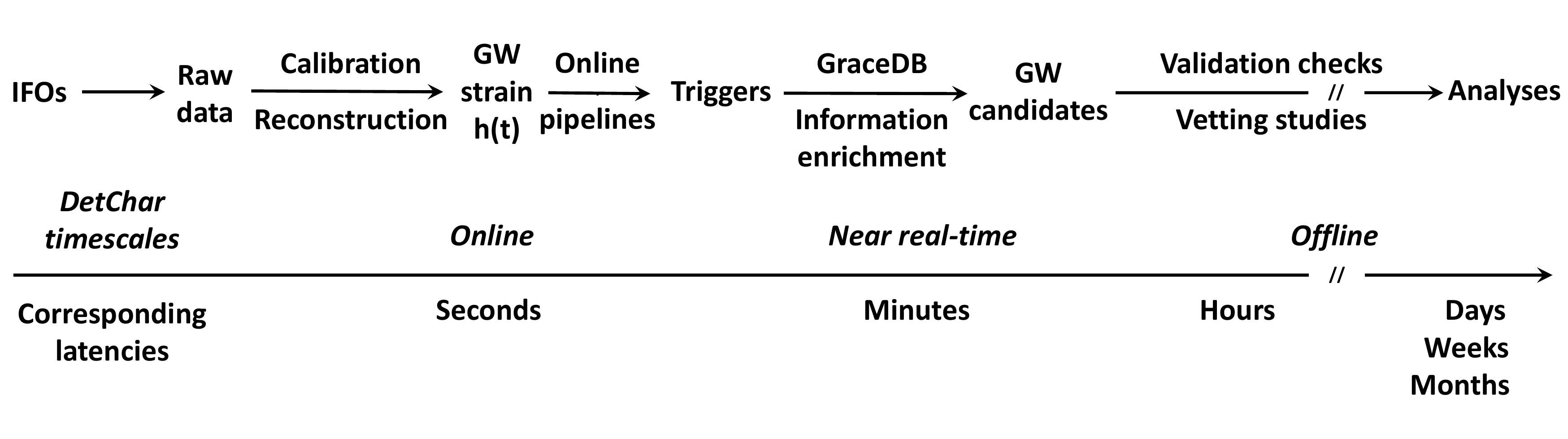}
  \caption{Dataflow from the interferometers (labelled "IFOs" on the left) to the offline validation of \ac{gw} candidates and the completion of the final dataset (right). It focuses on the generation and the vetting of the public alerts that are a key product of the LIGO-Virgo observing runs. It shows the three main timescales at which the Virgo DetChar group operates: online, near real-time and offline (see text for details).}
  \label{fig:DataflowDetChar}  
\end{figure}

Figure~\ref{fig:DataflowDetChar} shows the flow of data, from the interferometers (IFOs, on the left), to the physics analyses (on the right). While focusing on the \ac{gw} candidates, this schematic highlights the three main pillars of DetChar activities during a run.

\begin{itemize}
\item The first timescale on which DetChar activities take place is online (latency: $\mathcal{O}(\textrm{s})$). Quick automated checks are run on live data to mark out (good or bad quality) the data stream used as input by the ``pipelines''---that is the algorithms that scan the network data in real time, as soon as they become available. Initial data quality information is indeed shipped alongside the reconstructed \ac{gw} stream, as explained in Section~\ref{sec:onlinedq}.
\item The second timescale is near real-time (latency: $\mathcal{O}(\textrm{min})$), crucial to assess the quality of the \ac{gw} candidate public alerts. Thanks to a dedicated framework that is described in Section~\ref{section:public_alerts}, the data around a significant candidate are vet for each detector and a global decision is then taken: either to confirm the public alert sent to the telescopes or to retract it (see Section~\ref{section:oncall_RRT} below for a description of the procedure).
\item Finally, the last timescale is offline (high latency: up to months after the data taking). The goals of these studies are twofold: first, to finalize the dataset that all offline analyses will use, regardless of whether they look for transient or continuous signals; then, to validate the events that will be included in the final publications and whose parameters will be used to extract astrophysical information.
\end{itemize}

To ensure a continuous monitoring of the data quality, DetChar shifts were organized during the entire O3 run on a weekly basis, with two people (working onsite or remotely) on duty. The shifter crew changed every Tuesday morning, during the weekly maintenance of the Virgo detector. In addition  to attend all relevant meetings. DetChar shifters usually reported their findings at the weekly DetChar meeting on Fridays and at the weekly detector meeting on Tuesdays (thus at the end of their weekly shift).

\subsection{On call duty service and rapid response team meetings}
\label{section:oncall_RRT}

An on-call service was organized during the O3 run to ensure a 24/7 expert coverage for all the Virgo detector components, from hardware systems to online computing and DetChar. In case of a problem, the operator on duty would contact the relevant experts from the control room, plus the data taking coordinators if needed.

In addition, a joint LIGO-Virgo low-latency automated alert system was setup to contact the \ac{rrt} experts---specialists of data taking, data quality or \ac{gw} transient searches---that would meet remotely on short notice each time a public alert candidate had been identified in real time. They would vet that candidate, using all raw information available, plus the output of several data quality checks, triggered automatically by the generation of the signal candidate: the \ac{dqr}, see Section~\ref{subsection:DQR} for details. The outcome of an \ac{rrt} meeting could be twofold: either to confirm the public alert, or to retract it when the astrophysical origin of the candidate was questionable.

\subsection{Virgo O3 duty cycle}
\label{subsection:O3_perf}
Table~\ref{table:O3_locking} summarizes the performance of the global control acquisition procedure for the Virgo detector during O3. This performance has been stable over the whole run, showing the robustness of that procedure. As not all control acquisition {\em attempts} are successful, a global control  acquisition {\em procedure} is defined as a set of successive control attempts that leads to the global control of the instrument.

The median duration of a successful global control  acquisition attempt is 18 minutes: about 30\% of this time is spent reaching the detector working point (Michelson interferometer at the dark fringe, power recycling cavity and arm cavities resonant, \ac{ssfs} enabled); 50\% is spent to control the two \acp{omc} at the Virgo output port; the final 20\% are used to reach the lowest noise configuration at the level of the suspension actuation.The median number of attempts needed to complete a global control  sequence is 2 and the median duration of a successful global control acquisition sequence is 25~min, during O3 the quickest sequence took about 13~min.

\begin{table}[htbp!]
\centering
\caption{\label{table:O3_locking}Summary of the Virgo global control acquisition performance during O3: the control is acquired after a successful control acquisition {\it sequence} that counts one or more control acquisition {\it attempts}.}
\begin{tabular}{cr|c}
\toprule
\multicolumn{3}{c}{\textbf{Global control acquisition attempt}} \\\midrule
~ & Median duration & 18 minutes \\
\hline ~ & Distribution of this time & ~ \\
~ & Reaching the detector working point & ${\sim}$30\% \\
~ & Controlling the two \acp{omc} & ${\sim}50$\% \\
~ & Acquiring the lowest noise configuration & ${\sim}20$\% \\
\bottomrule
\multicolumn{3}{c}{\textbf{Global control acquisition sequence}} \\\midrule
~ & Median number of attempts & 2 \\
~ & Median duration & 25~min \\
\bottomrule
\end{tabular}

\end{table}

\begin{table}[htbp!]
\caption{\label{table:O3_perf}Summary of the O3 data taking performance of the Virgo detector. The last three rows of the table provide duty cycles for different configurations of the 3-detector LIGO-Virgo global network: the fraction of the time during which at least one the three instruments is taking data, at least two are and finally all three are.}
\centering
\begin{tabular}{lr|ccc}
\toprule
~ & ~ & O3a & O3b & O3 \\ \midrule
\multirow{2}{*}{Virgo global control segments} & Mean [hr]   & 6.1 & 6.4 & 6.3 \\
 & Median [hr] & 2.7 & 1.8 & 2.2 \\ \midrule
\multirow{2}{*}{Virgo Science segments} & Mean [hr] & 5.0 & 4.0 & 4.5 \\
 & Median [hr] & 2.6 & 1.4 & 1.9 \\ \midrule
\multirow{4}{*}{Duty cycles} & Virgo [\%] & 76.3 & 75.6 & 76.0 \\
 & Network---at least 1/3 [\%] & 96.8 & 96.6 & 96.7 \\
 & Network---at least 2/3 [\%] & 81.9 & 85.4 & 83.4 \\
 & Network---3/3 [\%]          & 44.5 & 51.0 & 47.4 \\
 \bottomrule
\end{tabular}

\end{table}

Table~\ref{table:O3_perf} details the control stability of the Virgo detector, separately for the sub-runs O3a and O3b, and averaged over the whole O3 run. The ``global control segments'' are stretches of data during which Virgo is controlled in its nominal low-noise configuration, while the ``Science segments'' are the subset of the global control segments during which Virgo is taking data of good, science-compatible, quality. The difference of duration between the global control and Science segments is dominated by limited disruptions of the data taking, that usually stop the Science mode for a short time. The dominant source of these breaks is the frequency-independent squeezer that lost its nominal configuration about 240 times during the O3 run; the median time to restore it and switch back to Science data taking was about 140 seconds.

We note that the Virgo segment duration summary numbers listed here are lower than those reported by LIGO~\cite{Buikema:2020dlj,LIGO:2021ppb}. Yet, this difference has no significant impact on the duty cycle that is very similar for the three detectors of the global LIGO-Virgo network. The comparison between the O3a and O3b sub-runs shows that the impact of the winter season (larger sea seismic activity, wind, and more generally bad weather), although real, has been limited. Overall, the global network duty cycle has improved during O3, mainly due to the increase of the LIGO detectors duty cycle, while the Virgo one has been very stable. With an average of 76\%, the Virgo O3 duty cycle is lower than that measured during August 2017, the final weeks of the O2 run Virgo took part of: ${\sim}$85\%. Yet, the O3 performance has been achieved over 11 months spanning a whole calendar year and cannot be directly compared to the duty cycle of a 25-day run in Summer time, the most favorable period to operate an instrument like Virgo. Running one full year instead of one month is also more complex person-power wise, and the Virgo organization implemented during O3, although perfectible, held on during the whole run. This experience represents a good base on which to build upon in order to improve the Virgo performance for the O4 run and beyond.

\begin{figure}
  \center
  \includegraphics[width=\textwidth]{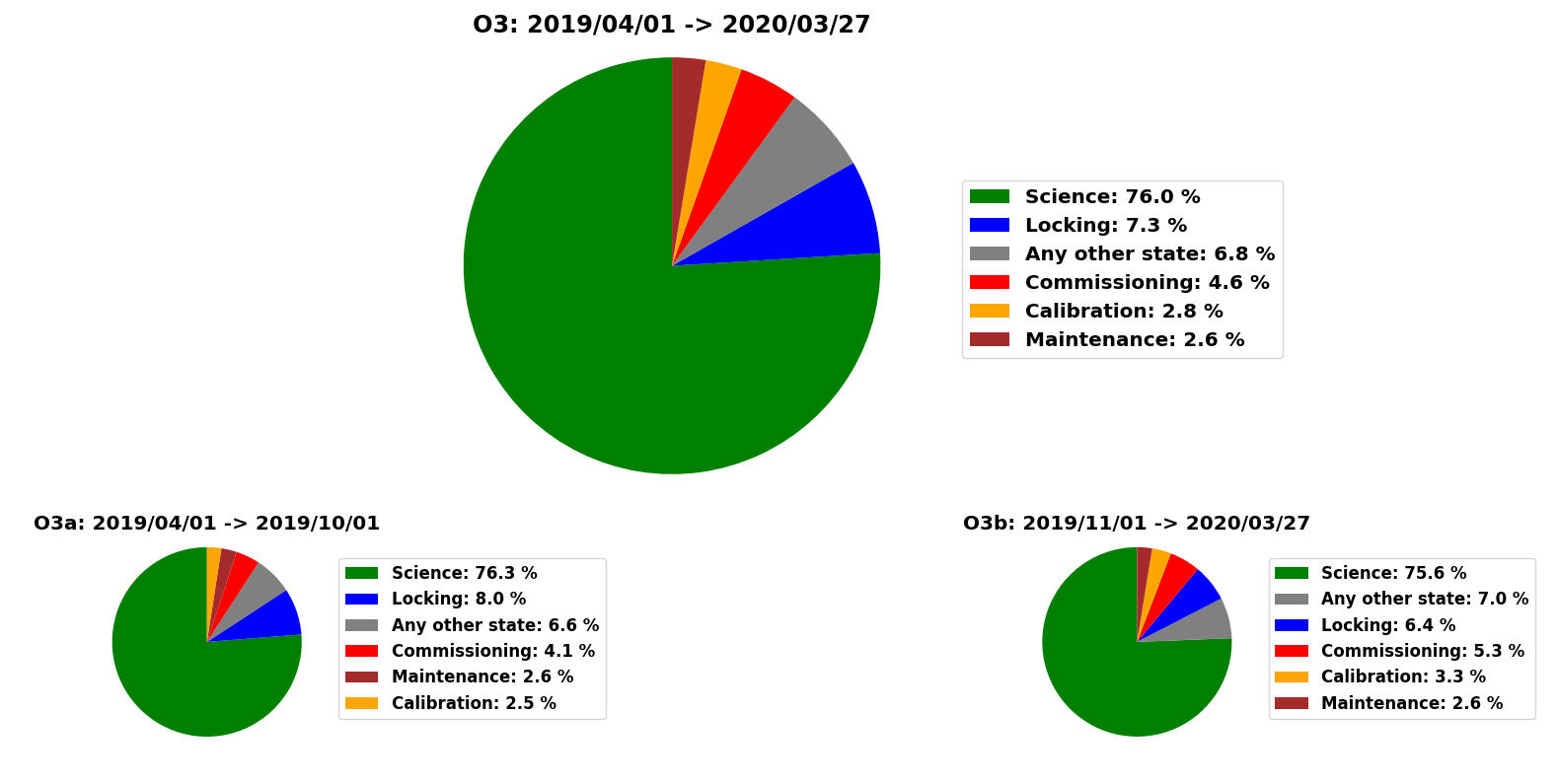}
  \caption{Breakdown of the time spent in different modes by the Virgo detector during the O3 run (larger, top-middle pie chart) and separately during the O3a and O3b sub-runs (pie charts at the bottom). The {\em Locking} mode corresponds to periods when the control of the detector is being acquired. The regular {\em maintenance} and {\em calibration} periods have been described in Section~\protect\ref{subsection:data_taking}. Finally, the {\em Any other state} category includes all the other situations encountered during the whole run: troubleshooting periods, various kinds of tuning, etc. These results exclude the 1 month-long commissioning break that took place in October 2019, in between the O3a and O3b sub-runs. In each pie chart, the modes are sorted by decreasing percentage.}
  \label{fig:O3_Virgo_status}  
\end{figure}

Figure~\ref{fig:O3_Virgo_status} shows the breakdown of the time spent in different modes by Virgo during O3. Overall, the O3a and O3b distributions are quite consistent. Breaking these 11 month-averaged duty cycle figures down to a 24-hour period, Virgo took data during about 18 hours, with the remaining six hours roughly divided into three blocks of the same duration: ${\sim}2$~hours for controlling the detector (Locking), ${\sim}2$~hours for recurring activities (Calibration, Commissioning and Maintenance) and ${\sim}2$~hours for solving issues (Any other state).

The analysis of these pie charts shows that increasing the duty cycle during future runs will not be straightforward. 
The room for improvement is limited in each area and so any significant duty cycle gain will likely stem from a combination of various small progresses, each made possible by the redesign or the optimization of a particular process.

\begin{figure}
  \center
  \includegraphics[width=\textwidth]{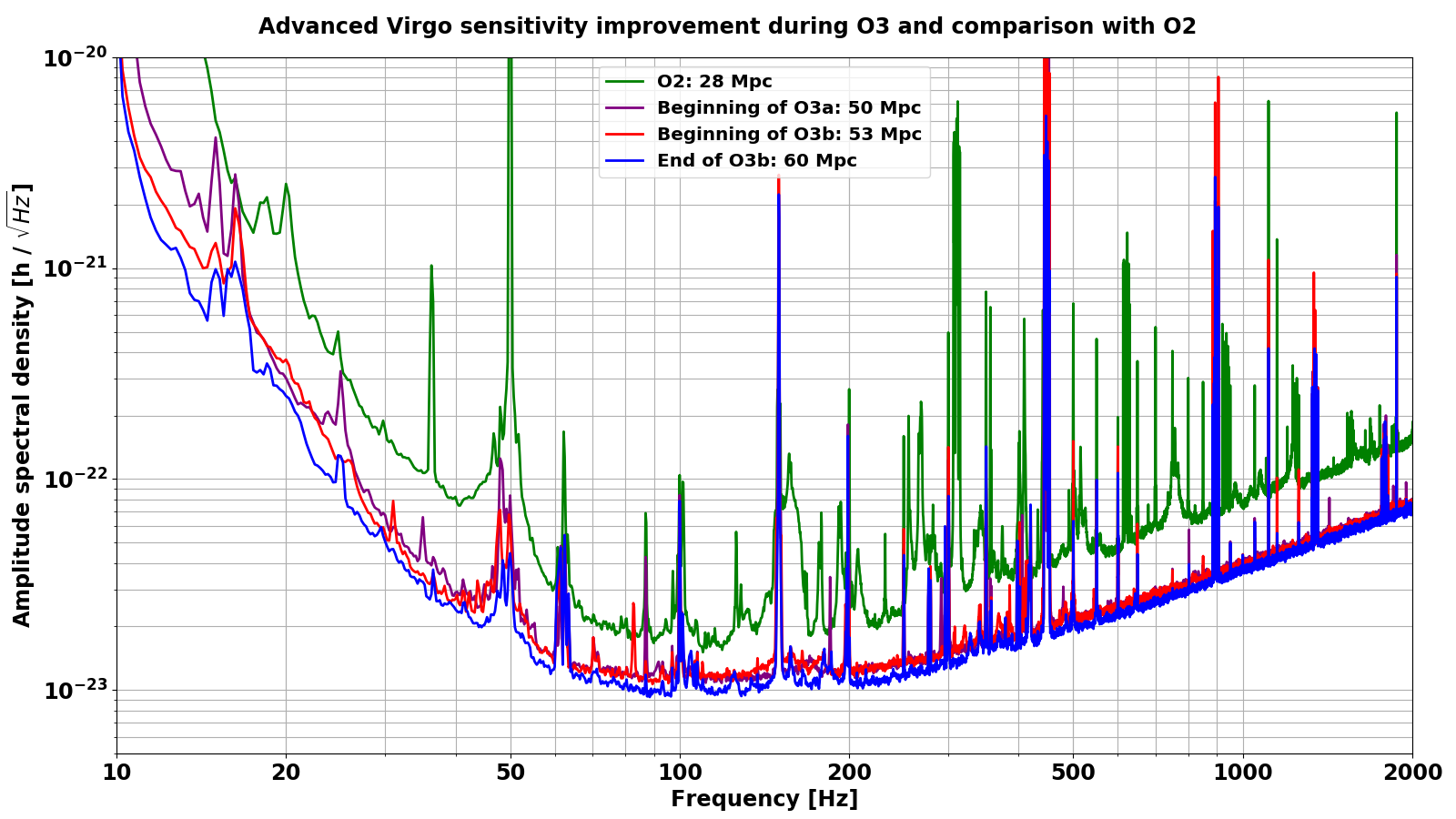}
  \caption{Comparison between four sensitivity curves of the \ac{adv} detector: during O2 (green trace), at the beginning of O3a (purple), at the beginning of O3b (red) and at the end of O3b (blue). The caption provides the corresponding estimated \ac{bns} ranges.}
  \label{fig:O3_sensitivities}  
\end{figure}

To conclude this overview, Figure~\ref{fig:O3_sensitivities} summarizes the improvement of the sensitivity of the \ac{adv} detector. The \ac{bns} range associated to each curve is given in the legend. From O2 to O3b, the \ac{bns} range has more than doubled from 28 to 60~Mpc, with a continuous improvement of the sensitivity in the whole bandwidth of the detector. Many spectral features of the residual noise structures have either been removed or significantly reduced over time.

\subsection{The Virgo O3 dataset}
\label{susbsection:O3_dataset}
The final Virgo O3 dataset consists of more than 250 days of data recorded during the O3a and O3b sub-runs and whose quality has been checked and validated (described in Section~\ref{sec:dq_offline}). 
It is built upon and supersedes the online good-quality Science dataset that was used as input by the analysis pipelines that looked for \acp{gw} in real time (see Section~\ref{sec:onlinedq}). Dedicated studies have been performed offline to refine the quality assessment of the data. In addition to running more in-depth analyses, new checks have been added during the run, as potentiel flaws got discovered in the existing analyses, or new problems identified at the detector level. Moreover, small sets of good data that had not been automatically included in the dataset (either because they were incorrectly labeled or because part of their data quality information was missing) were added by hand.

The main categories of checks applied to assess the quality of the Virgo data are the following.

\begin{itemize}
\item Are key components of the Virgo hardware (suspensions and photodiodes) having transient problems? \\ These checks, described in Section~\ref{sec:onlinedq:cat1}, were fast enough to be performed online on live data.
\item Is the reconstruction of the \ac{gw} strain time series $h(t)$ nominal? \\ This is a prerequisite for any further use of the Virgo data. The online reconstruction of the Virgo data was satisfactory: only about three weeks at the end of O3a were reprocessed offline to increase the sensitivity by a few percents~\cite{VIRGO:2021umk}. Yet, during periods of high seismic activities (bad weather, high wind or the passing of seismic waves from strong and distant earthquakes), it could be replaced~\cite{o3virgoenv}) by a more robust control configuration--- the so-called "earthquake (EQ)-mode"~\cite{VIRGO:2021umk}. Although that procedure saved some lock losses whose recovery would have costed time, it could not be validated against the nominal reconstruction of the $h(t)$ strain stream until the final two months of O3b. Therefore, during most of the O3 run, data taken in these peculiar conditions had to be excluded from the final dataset. 
\item Do the data suffer from known problems? \\ Tailored checks were run offline to identify and isolate periods during which the detector was not behaving nominally, although it was still controlled.One example of such studies is the fact that the North Input mirror suspension was randomly suffering from a transient (a few second-long) loss of data. This was usually enough to lose the control of the entire detector, and hence to lose at least about 20-30 minutes of data: the time to reacquire the locked state and to restore Science data taking. Therefore, a patch was developed by experts to detect the data loss and switch to a less robust---but still available---control until the missing data were back. This saved hours of running time for Virgo overall, but a dedicated scan of the data had to be performed offline to identify the occurrences of these control switches (potentially inducing transients and artifacts of instrumental origin in the data) and to remove them from the final dataset.
\item Are the data consistent? \\ For instance it was decided to remove offline the last few seconds of a segment preceeding a control loss of the detector as those data could be corrupted--- see Sec.~\ref{sec:dq_offline} for details
\item Is the dataset complete? \\ For example there could be segments with missing or corrupted $h(t)$ channel that would require a limited reprocessing. Or there could be segments with missing missing data segments due to problems in the DAQ, etc..
\end{itemize}

Data segments that fail one of the checks defined above are classified as "Category 1" (CAT1) vetoes and must be excluded from all analyses. Overall, only about 0.2\% of the Virgo O3 Science dataset have been CAT1-vetoed.

\begin{figure}
  \center
  \includegraphics[width=\textwidth]{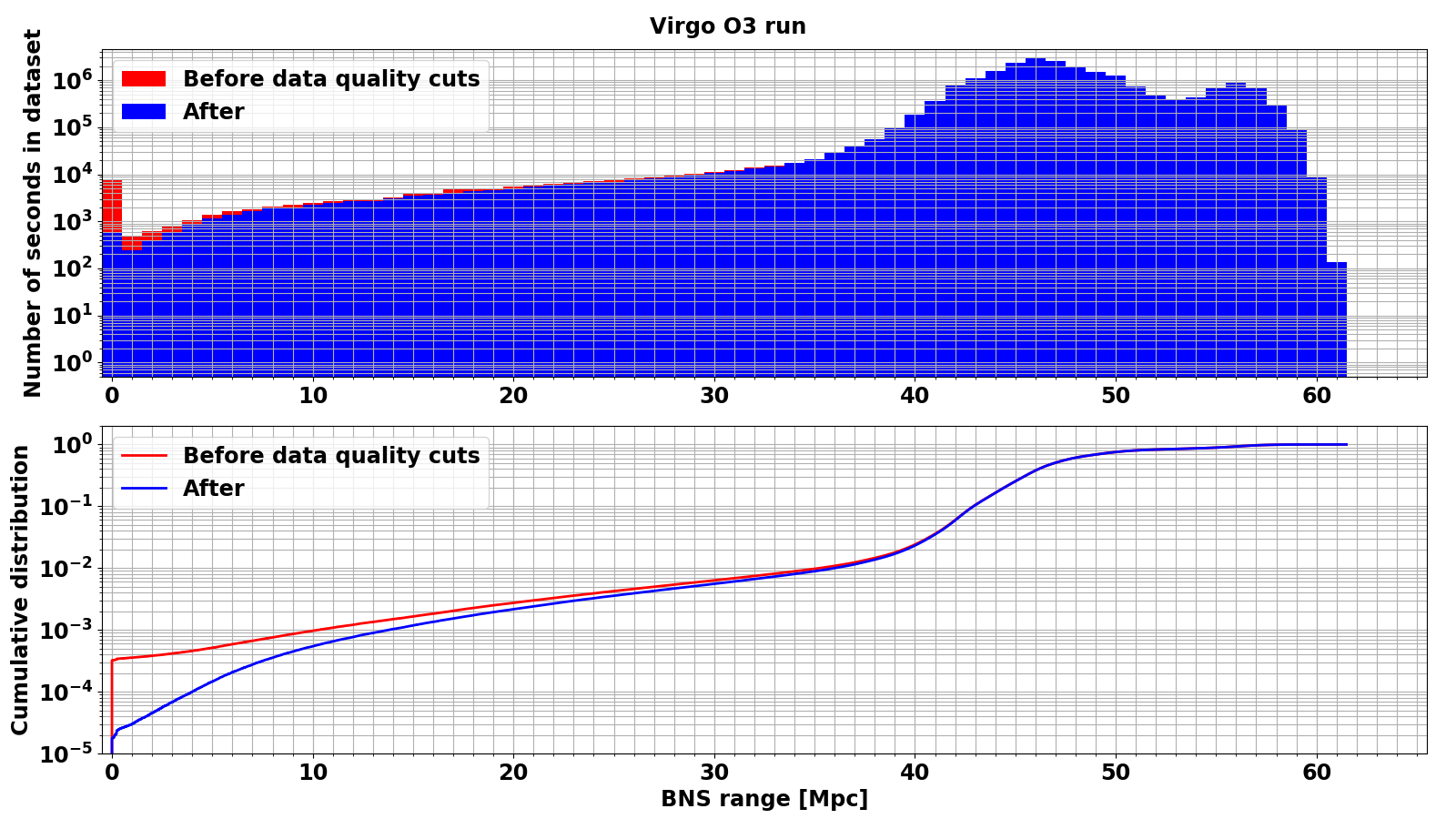}
  \caption{Distributions of the Virgo \ac{bns} range during O3 for the online (red histogram and trace) and offline (blue) dataset. The top (bottom) plot compares the histograms (cumulative distributions).}
  \label{fig:BNS_range_O3}
\end{figure}

To conclude this overview of the Virgo performance during the O3 run, Figure~\ref{fig:BNS_range_O3} compares the Virgo \ac{bns} range distributions before (red) and after (blue) applying data quality cuts 
to determine the final O3 dataset. As expected, data quality requirements remove periods of low \ac{bns} range, i.e. when the sensitivity was poor. Yet, about 1\% of the data have a \ac{bns} range lower than 35~Mpc, that is significantly below the typical values achieved during O3 for that sensitivity estimator. While these data have not been flagged as bad by the various checks run on the dataset, they correspond to periods during which the detector was less accurately controlled, in particular due to bad weather. 

\section{Tools for detector characterization and data taking monitoring}
\label{section:tools}
\markboth{\thesection. \Sectionname}{}
All DetChar analyses rely on dedicated software frameworks, called generically {\em tools} in the following. 
Most of these have been developed within Virgo. In addition, 
thanks to the long-lasting collaborations among the Virgo, LIGO and now KAGRA DetChar groups, we benefit from additional tools or methods that have been developed partly or totally by colleagues.

More than 100 servers have been running in real-time during O3 to monitor the Virgo detector, run various data quality checks and perform specific DetChar tasks. Data are processed by the tools described in the following subsections and whose outputs are included in the live data streams or stored on disk. Finally, the end products of these analyses are converted into information for the control room and summary plots that are updated with a latency of a few minutes at most and regularly archived for offline analyses.

All these processes are steered using the \ac{vpm} software interface, that allows to configure, start/stop and monitor processes running on Virgo online servers. These include detector control, data transfer to and from Virgo, and the analysis of the reconstructed $h(t)$ stream by the online \ac{gw} searches running in the EGO computing center. All actions performed using the \ac{vpm} interface are logged and recorded, in order to reconstruct as accurately as possible the software running at any given time, should this need arise.

The most important DetChar tools used during the O3 run are described in the following. They have been classified in a few categories depending on their usage or target: monitoring, generic data analysis, glitches, spectral noise or databases. Yet, they are not independent: they are often combined to characterize some features of the detector, or to provide a complete overview of the quality of the Virgo data. 
The flowchart in Figure~\ref{fig:tools_flowchart} represents the main analyses carried out by the DetChar group with the tools presented in this section.
The arrows follow the data-flow, which starts from the detector raw data that is analyzed by the various tools, whose data products are then saved to disk and used to generate DQ flags and reports.
The latter are used by both GW search pipelines and by commissioners and operators that control the status of the detector.

\begin{figure}
  \centering
  \includegraphics[width=\textwidth]{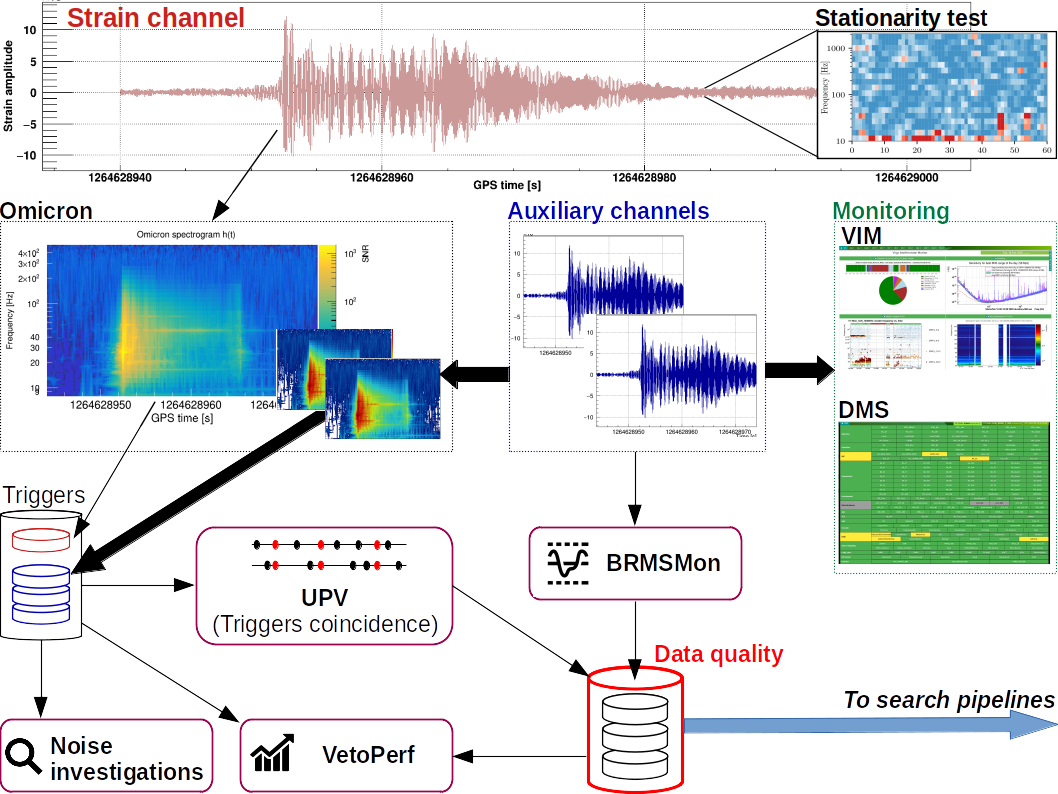}
  \caption{The Virgo strain data and auxiliary channels are analysed by DetChar tools to monitor and investigate transient noise. The noise stationarity is monitored with dedicated tools (Section~\ref{sec:tools:Bristol}). The data is analysed with \texttt{Omicron} (Section~\ref{sec:tools:glitch:omicron}) and transient triggers are saved to disk for further noise investigation. In particular, the \texttt{UPV} algorithm (Section~\ref{sec:tools:glitch:upv}) isolates coincidences between triggers from the strain and auxiliary channels. \texttt{BRMSMon} (Section~\ref{sec:tools:monitoring:brmsmon}) detects transient noise excesses in axiliary channels. Both \texttt{UPV} and \texttt{BRMSMon} generate data quality segments used to reject transient noise found by GW searches. The performance of these data quality segments is evaludated by a tool called \texttt{VetoPerf} (Section~\ref{sec:tools:glitch:vetoperf}). The transient noise is also monitors with web tools like \texttt{VIM} (Section~\ref{sec:tools:VIM}) and the \texttt{DMS} (Section~\ref{sec:tools:dms}).}
  \label{fig:glitch_flowchart}  
\end{figure}

\begin{figure}[th!]
  \centering
  \includegraphics[trim=0 10 40 0, clip, width=\textwidth]{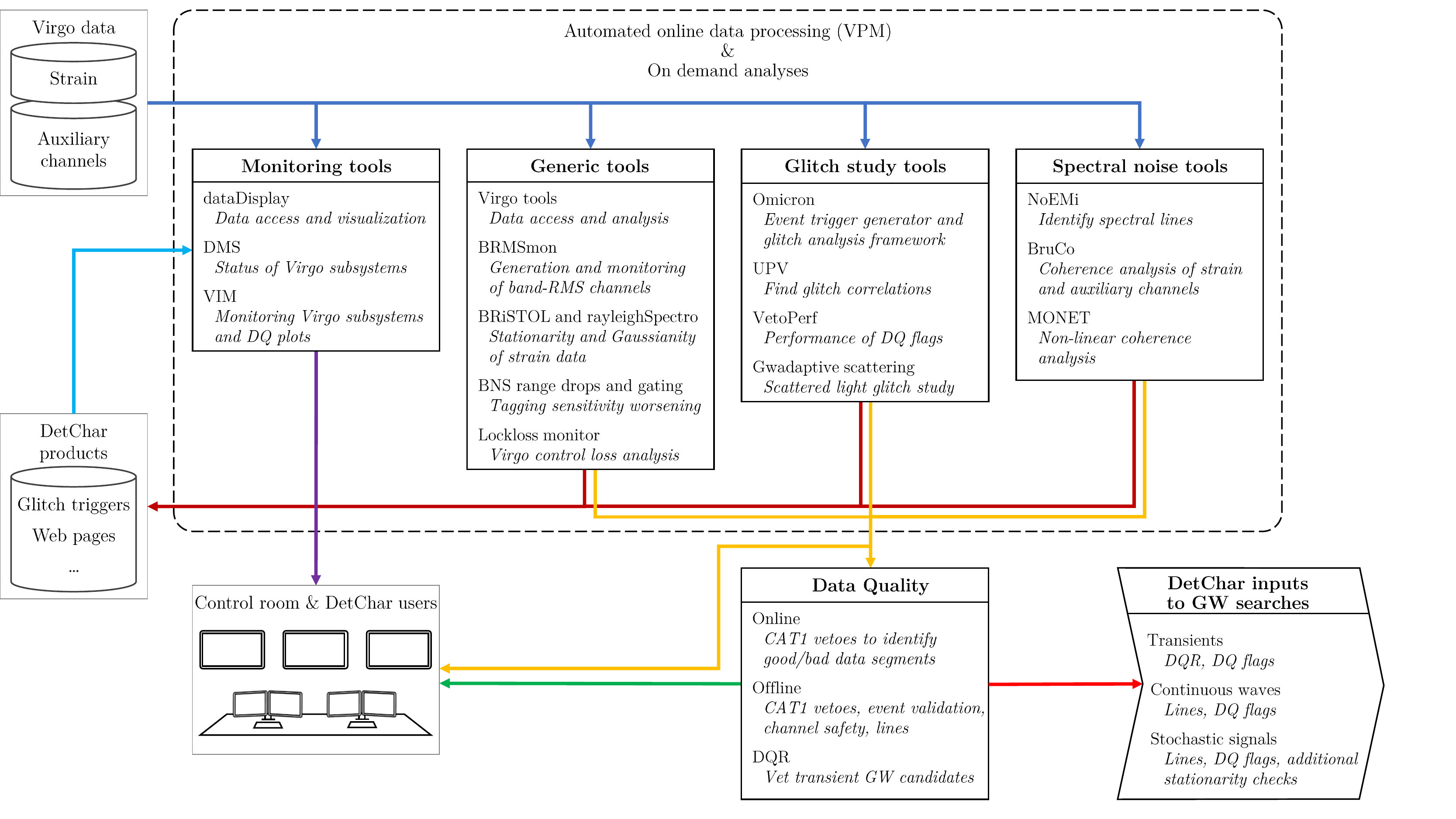}
  \caption{Flowchart of the various tools and monitors used for detector cha\-rac\-ter\-ization that are presented in Section~\ref{section:tools}. The data-flow starts from the raw data acquired from the detector, which is analyzed by the various tools and used to produce processed data and data visualizations used for monitoring purposes. Some of the outputs of the various tools are used to generate DQ products, to be used by GW search pipelines and commissioners to control the interferometer.}
  \label{fig:tools_flowchart}
\end{figure}

Figure~\ref{fig:glitch_flowchart} describes a specific example of joint application of various analysis tools and monitors to the study of transient noise.

\subsection{Monitoring tools}
\label{section:monitoring_tools}

\subsubsection{dataDisplay}
\begin{figure}
  \center
  \includegraphics[width=0.8\textwidth]{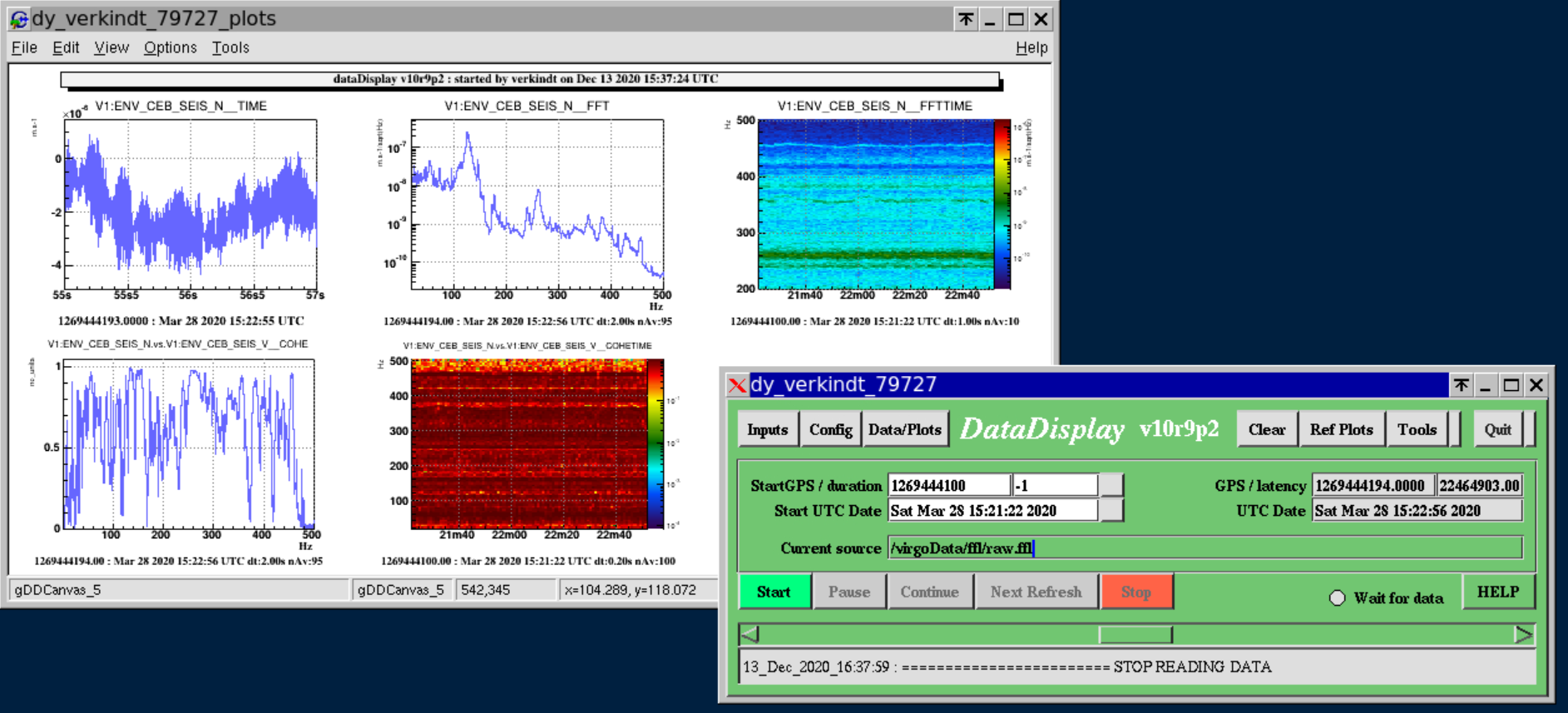}
  \caption{Example of the plots produced by the \texttt{dataDisplay} (left) and main panel of the dataDisplay graphical user interface (right).}
  \label{fig:dy}
\end{figure}

The \texttt{dataDisplay} software~\cite{datadisplay} allows the user to read (online or offline) Virgo data and to visualize various types of plots for all the channels available from the DAQ.
For instance, it helps to investigate
quickly the time evolution of a noise artifact, the coherence between two control signals or the
time-frequency characteristics of a transient noise.
It has been used extensively during the O2 and O3 runs and all over the \ac{adv} detector commissioning in between.
Figure~\ref{fig:dy} shows an example of the \texttt{dataDisplay} interface and output.

\subsubsection{DMS: the Detector Monitoring System}\label{sec:tools:dms}
The \ac{dms}~\cite{dms1,dms2} provides a detailed live status of all the components that make the Virgo detector operate, from the hardware parts to the online software used to control the instrument and take data. It also includes the monitoring of environmental data from around the experimental areas. 
Each of the many \ac{dms} monitors uses a set of \ac{daq} channels, combines them by performing mathematical and logical operations on their outputs and produces a flag whose value can take four severity levels, each associated with a color for visual display. A web interface is used to display and browse the \ac{dms} monitor flags with a few second-latency, both in the Virgo control room and remotely.

\begin{figure}
  \centering
  \includegraphics[width=\textwidth]{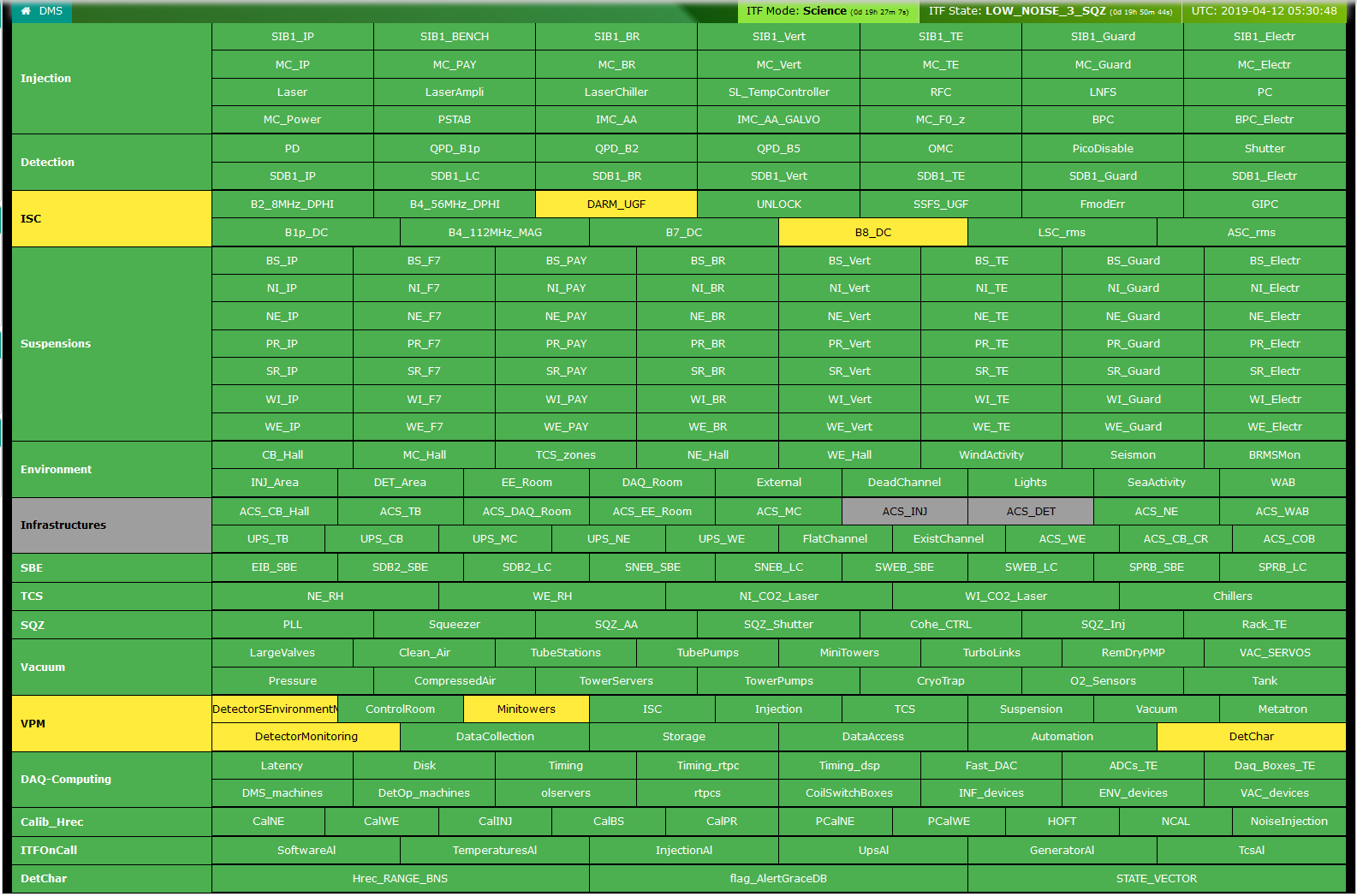}
  \caption{DMS snapshot closest in time to the GW190412 \ac{gw} event, showing the detailed status of the Virgo detector about four seconds after the arrival of that signal.
The DMS web interface looks like a checkerboard. Each row, labelled in the most-left column, corresponds to a different part of the instrument (mirror suspensions, vacuum system, etc.). That part is broken down in smaller sets that are each associated with a cell on the web interface. Each cell can contain many DMS flags and its color reflects the highest severity among all these flags (green $\leftrightarrow$ no alarm; yellow $\leftrightarrow$ warning; red $\leftrightarrow$ alarm (not present on that particular snapshot); grey $\leftrightarrow$ some information is missing). Clicking on a cell gives access to the flag individual information: their values and associated severities.
}
  \label{fig:DMS_playback_GW190412}  
\end{figure}

In addition, a new \ac{dms} archival system has been set up for the O3 run: complete \ac{dms} snapshots are taken every ${\sim} 10$~seconds and archived. They can be retrieved later at any time, by running a playback application that uses the same interface as the live \ac{dms}.
This functionality is particularly convenient to check the status of the detector a posteriori, when a \ac{gw} candidate or a particular feature in the data have been identified.
 For instance, Figure~\ref{fig:DMS_playback_GW190412} shows the Virgo detector status about four seconds after the detection of the \ac{gw} event GW190412~\cite{LIGOScientific:2020stg}.

\subsubsection{VIM: the Virgo Interferometer Monitor}\label{sec:tools:VIM}
\begin{figure}
  \center
  \includegraphics[width=0.95\textwidth]{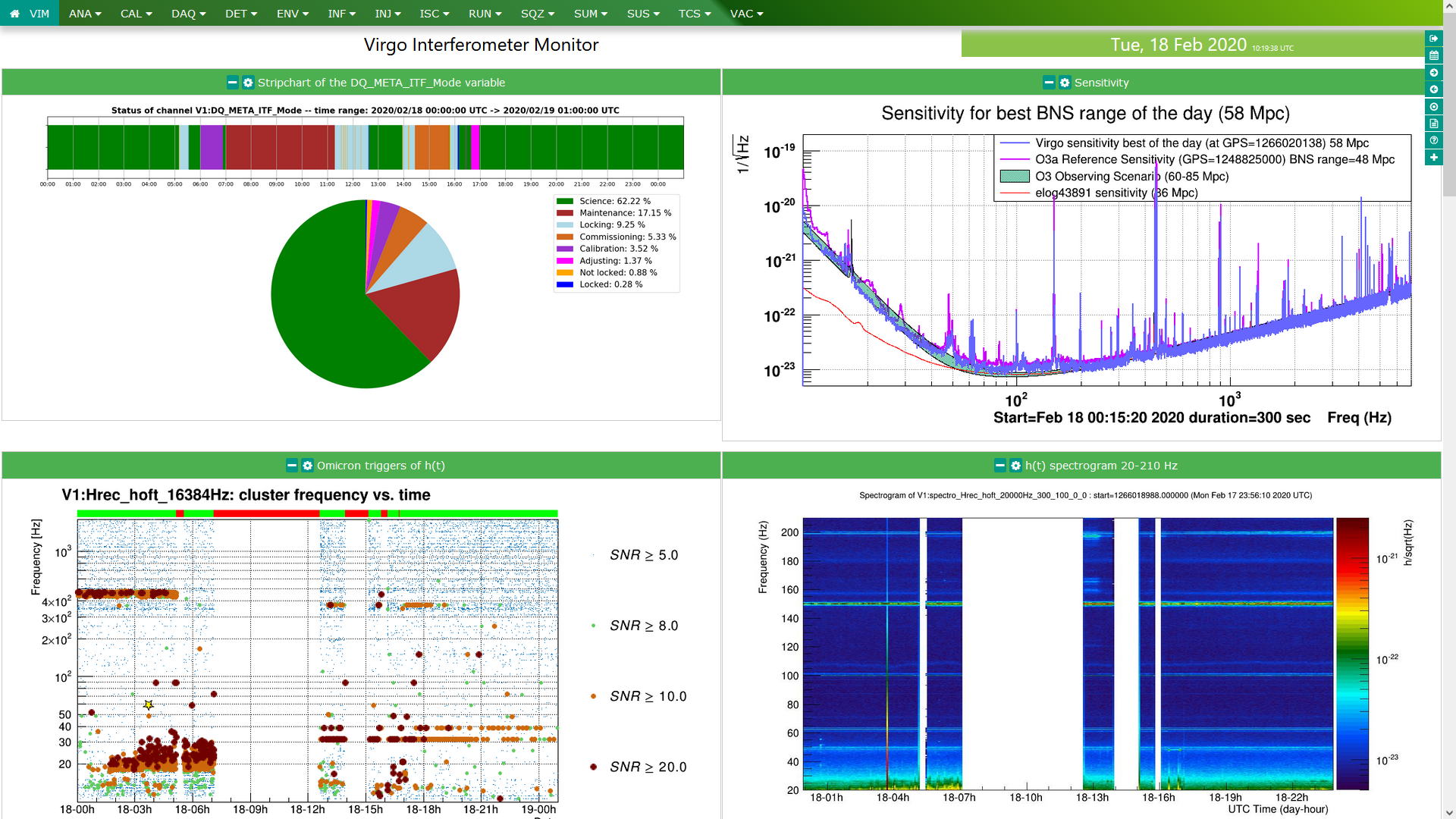}
  \caption{Screenshot of a \ac{vim} webpage displaying information about the Virgo detector on Tuesday February 18, 2020. Top left plot, stripchart of the detector status: the weekly maintenance, preceded by a planned calibration and followed by a short commissioning period, interrupts the data taking that restarts in the evening. Top right plot: daily sensitivity compared with references. Bottom left plot: glitch monitoring provided by the \texttt{Omicron} analysis described in Section~\ref{sec:tools:glitch:omicron}. Bottom right plot: spectrogram of the \ac{gw} strain $h(t)$ in the 20-210~Hz frequency range.}
  \label{fig:o3_vim}  
\end{figure}

The \ac{vim}~\cite{hemming2016,verkindt2019} manages a collection of automated scripts that update every few minutes a wide set of plots and tables; all these monitoring products are archived on a daily basis. A web interface allows users to browse that database, both for live monitoring of the experiment and for offline investigations.
\ac{vim} is an essential tool that provides a direct access to a detailed status of the various Virgo detector components and of related frameworks, such as calibration and online data processing, data transfer or online data analyses.
A snapshot of the \ac{vim} web interface is shown in Figure~\ref{fig:o3_vim}.

\subsection{Generic tools}

\subsubsection{The VirgoTools utilities}
In-depth studies of a particular feature observed in the data or analyses scanning a significant fraction of the dataset require the use of dedicated software. Common and key building blocks of these codes are access to the \ac{daq} channels and to the detector component configurations. Thus, dedicated packages have been developed over the years to provide simplified and generic interfaces to these data: they rely on low-level core packages like the \texttt{FrameLib} software library~\cite{FrameLib} but calls to these functions are hidden to the users. These packages interact with the software, hardware and data of the Virgo interferometer: they are widely used within the collaboration, from daily use in the control room to DetChar studies. The two main collections of such functions are \texttt{PythonVirgoTools}~\cite{PythonVirgoTools} and \texttt{MatlabVirgoTools}, targeting Python and Matlab developers respectively.

\subsubsection{Computing Band-limited RMS}\label{sec:tools:monitoring:brmsmon}
\ac{brms} of \ac{daq} channels in specific frequency ranges are useful indicators for transient disturbances or new features in the data. For instance, low-frequency \ac{brms} of seismometer data allow to separate different contributions to the seismic noise at EGO~\cite{o3virgoenv}. Going from low to high frequencies, one can isolate successively: distant and potentially strong earthquakes; sea activity on the Tuscany coastline; anthropogenic contributions with day/night and weekly periodicities; finally, on-site activities. In addition, \ac{brms} are used to monitor the excitation of the violin modes, the resonances of the mirror suspensions.

In Virgo, various software frameworks can compute \ac{brms}. One worth-mentioning is \texttt{BRMSMon}, a dedicated software that is widely used by the environmental monitoring team and in data quality studies. In addition to generating \ac{brms}, \texttt{BRMSMon} can compare their values to thresholds (either fixed or adaptive) and logically combine the outputs of these comparisons into binary channels called {\em flags}. For instance, assuming a collection of 9 sensors installed in
different EGO buildings, one can create a flag that is active (value equal to 1) if at least 5 of these 9 sensors exceed their own threshold and inactive (value 0) otherwise. The \texttt{BRMSMon} output channels, sampled at 1~Hz, are included in the DAQ.

\label{subsubsection:BLRMS}

\subsubsection{Testing stationary and Gaussianity}\label{sec:tools:Bristol}
Several analysis tools have been implemented to perform statistical tests to verify the stationarity and Gaussianity of the data.
These properties are indeed the typical assumptions at the base of most of the statistical analyses, and in particular
of the matched filter technique~\cite{Davis1989,Abbott_2020}, which modeled \ac{gw} searches such as \texttt{MBTA}~\cite{Aubin:2020goo}, \texttt{PyCBC}~\cite{PyCBCLiveO3} and \texttt{GstLAL}~\cite{messick2017analysis} are based on.
Moreover, the onset of a non-stationary behavior of the detector can be the symptom of some hardware malfunction
or some contamination from environmental noises. In any case, it requires prompt investigations of the causes and, possibly, the actuation of adequate mitigation strategies.

\ac{bristol} provides a multi-band stationarity test based on the empirical distribution of the signal \ac{brms}~\cite{DiRenzo:2020}.
Stationarity is tested dividing these \ac{brms}' into chunks and verifying the compatibility of their empirical distribution functions by means of a two-sample Kolmogorov--Smirnov test~\cite{kolmogoroff1941confidence}.
This provides $p$-values that, compared to a previously decided significance level, indicates where in a time--frequency map the hypothesis of stationary should be rejected.
The resolution of this map is given by the duration of each chunk and that of the \ac{brms} estimates, typically one minute and one second respectively; that in frequency is determined by the band division of the spectrum for computing the \ac{brms}', which is conveniently done choosing exponentially spaced frequency intervals.
The typical output of this tool is reported in Figure~\ref{fig:bristol}, while further details about the definition of the test statistic are discussed in~\ref{appendix:BRiSTOL}.

\ac{bristol} mainly targets slow non-stationarities, that is, changes in the statistical properties of the data over time scales longer than a second; for faster transients, namely glitches, other strategies are typically used and will be described in Section~\ref{sec:tools:glitch}.

This tool has been developed in the commissioning phase preceding O3, and has been used during the run to assess the quality
of the data as part of the event validation procedure (refer to Section~\ref{sec:offlinedq:validation} for more details).

\begin{figure}[!t]
	\centering
	\includegraphics[width=.6\textwidth]{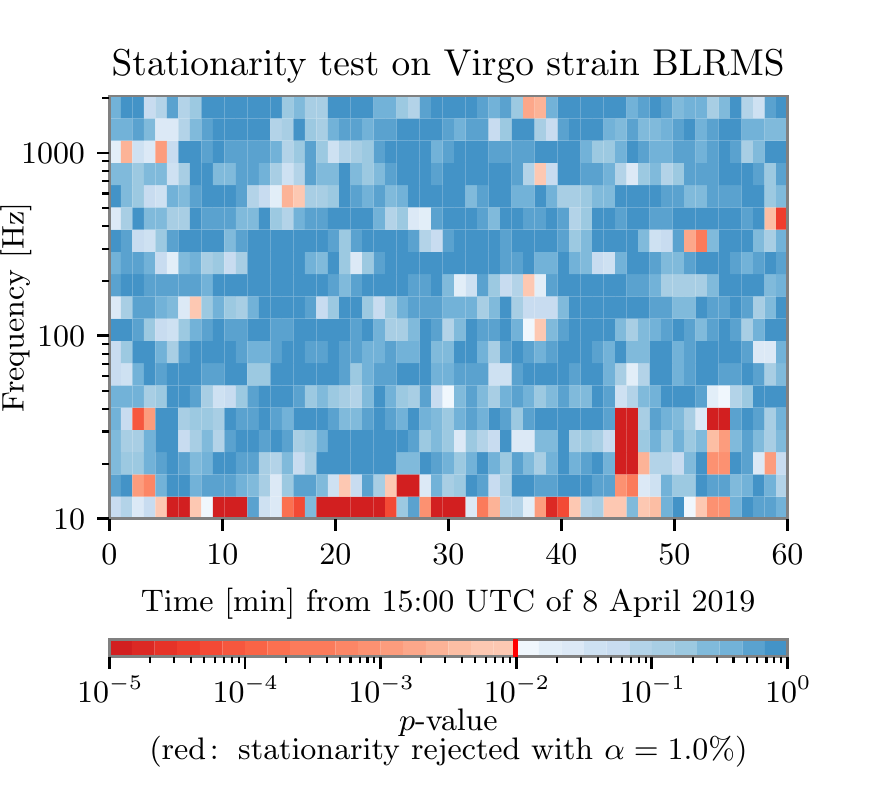}
	\caption{Example of stationarity time--frequency map obtained with \ac{bristol}, where a significance $\alpha=1\%$ has been chosen and regions rejecting the stationarity hypothesis are colored in shades of red.
		\label{fig:bristol}}
\end{figure}

\texttt{rayleighSpectro}~\cite{verkindt_spectro} is a tool to test the hypothesis of Gaussianity of the data at each frequency of its spectrum. 
This is based on the Rayleigh test~\cite{finn2001}, which is a consistency test of the \ac{asd} estimates on various data intervals.

\begin{figure}[!t]
	\centering
	\includegraphics[width=.67\textwidth]{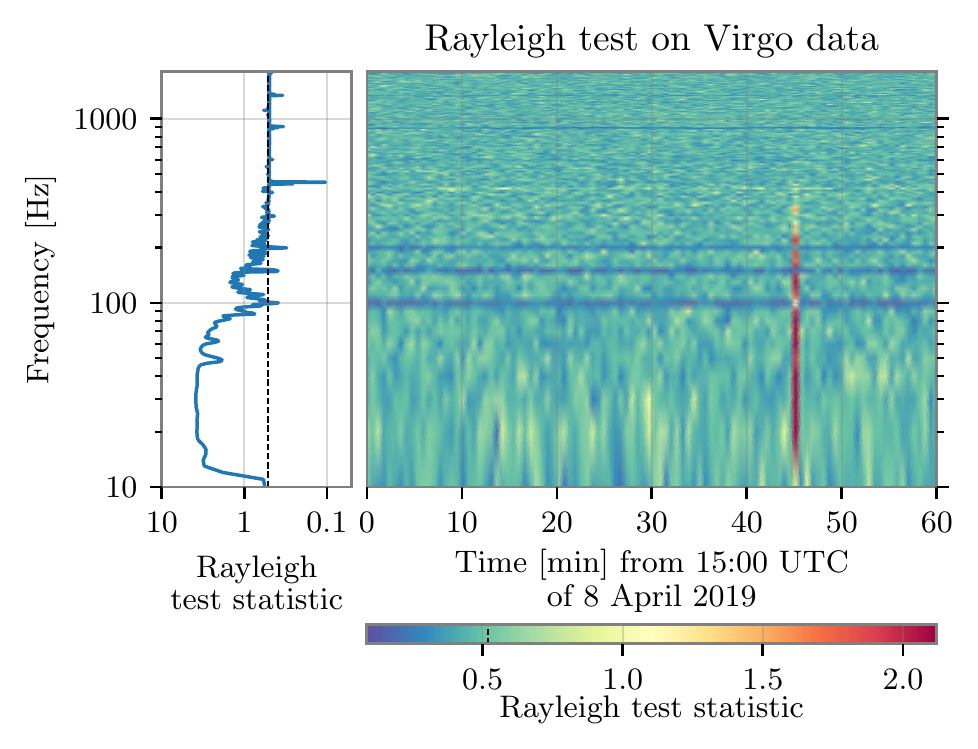}
	\caption{Example of application of the Rayleigh test, where the blue line in the left panel is the test statistic estimated over the entire hour of data, while the colormap corresponds to \ac{asd} estimates over $10$ seconds of data. The vertical black line on the left plot indicates the limit value 0.52. 
	See text for details.
		\label{fig:rayleigh}}
\end{figure}

If the data is stationary and Gaussian, the \ac{asd} estimate is drawn from a Rayleigh distribution at every frequency, and the ratio of its standard deviation and mean should be asymptotically equal to 

\begin{equation}\label{eq:Rayleigh_test_statistic}
\frac{\sqrt{4-\pi}}{\sqrt{\pi}}\simeq0.52
\end{equation}

That constitutes the test statistic. 
Deviations from this value can be both a symptom of \ac{asd} misestimation, due for example to non-stationary data, or to regions of the spectrum where the data is not compatible with a Gaussian distribution, as for example regions corresponding to spectral lines.
More details about this test are presented in~\ref{appendix:Rayleigh}.

This tool can be used complementary to \ac{bristol} to independently test stationarity and Gaussianity. 
It is included in \ac{vim} and also used in the \ac{dqr} for event validation (see  Sections~\ref{sec:tools:VIM} and~\ref{section:public_alerts}).

Figures~\ref{fig:bristol} and~\ref{fig:rayleigh} show examples of application of these two tools to one hour of data at the beginning of O3a. In the former, \ac{bristol} highlights many slow non-stationarities at frequencies up to about $20~\mathrm{Hz}$, most likely due to high microseismic activity, as well as a loud glitch at about 15:45~UTC. The latter is clearly identified by the Rayleigh test with values of the test statistic larger than what is expected for stationary and Gaussian noise. Moreover, in the colormap of Figure~\ref{fig:rayleigh}, spectral lines, in particular those 
associated with the $100$, $150$ and $200~\mathrm{Hz}$ harmonics of the mains (the European power grid frequency is 50~Hz), 
are highlighted in blue, corresponding to values of the test statistic smaller than the reference one of Equation~\eqref{eq:Rayleigh_test_statistic}. 
In the left-hand side panel of the same image, the $450~\mathrm{Hz}$ frequency of the main test masses violin modes, and its first harmonic at about $900~\mathrm{Hz}$, are highlighted as well.

\begin{figure}[!htb]
	\centering
	\includegraphics[width=.9\textwidth]{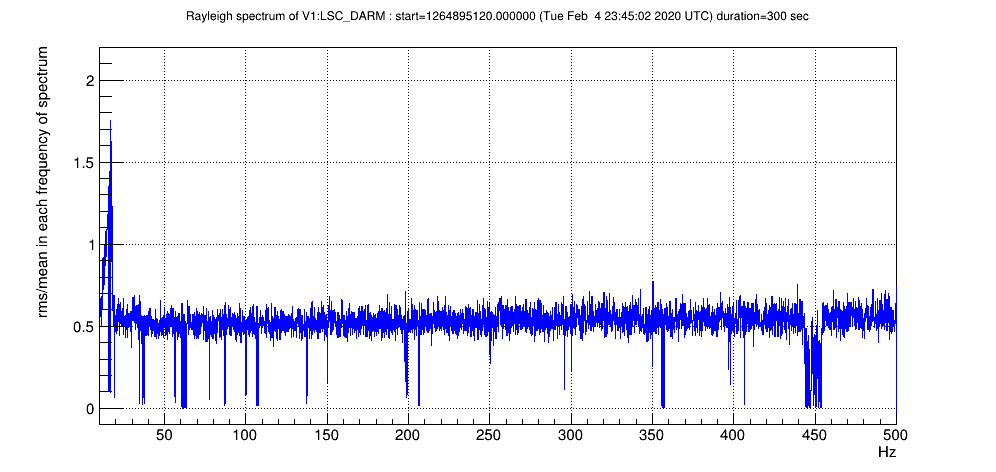}
	\caption{Example of Rayleigh spectrum averaged over 300~s, where any bin above 0.52 is a potential non-stationary or non-Gaussian noise present during those 300~s. Values well below 0.52 correspond to persistent frequency lines. 
	}
	\label{fig:rayleigh_dv}
\end{figure}

Figure~\ref{fig:rayleigh_dv} shows another example of a Rayleigh spectrum around an O3b time where transient noise was present for several minutes between 10~Hz and 20~Hz.

\subsubsection{Monitoring BNS range drops and gating data}
Two useful high-level data quality monitors are based on \ac{bns} range downwards excursions: one tags \ac{bns} range {\em drops}, that are significant dips in that quantity, while the other automatically generates (logical) {\em gates} that are applied on the \ac{gw} strain channel to smooth out to zero the data that are affected by a strong noise transient.

{\bf \ac{bns} range drops}

A \ac{bns} range drop means that the live sensitivity of the detector is degrading significantly, at least in a given frequency band, possibly in the entire bandwidth of the instrument. 
Therefore, it is important to identify transient sensitivity worsenings and investigate their causes.
\ac{bns} range drops are very diverse: the decrease goes from a few percents to almost the full range, while the drops can last from a few seconds to minutes.
 
\begin{figure}
  \center
  \includegraphics[width=0.8\textwidth]{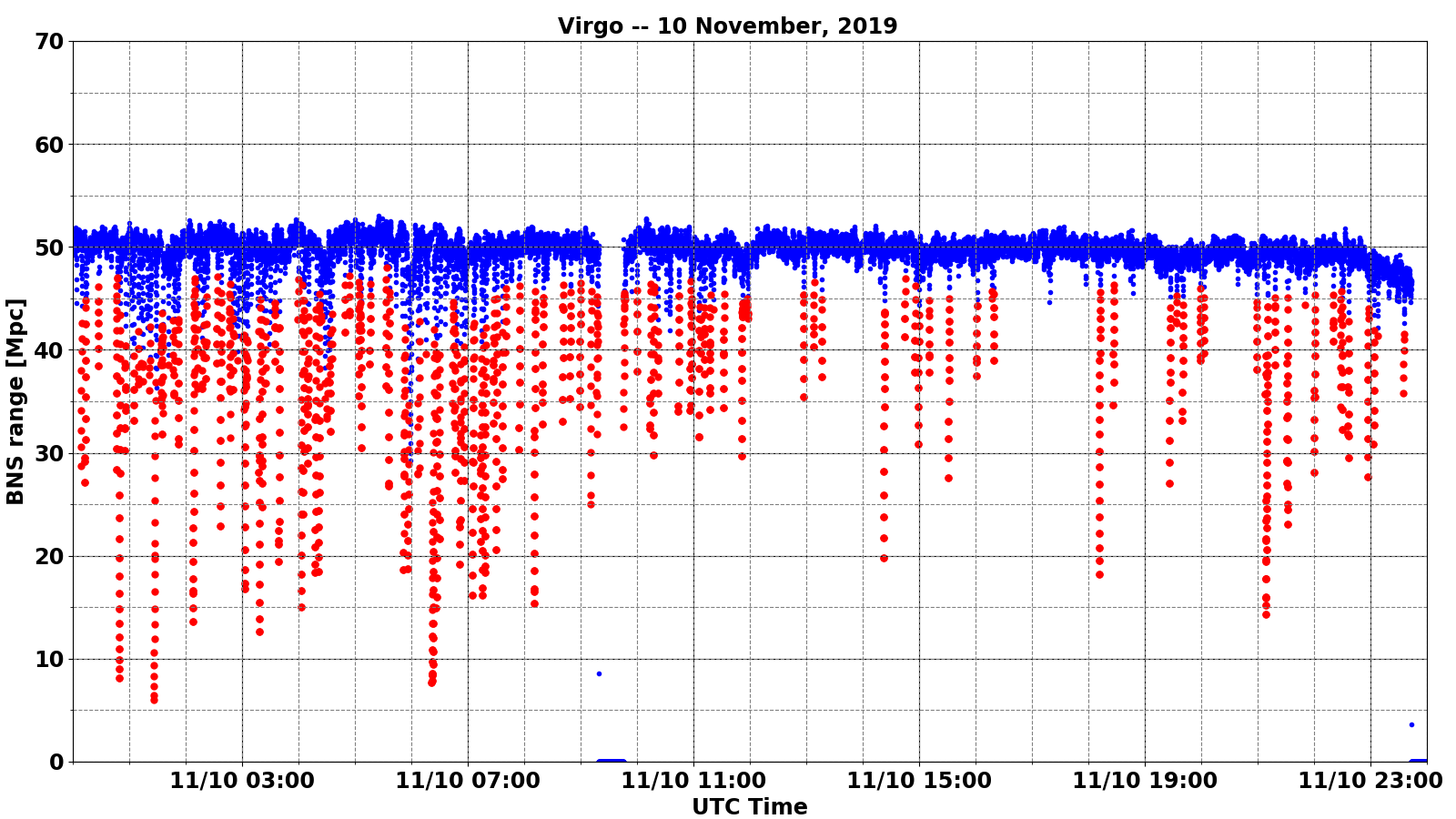}
  \includegraphics[width=0.8\textwidth]{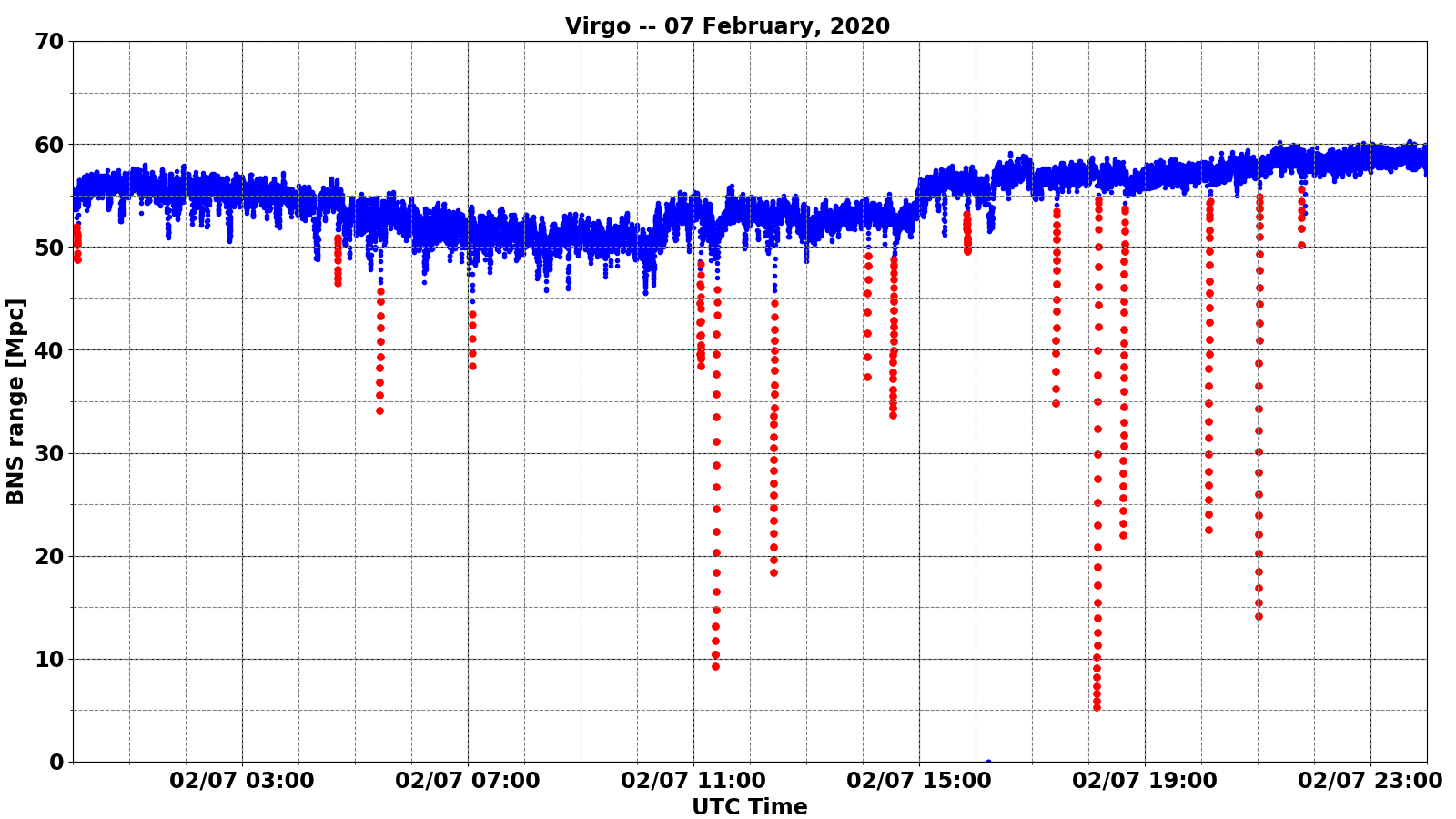}
  \caption{Performance of the \ac{bns} range drop locator during two days of O3. Top plot: November 10, 2019, a day during which the duty cycle was quite high but the data taking conditions were not stable; many glitches and consequently \ac{bns} range drops were observed, mostly due to the laser power stabilization system in the morning and to a worsening of the weather conditions starting from the afternoon. Bottom plot: February 7, 2020, a day with no global control loss but a \ac{bns} range baseline varying over time; actions took place during the afternoon to improve the Virgo performance, leading to visible improvements of the \ac{bns} range in steps.
The blue traces show the range vs. time, while the red dots show the drops that have been identified.
In both cases the \ac{bns} range drop locator is able to identify most, if not all, significant drops.
}
  \label{fig:BNS_range_drops}
\end{figure}

During O3, \ac{bns} range drops were detected using an absolute threshold on the live value of that quantity. After the end of the run, adaptative methods able to follow the natural evolution of the \ac{bns} range and to locate all significant drops have been developed.
Figure~\ref{fig:BNS_range_drops} shows examples of the output of the adaptative \ac{bns} range drop locator running on O3 data.

{\bf Gates}

If not removed from data, noise bursts can pollute the estimation of the noise spectrum for several seconds, hence limiting the sensitivity of the \ac{gw} search algorithms during that period. In Virgo, this problem is mitigated online by gating out (meaning zeroing) glitchy chunks of data. The gating algorithm triggers on significant \ac{bns} range drops: at least 40\% below its median value, computed over the last 10 seconds.
On both sides of the gate, a weight is applied on the $h(t)$ strain channel during 10/32$^{th}$ of a second, varying smoothly from 1 to 0 (0 to 1) before (after) the gate. The online gated $h(t)$ strain channel is included in the DAQ alongside the ungated one and GW searches are free to use one or the other stream as input.

As gating is based on $h(t)$ variations, gated data cannot simply be removed from the physics-analysis  dataset as this procedure could flag real \ac{gw}, for instance loud high-mass binary black hole mergers. On the other hand, gating information can be used in a statistical way to help identify potential periods of bad data quality characterized by frequent gating usage. This can be measured using both the density of gates (number per time unit) and the fraction of the wall-clock time that is gated out.

During O3, the gating algorithm has produced more than 13,000 gates (corresponding to a few tens per day in average), adding up to about 4~hours of gated data in total. The gate mean duration is around 1.1~s while the median is around 0.8~s, meaning that most gated glitches are very short as 20/32$^{th}$~s are always added to the measured glitch duration to transition from non-gated data to the gate itself and back. The longest gate is about 10~s.

Excluding from this online Science dataset the segments that have been vetoed for offline data analyses (see Section~\ref{susbsection:O3_dataset}) leads to a removal of about 20\% of the gates and of about 30\% of their total duration---although this procedure only removes about 0.2\% of the Virgo O3 dataset. As expected, gates are most likely when the data are bad. Going one step further by requesting in addition that the baseline \ac{bns} range be greater than 35~Mpc, one excludes more than 50\% of the remaining gates and more than 60\% of the gated times while that cut would remove about 1\% of the data from the final dataset. Gates are generated more often when the data taking conditions are sub-optimal.

Finally, one can associate all gates with a glitch detected by \texttt{Omicron} (see Section~\ref{sec:tools:glitch:omicron}) whereas the opposite is not true: there are many glitches that have no impact on the \ac{bns} range. These glitches have a frequency range that is outside of the Virgo bandwidth for \ac{bns} \ac{gw} waveforms: either because there is no significant signal contribution expected in this frequency range, or because the noise level is high enough to make that range contribute little if anything at all to the \ac{bns} range.

\subsubsection{Monitoring global Control losses}
Losses of the global control of the Virgo interferometer do not just interrupt the data taking: they decrease the overall duty cycle as few tens of minutes are needed after each such event to restore the conditions for taking good-quality data sensitive to the passing of \acp{gw} (see discussion in Section~\ref{subsection:O3_perf}). Therefore, categorizing control losses is important to understand their main causes and to get alerted when a new family appears, or when a known category becomes more frequent.

An extensive offline study of the global control losses in science data-taking mode during the O3 run has lead to the identification of the root cause of the control losses in most cases~\cite{o3virgoenv}. The experience gained with 
this work will be useful for the pre-O4 commissioning phase (noise hunting) and the subsequent data taking periods in two ways. First, the categories identified during O3 will be reused as a starting point to investigate new control losses. Then, an online monitor will analyze these global control losses within minutes of their occurrence; it will automatically provide a set of automated plots for further human diagnosis and possibly point out their probable cause. This framework is currently under development and will reuse the approach (if not the proper software infrastructure) of the \ac{dqr}--- see Section~\ref{subsection:DQR}.

\subsection{Glitch identification and characterization tools}\label{sec:tools:glitch}

\subsubsection{Omicron}\label{sec:tools:glitch:omicron}
To detect and characterize transient noises, we use a search algorithm called \texttt{Omicron}~\cite{omicron-softx}. The data is processed using the $Q$ transform~\cite{Brown:1991} which consists in decomposing a time series $x(t)$ onto a generic basis of complex-valued sinusoidal Gaussian functions centered on time $\tau$ and frequency $\phi$:
\begin{equation}
  X(\tau, \phi, \sigma_t) = \int_{-\infty}^{+\infty}{ x(t) \frac{W}{\sigma_t\sqrt{2\pi}}\exp{\left[-\frac{(t-\tau)^2}{2\sigma_t^2}\right]} e^{-2i\pi\phi t}dt}.
  \label{eq:qtransform}
\end{equation}
This transformation is a modification of the standard short Fourier transform in which the analysis window size $\sigma_t$ varies inversely with the frequency and is characterized by a quality factor $Q$: $\sigma_t=Q/(\sqrt{8}\pi\phi)$. The parameter space $(\tau, \phi, Q)$ is tiled to guarantee both a high detection efficiency and an optimized processing speed. The noise of the input signal $x$ is whitened prior to the $Q$ transform such that all noise frequencies have the same weight. This is done through the normalization factor $W$ which includes an estimate of the local stationary noise such that the $Q$ transform coefficient $X$ directly measures the \ac{snr} associated to each individual tile $(\tau, \phi, Q)$. A glitch in the data is detected by \texttt{Omicron} as a collection of tiles with high-\ac{snr} values. An \texttt{Omicron} glitch is characterized by a set of parameters $(\tau, \phi, Q)$ given by the tile with the highest \ac{snr} value. \texttt{Omicron} offers a two-dimensional representation of glitches where the \ac{snr} distribution of tiles is plotted in one or several $Q$ planes. Examples of spectrograms are given in Figure~\ref{fig:glitch_flowchart} and Figure~\ref{fig:S200303baOmicron}

\subsubsection{Use-percentage veto}\label{sec:tools:glitch:upv}
The \ac{upv} algorithm~\cite{Isogai:2010zz} was developed to detect and characterize noise correlations between two glitch data samples; one derived from the gravitiational-wave strain channel $h(t)$ and the other derived from an auxiliary channel. The algorithm tunes, considering \texttt{Omicron} triggers of a given auxiliary channel, a signal-to-noise ratio threshold such that, when a trigger is above threshold, there is a high probability to find a coincident glitch in $h(t)$ data. In O3, the Vigo data were processed with the \ac{upv} algorithm on a daily basis to support the noise characterization effort; some auxiliary channels were identified by \ac{upv} as exhibiting glitches correlated with $h(t)$ glitches, providing hints about the noise coupling in the detector.

\subsubsection{VetoPerf}\label{sec:tools:glitch:vetoperf}
The \texttt{VetoPerf} analysis tool measures the performance of a data quality flag. A data quality flag is defined as a list of time segments targeting transient noise events. VetoPerf counts the number of $h(t)$ triggers detected by \texttt{Omicron} which are coincident with the data quality flag time segments. From this, it derives performance numbers and produces diagnostic plots characterizing that data quality flag.

\subsubsection{Scattered light monitor}
Scattered light is a non-linear, non-stationary noise affecting the sensitivity of the interferometer in the \ac{gw} detection frequency band. As adaptive algorithms such as Empirical Mode Decomposition (EMD)~\cite{huang1998empirical,huang1999new,yang2009analysis} are suitable for the analysis of non-linear, non-stationary data, they can be used to quickly identify optical components which are sources i.e., culprits, of scattered light ~\cite{Valdes_2017}. As part of the detector characterisation effort, a tool was developed and applied to Virgo O3 data with the aim of identifying culprits of scattered light in the \ac{darm} \ac{dof} of the detector~\cite{Longo_2020sc}. The tool employs the recently developed time varying filter EMD algorithm (tvf-EMD)~\cite{Li_2017} as it was found to give more accurate results compared to EMD ~\cite{Longo_2020sc}. When scattered light is affecting the detector, arches show up in \ac{darm} spectrograms. The arches frequency and their time of occurrence is given by the so called predictor (measured in Hz)
\begin{equation}
f_{arch}(t)=2\frac{|v(t)|}{\lambda},
\label{predictor}
\end{equation}
where $v(t)$ is the velocity at which the optical component is moving and $\lambda$ is the laser wavelength. Equation~(\ref{predictor}) is computed using the position data of several optics of the detector, such as for example the \ac{sweb}. Having obtained predictors for several optical components the tool computes the instantaneous amplitudes $IA(t)$ i.e., the envelope of \ac{darm}'s oscillatory modes which are extracted by tvf-EMD. $IA(t)$ can be correlated with the list of predictors. The optical component with the highest correlation among its predictor and the $IA(t)$ of \ac{darm} is considered to be the culprit of the scattered light noise witnessed in \ac{darm}. Visual counterproof can be performed (see Figure~6 of~\cite{Longo_2020sc}) overlapping the culprit's predictor on the \ac{darm} spectrogram~\cite{Chatterji_2005}.  
The methodology of~\cite{Valdes_2017,Longo_2020sc} was extended and integrated in the \emph{gwadaptive-scattering} pipeline, an automated Python code which allowed to characterise the origin of scattered light glitches in LIGO during the O3 run~\cite{bianchi2021gwadaptive_scattering}. Furthermore, adaptive analysis can be used to daily monitor the onset and time evolution of scattered light noise in connection with microseismic noise variability ~\cite{Longo:2021avq}. So called daily analysis have been integrated in the \emph{gwadaptive-scattering} pipeline as well.

\subsection{Spectral noise identification and characterization tools}

\subsubsection{Spectrograms and injected lines identification}
\label{sec:spectro}

\begin{figure}
  \center
  \includegraphics[width=0.9\textwidth]{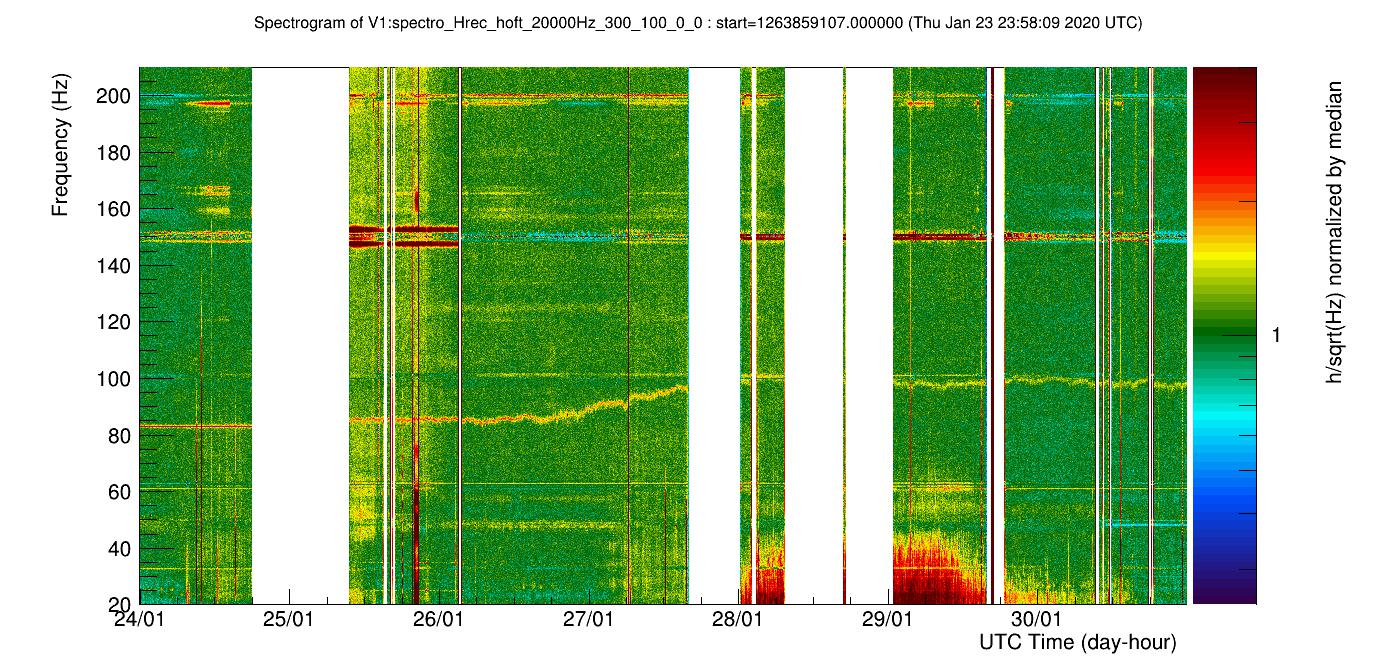}
  \caption{A typical 7-day spectrogram of the $h(t)$ channel, generated by the Spectro tool and allowing the monitoring of a wandering spectral line whose main frequency was first stable around 83~Hz before increasing up to around 100~Hz in about a day.}
  \label{fig:o3_spectro} 
\end{figure}

Within the \ac{vim} (see Section~\ref{sec:tools:VIM}), spectrograms spanning periods from one day to a week are regularly updated using the custom Spectro software~\cite{verkindt_spectro}. This framework is based on a set of ROOT~\cite{BRUN199781,rene_brun_2019_3895860} scripts that provide various indicators (\ac{brms}, Rayleigh spectra, etc.), useful to help the investigation of non-stationary spectral lines or intermittent noises. The Spectro tool has also been used during O3 to probe the time-frequency pattern of the glitches associated with \ac{bns} range drops. 
An example of time-frequency plot provided by this tool during O3 and discussed in section \ref{sec:wanderingline83} is shown in Figure~\ref{fig:o3_spectro}.

\subsubsection{NoEMi and the (known) lines database}
\label{sec:noemi}
The \ac{noemi} tool~\cite{Accadia_2012,noemiwui} tracks on a daily basis spectral lines, both stationary and wandering ones, and searches for coincidences between the lines found in a main channel---typically the GW channel $h(t)$---and in a list of auxiliary channels. The \ac{noemi} configuration defines several parameters and thresholds, like for instance: the threshold on the critical ratio\footnote{Defined as the number of standard deviations a given peak amplitude is different from the mean of the peaks amplitude distribution.} for peak selection in the spectra, the frequency resolution (linked to the time length of the data segments over which the \ac{fft} is computed), the name of the main channel, the list of auxiliary channels to search for coincidences. During O2, the \ac{noemi} software produced daily results and looked for peaks in the spectra using a frequency resolution of 1 mHz. With this configuration \ac{noemi} looked for coincident spectral peaks between the \ac{darm} channel and approximately 40 auxiliary channels.

During the break between the O2 and O3 runs, the \ac{noemi} software has been intensively modified, resolving the main issues identified in the old version. The original 
code was not well-structured (and hence difficult to modify) and also not fully-efficient CPU-wise.
Furthermore, the original version produced several static files which were unessential for the final output. As a further improvement, the MySQL database which stores all parameters of each spectral line found during the run has been normalized, meaning that useless or redundant data have been removed and that the data storage is now more coherent. The database scalability has been improved as well, in order to allow storing more data and handling a higher load of requests. Additionally, a more dynamic interaction with the web interface used to browse the results has been introduced. The new version of the code has been used for the first time in O3.

\begin{figure}
  \center
\includegraphics[width=1\textwidth]{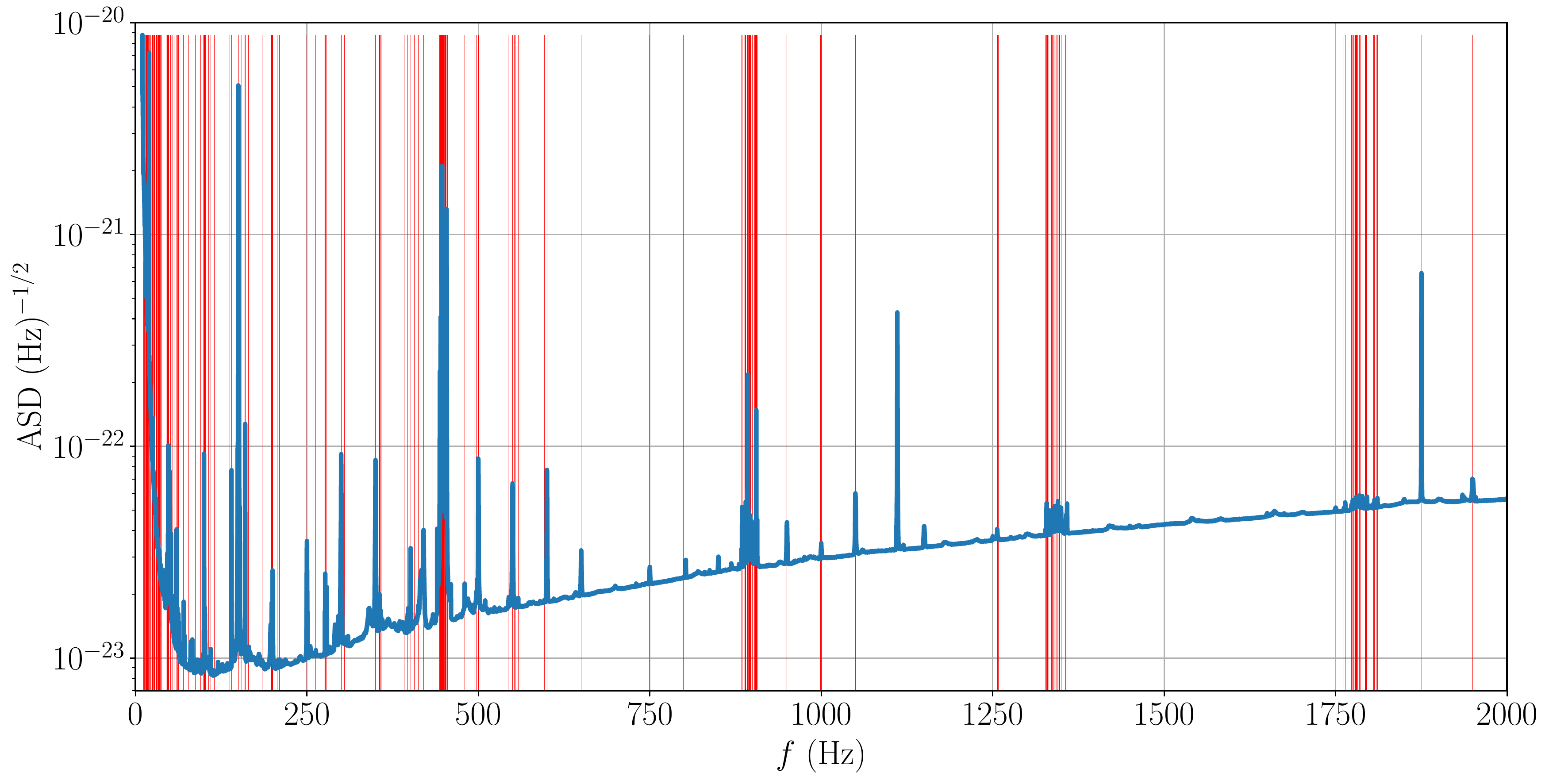}
\caption{Virgo spectral lines identified during O3. The blue curve is the  
estimated \ac{asd} of the Virgo strain signal during O3. Red vertical bars mark the frequency of the identified  
spectral lines. Lines parameters are listed in~\cite{GWOSC_O3_lines}.
Most of the lines have been found by \ac{noemi}.
}
 \label{fig:O3lines_fullband}  
\end{figure}

During O3, \ac{noemi} used the same set of ${\sim}$40 auxiliary channels as in O2, plus an additional set of ${\sim}$140 environmental channels, e.g. seismic, magnetic, and acoustic probes. 
The coincidence between a line in the GW strain signal and the signal of one of the environmental monitor, suggests that the noise line originates from a physical source such as a vacuum pump, a cooling fan, an electronic device, etc. This information helps to identify the instrumental origin of detected lines in the GW signal, and it has been included in the official Virgo-O3 line list publicly released by the \ac{gwosc}~\cite{GWOSC_O3_lines}. Figure~\ref{fig:O3lines_fullband} illustrates the lines identified in the Virgo GW strain signal during O3.

Internally, lines that have been identified are stored in a dedicated database that includes detailed informations about them: most notably their times of appearance, and links pointing to the associated documentation (logbook entries, studies, mitigation actions, etc.). The contents of the database can be compared with a new \ac{noemi} processing, to find out quickly which lines identified by \ac{noemi} are already known and which ones are not.

\subsubsection{Bruco}
\label{sec:bruco}
The \ac{bruco}~\cite{Vajente:2008bka, BRUCO_git} is a 
python-based tool designed to search for correlated noise
by computing the magnitude-squared coherence between a main channel (typically, but not necessarily, the strain signal $h(t)$) and all other non-redundant auxiliary channels (about 3,000 channels in O3).
Implementation details of the \ac{bruco} software at EGO during O3 are described in \ref{section:appendix.bruco}.

\begin{figure}[ht!]
    \centering
	{
	\includegraphics[width=1.0\textwidth]{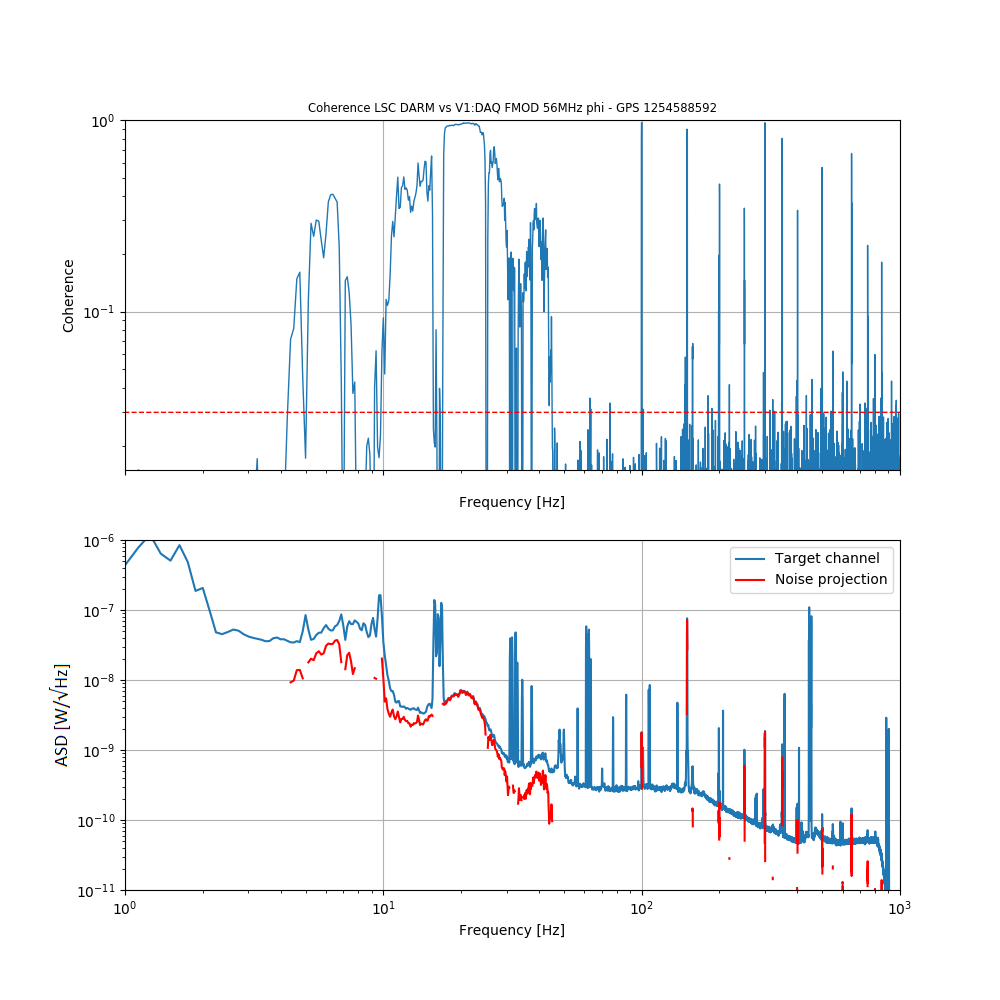}
	}
	\caption{Selection of \ac{bruco} \ac{vim} daily plots evidencing noise contaminating the Virgo strain signal during O3. The top plot shows the coherence between the \ac{darm} and the laser \ac{eom} that produces the 56~MHz signals used for the arms length control.
 The bottom plot shows the \ac{asd} of the \ac{darm} signal (blue line) and the corresponding projected coherence (red line) in the frequency ranges where it was found significant enough. The noise was then found to originate from back-reflected light onto the laser bench, most likely due to a damage on the \ac{eom} that component has been removed after O3.}
	\label{fig:bruco_1}
\end{figure}

\ac{bruco} main output is a table that contains, for each frequency bin, the ordered list of the auxiliary channels that are most coherent with the main channel. For each auxiliary channel in that list, the {\it projected coherence} (defined in~\ref{section:appendix.bruco}) is plotted and linked from the table. Assuming linear coupling, the projected coherence estimates the contribution of the noise witnessed by that auxiliary channel to the main one.
Figure~\ref{fig:bruco_1} shows \ac{bruco} daily plots illustrating one example of noise contamination spotted during O3, which triggered a more in-depth investigation~\cite{EnvHuntVirgoO3,Was2:elog}.

\ac{bruco} jobs were run regularly and automatically during the whole O3 run, with daily results
displayed on a dedicated \ac{vim} web page (see Section~\ref{sec:tools:VIM}). 
In addition, \ac{bruco} has often been used as an on-demand analysis tool 
to examine specific time periods.

\subsubsection{MONET}
\label{sec:monet}
The interferometer noise spectrum sometimes present some peculiar structures as a consequence of the non-linear couplings between different noise processes; these structures constitute two pairs of sidebands around known lines (see Sec. \ref{section:data_detchar}), which are not explained by means of the previously described linear coherence methods. One example of this kind of noise is bi-linear noise, generated by the coupling of two noise sources that jointly affect a third signal. In GW detectors, the main cause of this bi-linear noise is due to the upconversion of the low frequency seismic noise, that can affect the mirrors angular controls, which couples with some narrow-band noise processes, like power lines and calibration lines (see, e.g., \cite{2012CQGra..29o5002A,DiRenzo:2020}).

The \ac{monet},~\cite{MONET-tds}, is a python-coded tool designed to investigate these sidebands. The main hypothesis at the basis of this tool is that the sidebands are due to some coupling of a carrier signal with the low-frequency (up to a few~Hz) part of an auxiliary channel. Under this hypothesis, \ac{monet} searches for coherence between a main channel (typically, but not necessarily, the detector strain signal) and a new signal, created as the product in the time domain of the chosen carrier signal and a modulator signal. The modulator signal is constructed by applying a low-pass filter to the signal of an auxiliary channel. More details are reported in \ref{subsec:monet}, including a typical \ac{monet} output plot.

\ac{monet} has been successfully used during the commissioning phase between O2 and O3 and during O3, allowing to spot the auxiliary channels contributing to the observed sidebands; for instance, it allowed to investigate the sidebands observed around the 1111~Hz line (injected for the purpose of the laser frequency stabilization control loop)~\cite{mwas-monet:elog} and the 50~Hz harmonics~\cite{direnzo-monet:elog,fiori-monet:elog,fiori2-monet:elog,fiori3-monet:elog}.

\subsection{Common LIGO-Virgo tools}

\subsubsection{DQSEGDB}
For each data quality flag, the \ac{dqsegdb}~\cite{dqsegdb} stores the segments (integer GPS ranges) during which that particular flag is active, meaning that the set of conditions it is based on is fulfilled. For instance one such flag tags the GPS segments during which the Virgo detector is taking data in science mode, meaning that the data acquired live is expected to meet the quality criteria for physics analysis. There are two ways to fill this database with Virgo flags:
\begin{itemize}
\item online, during the data taking, through the SegOnline server that is compiling information provided by various data streams;
\item offline, by completing or fixing existing segment sets, or adding new data quality flags to monitor additional conditions.
\end{itemize}
A versioning system is used to keep track of changes in segment lists that can modify a particular flag, i.e. that impact offline analyses, by changing the contents of the dataset they are processing. By convention, the highest version number corresponds to the best (most recent) segment list and is the one queried by default.

\subsubsection{GraceDB}
During O3, the \ac{gracedb}~\cite{gracedb} has been the central place where informations about transient \ac{gw} candidates was uploaded and stored: online search triggers, source localisation estimates in the sky, data quality information, other metadata, etc. In particular, \ac{gracedb} triggered automatically frameworks like the \ac{dqr} through the \ac{lvalert}~\cite{lvalert} when candidate events of interest were identified; and, consequently, \ac{dqr} results (see Section~\ref{subsection:DQR}) got uploaded back to \ac{gracedb} as soon as they became available. \ac{gracedb} has a public-faced portal that provides information about the public alerts shared with the astronomer community, while most of its data are private and reserved to the LIGO, Virgo and KAGRA collaborations.

\section{Real-time data quality}
\label{sec:onlinedq}
\markboth{\thesection. \Sectionname}{}
Online data quality was a key challenge to tackle for DetChar during the O3 run. The availability and the reliability of that information, supporting the data taking, had to be high in order to allow the real-time transient \ac{gw} searches to make the best use of the Virgo data. Significant candidates identified by those analyses--- usually found in data from at least two of the three detectors of the global network, but sometimes significant in a single instrument--- would then lead to public alerts, used by telescopes to search for counterparts of potential \ac{gw} signals.

In this section, we first describe the different blocks of the Virgo online data quality architecture, in use at EGO during the O3 run. This framework matches the dataflow shown in Figure~\ref{fig:DataflowDetChar} and is complemented by the vetting of the most significant triggers identified in low latency, described in the following Section~\ref{section:public_alerts}. In summary, real-time information about the detector status was combined with fast data quality estimators to produce a single integer channel sampled at 1~Hz, the {\em Virgo state vector}. That state vector was shipped alongside the \ac{gw} strain channel $h(t)$ to computing centers where data were analysed in real time. Its integer value was constructed by gathering several binary information (schematically: good vs. bad) encoded as bits; that bit pattern would later be decoded by the analysis frameworks to discard any bad data. Parallel to this data analysis stream, this information---the detector status plus the real-time assessment of the data quality---was automatically uploaded by a dedicated online process (called \texttt{SegOnline}) to \ac{dqsegdb}.

Finally, we present the experience gained during O3 with additional data-quality inputs, called {\em veto streams} whose aim is to help searches to reduce their false alarm rate by identifying triggers that are very unlikely to be of astrophysical origin.

\subsection{The Virgo O3 online data quality framework}\label{sec:onlinedq:architecture}

\begin{figure}
  \center
  \includegraphics[width=0.95\textwidth]{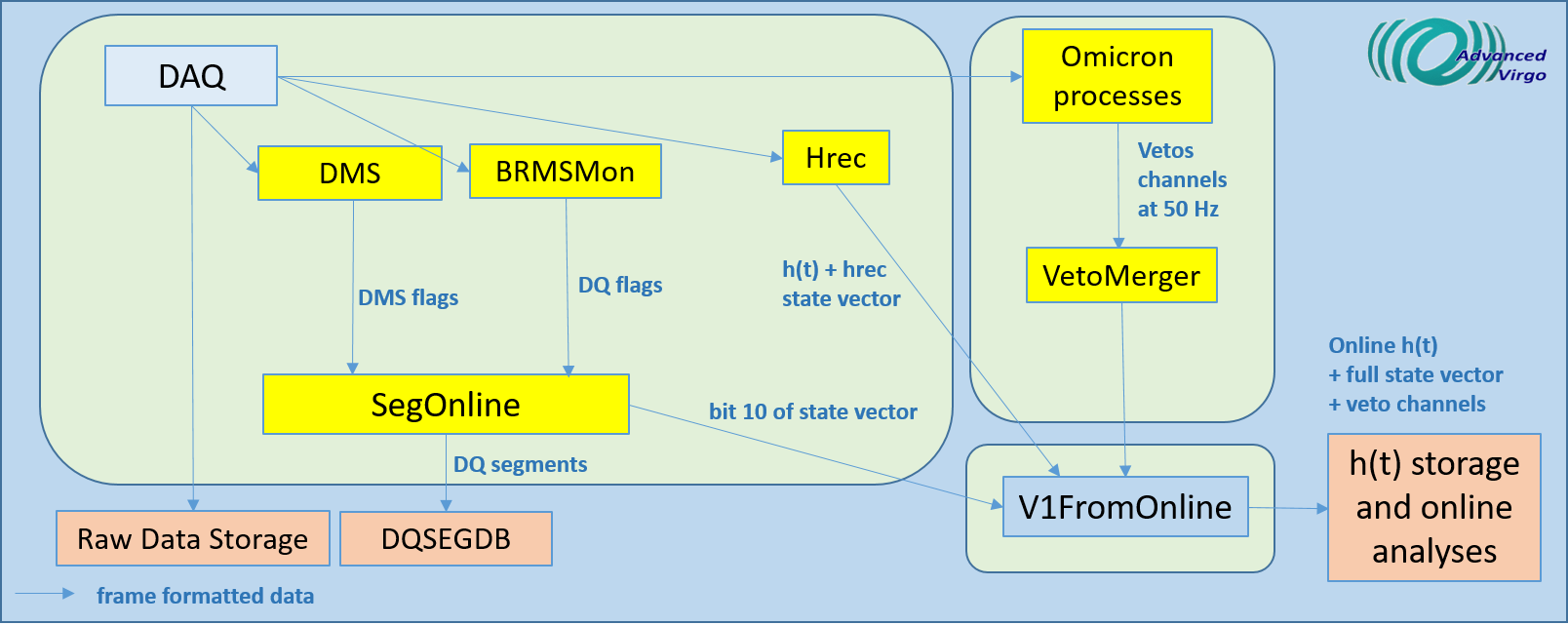}
  \caption{Online architecture to produce data quality products during the O3 run. The status of the interferometer is monitored by a dedicated \texttt{Metatron} server (see Section~\ref{subsubsection:metatron}). Data quality flags are generated by dedicated servers: the \ac{dms}, the \texttt{BRMSMon} process (environment), the \texttt{VetoMerger} process (large deviations in auxiliary signals), the Hrec process ($h(t)$ reconstruction), and the \texttt{Omicron} algorithm (glitches in $h(t)$ and auxiliary signals). Data quality segments are then generated by the \texttt{SegOnline} process and saved in the LIGO-Virgo segment database, while the online $h(t)$ stream, the state vector and the veto channels are sent to online data analysis pipelines through the \texttt{V1FromOnline} server. See text for additional informations.}
  \label{fig:onlinedq}  
\end{figure}

The online data quality architecture is designed to deliver data quality products to online transient searches.
It is based on a set of servers connected to the DAQ and providing
relevant information about the quality of the data (raw data plus the reconstructed $h(t)$ stream).
In the following, the main elements of this architecture, summarized in Figure~\ref{fig:onlinedq}, are presented.

\subsubsection{State vector}\label{sec:onlinedq:statevector}
Table~\ref{table_state_vector} defines the 16 bits of the Virgo {\em state vector} integer channel
in use during the O3 run.
A bit is said to be {\em active} when its value is 1, meaning that the corresponding check is passed. A value at 0 means instead that a problem, or a non-nominal state, has been detected. The information provided by these bits is on purpose partially redundant, in the sense that several bits can be at 0 when proper data taking conditions are not met. During O3, the bits 0, 1 and 10 were required to be active to have the 1~s data frame processed by real-time analyses.

\begin{table}[htbp!]
\caption{\label{table_state_vector}Definition of the bits of the Virgo state vector during the O3 run (see text for details).}

\begin{tabular}{cc}
\toprule
Bit number & Active when \\
\midrule
0   & $h(t)$ successfully computed. \\
\hline
1-2 & Science mode enabled. \\
\hline
3   & $h(t)$ successfully produced by the calibration pipeline. \\
\hline
\multirow{2}{*}{4-7} & Bits irrelevant for the present discussion: \\ 
                     & either redundant with other bits or unused during O3. \\
\hline
\multirow{2}{*}{8}   & No DetChar-related hardware injection \\ 
                     & (see Section~\ref{subsection:safety} for more details). \\
\hline
\multirow{3}{*}{9}   & No continuous wave hardware injection \\ 
                     & (the only type of non calibration-related injections \\ 
                     & performed for a short period during O3, while taking nominal data).\\
\hline
10  & Online data quality is good (no CAT1-type veto). \\
\hline
\multirow{2}{*}{11}  & Virgo interferometer fully controlled, \\ 
                     & with a nominal working point or close to it.\\
\hline
12-15 & Not used.\\
\bottomrule
\end{tabular}

\end{table}

\subsubsection{Online CAT1 vetoes}\label{sec:onlinedq:cat1}
During the O3 run, the problems detected online and leading to CAT1 vetoes are listed below. These saturation checks were combined using a logical OR to produce CAT1 vetoes with a 1~s granularity. Section~\ref{sec:dq_offline} describes the corresponding set of {\em offline} CAT1 vetoes, used by all analyses processing the final O3 Virgo dataset--- and obviously including these {\em online} CAT1 vetoes.

\begin{itemize}
\item No saturation of any of the 4 dark fringe photodiodes, using the `DC' (from 0 to a few~Hz) and `Audio' (from a few~Hz to 10-50~kHz) demodulated signals.
\item No saturation of the correction signal of any of the 16 suspension stages monitored.
\item No saturation of the rate of glitches reported by the online \texttt{Omicron} framework for the \ac{darm} correction channel\footnote{A more correct way to monitor the glitch rate would have been to scan $h(t)$, but the latency added by that check would have made the strain channel available too late for online processing. The offline equivalent version of that check did use $h(t)$, as latency was not an issue anymore in that case.}.
\end{itemize}

\subsubsection{SegOnline}\label{sec:onlinedq:segonline}

Any channel provided by the \ac{daq} or by the online processing (for instance \ac{dms} monitors or \texttt{BRMSMon} process)
can be used by the \texttt{SegOnline} process to build segments of data quality flags which are sent online to
\ac{dqsegdb}.

\texttt{SegOnline} writes down segments into XML files with a latency of about 10~s and those XML files are then read by a {\em rsync} process
to upload the segments into \ac{dqsegdb} every 5~min. Such data quality segments can be then used by any analysis
or can be viewed and downloaded through a dedicated web interface~\cite{dqsegdb_web}.

\subsection{Veto streams}\label{sec:onlinedq:vetostreams}
Low-latency transient searches are limited by glitches in the $h(t)$ data. Each search pipeline is sensitive to specific families of glitches. The online data quality architecture is designed to deliver a channel to flag glitches relevant to a given low-latency pipeline. These channels are called veto streams. A veto stream is a time series which can only take two values: 0 means good quality and 1 means bad quality. A veto stream is generated by the \texttt{VetoMerger} process which combines information from many online data quality processes, which must be carefully selected to target the glitches limiting the search of interest.

Some \texttt{Omicron} processes (Section~\ref{sec:tools:glitch:omicron}) are configured to select triggers detected in auxiliary channels with a signal-to-noise ratio above a threshold tuned with the \ac{upv} algorithm (Section~\ref{sec:tools:glitch:upv}). These triggers are known to witness glitches in the $h(t)$ channel. When this is the case, the veto channel is set to 1. \texttt{VetoMerger} also ingests the data quality flags generated by \texttt{BRMSMon} (Section~\ref{sec:tools:monitoring:brmsmon}) to veto environmental disturbances.

In O3, the veto stream system was experimented as an input to one of the low-latency searches for compact binary mergers, \texttt{PyCBC} Live~\cite{PyCBCLiveO2,PyCBCLiveO3}. The veto stream, named \texttt{DQ\_VETO\_PYCBC}, combined two elements: a veto channel delivered by \texttt{Omicron} to target scattered-light glitches, and a data quality flag produced by \texttt{BRMSMon} to tag occasional glitches associated to lightning strikes. A conservative approach was adopted to tune the vetoes: their thresholds were set at high values to reliably flag really limiting glitches, while keeping the rejected time low. As a result, only 0.05\% of the O3 science time was flagged by the \texttt{DQ\_VETO\_PYCBC} veto stream.
\texttt{PyCBC} Live used the veto stream to simply prevent the generation of a candidate event from Virgo data, or remove Virgo's contribution from a LIGO-Virgo candidate, during periods of active veto.
In future runs, the veto streams may be integrated in a more general framework based on auxiliary channels to discard or down-weight transient noise events.

\begin{figure}
  \center
  \includegraphics[width=0.48\textwidth]{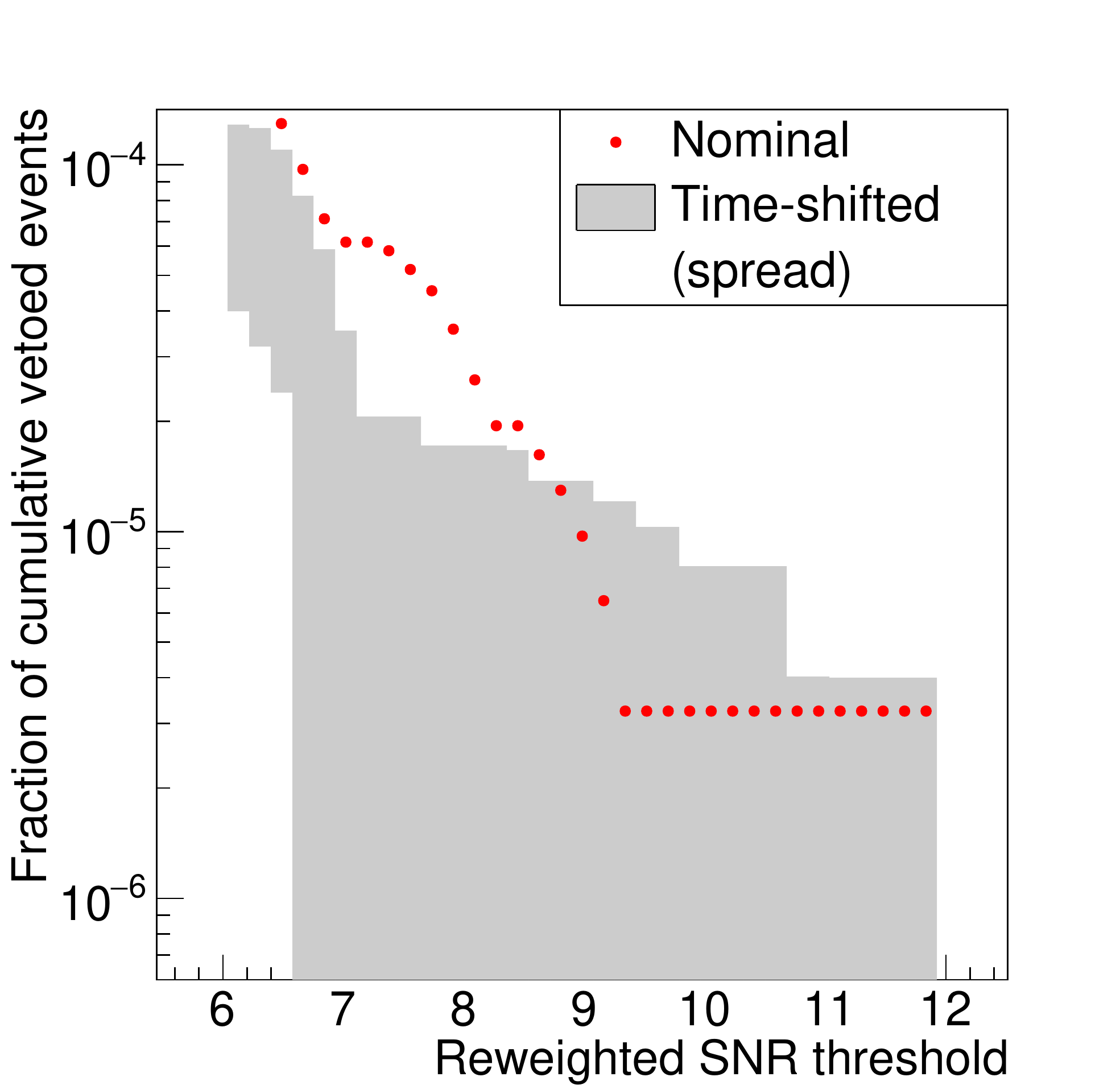}
  \includegraphics[width=0.48\textwidth]{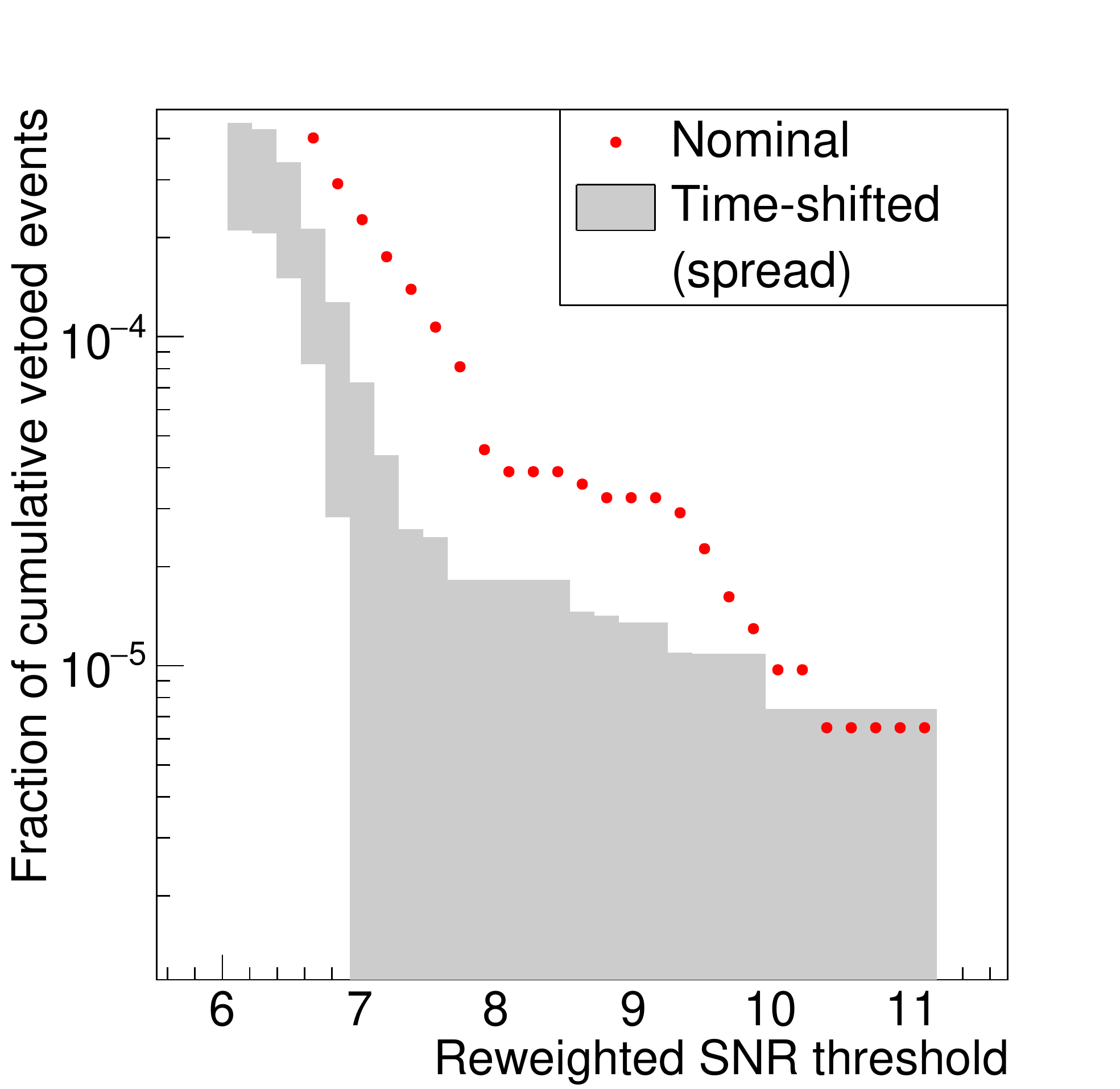}
  \caption{Cumulative fraction of vetoed \texttt{PyCBC} triggers with reweighted \ac{snr} higher than a threshold, as a function of the threshold value. The grey band shows the envelope of the fraction of rejected triggers from 1,000 time-shifted trials (covering from the minimum to the maximum obtained value), while the red points show the results obtained using the unshifted veto segments. The fractions are relative to the overall number of triggers generated by the offline \texttt{PyCBC} broad-space search. The left plot considers vetoes targeting scattered-light glitches, and the right plot considers vetoes associated with glitches from lightning.}
\label{fig:veto_streams_offline}
\end{figure}

The effect of the veto streams has also been evaluated on the \texttt{PyCBC} offline analysis, using Virgo single-detector triggers generated by the broad-space \texttt{PyCBC} search~\cite{PyCBCOfflineO3} during the period from April 1 to May 11, 2019. The triggers used for this study are ranked by the reweighted \ac{snr}, the signal-to-noise ratio returned by the matched filtering technique, weighted by the result of $\chi^2$ tests that quantify how well the time-frequency distribution of power observed in the data is consistent with the one expected from the matching template.  For practical reasons, only triggers with reweighted \ac{snr} higher than 6 are considered. To evaluate the impact of the vetoes on the offline search, triggers with a merger time belonging to a vetoed segment are removed.  The study is performed separately for vetoes targeting scattered-light glitches and glitches from lightning. The fraction of vetoed triggers is shown by the red points in Figure~\ref{fig:veto_streams_offline}, and in both cases is found to be of the order of $10^{-4}$, mainly affecting low reweighted \acp{snr}.

The statistical significance of the impact of the vetoes on the offline \texttt{PyCBC} triggers is assessed by performing a time-shifted analysis, and using it to calculate the probability that the fraction of vetoed triggers obtained from the unshifted analysis corresponds to triggers that have no correlation with the veto segments.
To this end, we shift the veto segments by a constant time offset and recompute the fraction of vetoed triggers, obtaining a ``null sample'' that we can compare to the fraction obtained using the unshifted segments.
Note that the null fraction is rescaled to account for the overlap between the science mode segments and the time-shifted veto segments.
We construct 1000 such null samples by repeating the time-shifted analysis with time offsets covering the range $[-50000, +50000]$~s in steps of 100~s.
The spread (range between minimum and maximum) of the obtained fractions of rejected triggers is shown in gray in Figure~\ref{fig:veto_streams_offline}.
At a reweighted-\ac{snr} threshold of 6, the unshifted fraction is higher than any time-shifted fraction, for both scattered-light and lightning vetoes.
We conclude that the probability for the observed effect of the vetoes on the \texttt{PyCBC} offline triggers to be a statistical fluctuation is less than $10^{-3}$.
For higher reweighted-\ac{snr} thresholds of 8.5 (10.5), this probability is $2 \times 10^{-3}$ ($2 \times 10^{-2}$) for vetoes targeting scattered-light glitches, and less than $10^{-3}$ ($6 \times 10^{-3}$) for vetoes associated with lightning glitches.
It does therefore appear that scattered light and lightning strikes are correlated with a small population of \texttt{PyCBC} triggers, and that the veto streams can in principle be used to remove or down-weight these triggers.

\section{Public alerts}
\label{section:public_alerts}
\markboth{\thesection. \Sectionname}{}
As demonstrated with the extraordinary GW170817~\cite{TheLIGOScientific:2017qsa} event from the O2 run, public alerts sent by the LIGO-Virgo network are key deliverables targeting the astronomy community. Yet, how successful these are depends on the accuracy of the information provided and of the latency at which they are delivered. For O3, the main contribution of the DetChar group to this effort has been the design and the implementation of the \ac{dqr} framework. A \ac{dqr} is a set of data quality checks, automatically triggered by the finding of a new \ac{gw} candidate. Its output allowed the \ac{rrt} team to vet the associated data in a timely way. Moreover, its usage extended way beyond the data taking period, as it was the main tool used to assess the data quality of all \ac{gw} candidates identified by analyses, in some cases with a latency longer than a year (compared to when the corresponding data were acquired).

The implementation and the performance of the Virgo O3 \acp{dqr} are described below, before summarizing how Virgo contributed to the LIGO-Virgo public alerts during the O3 run.

\subsection{Data Quality Reports}
\label{subsection:DQR}

\subsubsection{Introduction}

The \ac{dqr} is a framework developed by LIGO and Virgo for the O3 run, in order to quickly gather enough information to vet the significant triggers found by the online transient \ac{gw} searches. The goal is to either confirm the associated public alert, or have it retracted at once. All 80 public alerts delivered during O3~\cite{public_alerts} (of which 24 retracted) have used this input.

A \ac{dqr} runs on a computing cluster where the $h(t)$ strain channel and the associated raw data auxiliary channels are available in low latency. Therefore, each collaboration (Virgo at EGO and LIGO for its two detectors) was responsible for the implementation, the operation, the monitoring and the upgrade of its own \ac{dqr} framework. There was however an agreement on a common format for the check outputs, originally developed by LIGO~\cite{dqr}.

The \ac{dqr} framework is triggered by \ac{gracedb} through the \ac{lvalert} protocol. A \texttt{JSON} payload received from \ac{gracedb} allows for the generation and the configuration of a new \ac{dqr}. The checks are then processed and their results are uploaded back to \ac{gracedb}, alongside all the records associated with that particular \ac{gw} candidate.
In order for the \ac{dqr} to be triggered, a candidate event is required to have a false-alarm rate below 1/day.
This is a conservative threshold, much higher than that required to release the candidate as a public alert, but still low enough to keep the computational cost of generating the \acp{dqr} under control. 
Therefore, in average, only a handful of \acp{dqr} were automatically processed on a daily basis during the ${\approx} 330$~days of the O3 run: not a high CPU load overall, but still about 20 times more \acp{dqr} than the number of public alerts that had to be vetted.

\subsubsection{Virgo implementation and contents}

\begin{figure}
  \center
  \includegraphics[width=\textwidth]{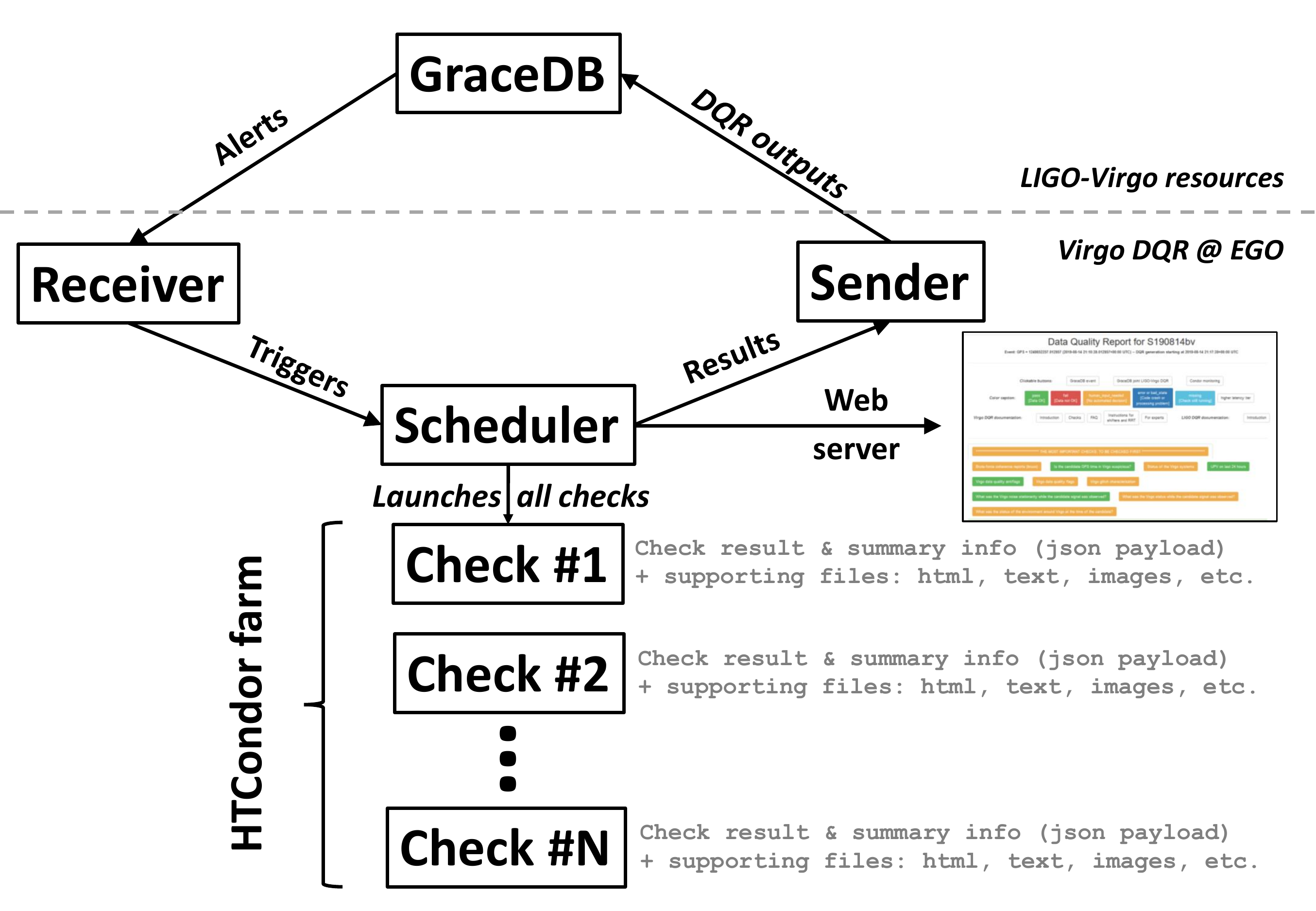}
  \caption{Schematics of the Virgo O3 \ac{dqr} architecture. See text for a description of this workflow.}
  \label{fig:DQR_architecture}  
\end{figure}

Figure~\ref{fig:DQR_architecture} summarizes the Virgo \ac{dqr} architecture used during the O3 run. When a trigger with a low-enough false alarm rate is received, a new \ac{dqr} is created and configured, using information from \ac{gracedb}. Then, the data quality checks are run in parallel on the EGO HTCondor~\cite{condor-hunter} farm. As soon as a given check is complete, its results are uploaded back to \ac{gracedb}. In parallel, the \ac{dqr} progress and results are immediately available for Virgo DetChar experts and on-duty people, through an EGO-internal web server. The \ac{dqr} format~\cite{dqr}, originally developed by LIGO, is lighter and more versatile than the \ac{gracedb} user interface: it ensures a direct access to the Virgo \ac{dqr} outputs. The \ac{dqr} webpage URL is automatically sent to the relevant internal mailing list as soon as the newly-created \ac{dqr} processing starts.

The Virgo \ac{dqr} framework has evolved quite significantly over the course of O3. Partly to tune and improve the workflow based on the experience accumulated when stressing the system during an actual run, but mainly to extend the scope of the \ac{dqr} by adding additional data quality checks. These new checks were either tests that had been foreseen but could not have been implemented by the start of O3, or new procedures that brought additional information that was found missing or useful when starting vetting real triggers.

Therefore, at the end of O3, the Virgo \ac{dqr} included 34 checks, for a total of 99 jobs. There are roughly three jobs per check: the first, to run the code and process the data; the second, to post-process the check results and convert them to the LIGO-Virgo common \ac{dqr} format; finally, the third to upload the results back to \ac{gracedb}. Some checks included a fourth job as an initial configuration phase while others, developed specifically for the \ac{dqr}, produced directly check outputs in the required \ac{dqr} output format, meaning that those checks required one job less.

The Virgo \ac{dqr} checks have been categorized in the following way.
\begin{itemize}
\item Key checks \\ They bring information mandatory to properly vet a candidate event. This includes: the top-level status of the detector at the time of the trigger; some time-frequency spectrograms of the \ac{gw} strain data at different timescales around that time; finally, the scan of the main data quality flags available online, in order to look for any obvious problem in the data.
\item Characterization of the Virgo detector noise around the time of the trigger \\ The noise transients (glitches) are inventoried and their potential overlap with the time-frequency extent expected for the candidate is probed--- if applicable. In addition, searches for noise correlations in the time domain and noise coherences in the frequency domain are run, such as tests of noise Gaussianity and stationarity.
\item Detailed Virgo status \\ Several different analysis contribute to this global picture of the instrument. All data quality flags available are checked. In addition, the \ac{dms} database is scanned to extract the snapshots closest in time to the trigger, to see what warnings or alarms were on--- if any. Also, the logfiles of all the online servers running in the \ac{daq} are scanned to spot errors that could be coincident with the trigger or impact it. Finally, various live data/reference comparison plots are generated to check the time series and distributions of a subset of the DAQ channels..
\item Digest of the environment status \\ This includes checking the seismic noise at EGO in various frequency bands corresponding to different sources (microseim related to sea activity on the Tuscany shoreline or local anthropogenic activities: see~\cite{o3virgoenv} for details), the sea activity and the weather.
\end{itemize}

\subsubsection{Performance of the Virgo O3 \ac{dqr} framework}

This section briefly summarizes the performance of the Virgo \ac{dqr}, via a statistical analysis using data from O3b that correspond to the final, most complete, version of that framework during the O3 run. As time distributions can include outliers due to occasional technical problems impacting the \ac{dqr} dataflow somewhere along its way, from \ac{gracedb} to the EGO HTCondor farm and back, the results presented in the following two tables include the $50^{\rm th}$ and $95^{\rm th}$ percentiles in addition to the mean values.

Table~\ref{table:DQR_perf_1} provides the measured latencies for the processing steps that occur upstream of the \ac{dqr}. The meaning of each row is reported below.

\begin{itemize}
\item The first figure is the difference between the time when the trigger is recorded in \ac{gracedb} and the time when the corresponding data were acquired.
\item The second measures the time needed for \ac{gracedb} to send the \ac{lvalert} and to have this message trigger the Virgo \ac{dqr} framework upon reception.
\item The third number reports the time needed to create and configure a new \ac{dqr} instance, until it is ready for processing. One should note that this duration includes a 300~s wait time, imposed in order to allow \ac{gracedb} to receive, process and gather all triggers found by the different online searches that analyse strain data in parallel and independently. The assumption is that, after these five minutes, the low-latency information available in \ac{gracedb} should be optimal and stable in the vast majority of cases. Therefore, the actual \ac{dqr} configuration phase only takes a few tens of seconds: the needed data are located in the low-latency streams just made available by the \ac{daq} and the 30+ check scripts are generated one after the other.
\item Finally, the last reported duration accounts for the time needed to start processing the \ac{dqr} on the EGO HTCondor farm. This depends on the occupancy of the farm and of the EGO internal network performance.
\end{itemize}

\begin{table}[htbp!]
\caption{\label{table:DQR_perf_1}Summary of the performance of the low-latency + Virgo \ac{dqr} dataflow during O3b, from the GPS time of a trigger to the start of the Virgo \ac{dqr} on the EGO HTCondor farm: see text for details.}
\footnotesize
\centering
    \begin{tabular}{r|ccc}
    \toprule
    \multirow{2}{*}{Operation} & \multicolumn{3}{c}{Time taken [s]} \\
    \cline{2-4}
    & Median & Mean & $95^{\rm th}$ percentile \\
    \hline
    Data acquired $\to$ Candidate on \ac{gracedb}          & 52  & 166 & 331 \\
    Candidate on \ac{gracedb} $\to$ \ac{lvalert} trigger   & 4   & 4   & 11 \\
    \ac{lvalert} trigger $\to$ Virgo \ac{dqr} configured   & 331 & 339 & 383 \\
    Virgo \ac{dqr} configured $\to$ Virgo \ac{dqr} started & 8   & 10  & 21\\
    \bottomrule
    \end{tabular}

\end{table}

We can see that the mean time elapsed between the recording of the data by the different detectors and the creation of a new record in \ac{gracedb} is under three minutes; the median time is even under one minute while the tail of the time distribution extends beyond five minutes. This includes the reconstruction of the \ac{gw} strain channels; the transfer of these data alongside the associated online data quality information to computing centers; the processing of these data by real-time \ac{gw} searches; the automated analysis of the results and the final transfer of trigger information to \ac{gracedb}. Then, the new alert is received at EGO a few seconds later, triggering the creation and the configuration of a new \ac{dqr} instance. Removing the compulsory wait time of 300~s, the \ac{dqr} configuration takes a few tens of seconds only. Finally, about 10 additional seconds are needed on average to have the first \ac{dqr} jobs processed on the EGO HTCondor farm.

Table~\ref{table:DQR_perf_2} summarizes the performance of the Virgo O3 \ac{dqr} framework in terms of running time. Each row corresponds to a category of checks. The quoted durations increase from one row to the next as each new set of checks includes the previous ones.

\begin{itemize}
\item The quick checks whose outputs are mandatory to vet a trigger take about 6 minutes to be all available, with a few minutes spread.
\item Adding information about the \texttt{Omicron} triggers around the candidate takes about 10 more minutes. During O3, this latency was dominated by the fact that \texttt{Omicron} triggers were computed in real time and stored internally by the online server: they were only written to disk every 600~s, in order to allow the framework to cope with the incoming data flow. Work will be done prior to O4 to optimize this latency and to make the \ac{dqr} aware of when the needed data have been written to disk, so that their processing can start immediately after.
\item \texttt{Omicron}-scanning all the available channels (more than 2,000 in total, with the vast majority of them sampled at 10~kHz) around the trigger time requires 15-20 additional minutes.
\item Finally, the full \ac{dqr} took from 1.5 to 2~h to complete. The longest checks were \ac{bruco} and \ac{upv}, plus the scan of all online logfiles described above. 
\end{itemize}

\begin{table}[htbp!]
\caption{\label{table:DQR_perf_2}Summary performance of the Virgo \ac{dqr} processing during the last $\sim$100 days of the O3b run. The quoted durations include the time to upload \ac{dqr} check results back to \ac{gracedb} that usually takes from $\sim$5 to $\sim$20~s.}
    \begin{center}
        \begin{tabular}{r|ccc}
        \toprule
        \multirow{2}{*}{Operation} & \multicolumn{3}{c}{Time from start [s]} \\
        \cline{2-4}
        & Median & Mean & $95^{th}$ percentile \\
        \hline
        Quick key checks & 374 & 383 & 619 \\
        Adding \texttt{Omicron} trigger distributions & 868 & 816 & 935 \\
        Adding full \texttt{Omicron} scans & 1740 & 2159 & 4690 \\
        End & 5185 & 4954 & 6330\\
        \bottomrule
        \end{tabular}
    \end{center}

\end{table}

Another key figure of merit of the \ac{dqr} framework is the number of (software) check failures per \ac{dqr} instance. Table~\ref{table:DQR_perf_3} shows the results of a statistical analysis based on the subset of the \acp{dqr} that were automatically processed in real time during O3b because the candidate false-alarm rate was below the 1/day threshold. Only 13\% (2\%) of the \ac{dqr} had 1 (2) failed checks. No exhaustive analysis of these failures has been performed, as most of these \acp{dqr} were never checked by hand because the associated trigger was not significant enough. The two main causes of problems were, however, incomplete handling of edge-cases with the input data and actual bugs in processing algorithms. These issues did not affect the rapid vetting of any public alert during O3, and this framework worked smoothly as well for the validation of the offline events found later by the archival \ac{gw} searches.
Both issues are being addressed as part of the upgrade of the \ac{dqr} framework for the O4 run.

\begin{table}[htbp!]
\caption{\label{table:DQR_perf_3}Percentages of the O3b Virgo \acp{dqr} with 0, 1 and 2 unsuccessful checks respectively.}
\centering
    \begin{tabular}{c|ccc}
    \toprule
    Number of unsuccessful checks & 0 & 1 & 2 \\
    Percentage of O3b automatically processed \acp{dqr} & 85\% & 13\% & 2\%\\
    \bottomrule
    \end{tabular}

\end{table}

\subsection{O3 public alerts}

\subsubsection{Public alerts retracted because of an issue with Virgo data}

\begin{figure}[htb!]
  \center
  \includegraphics[width=\textwidth]{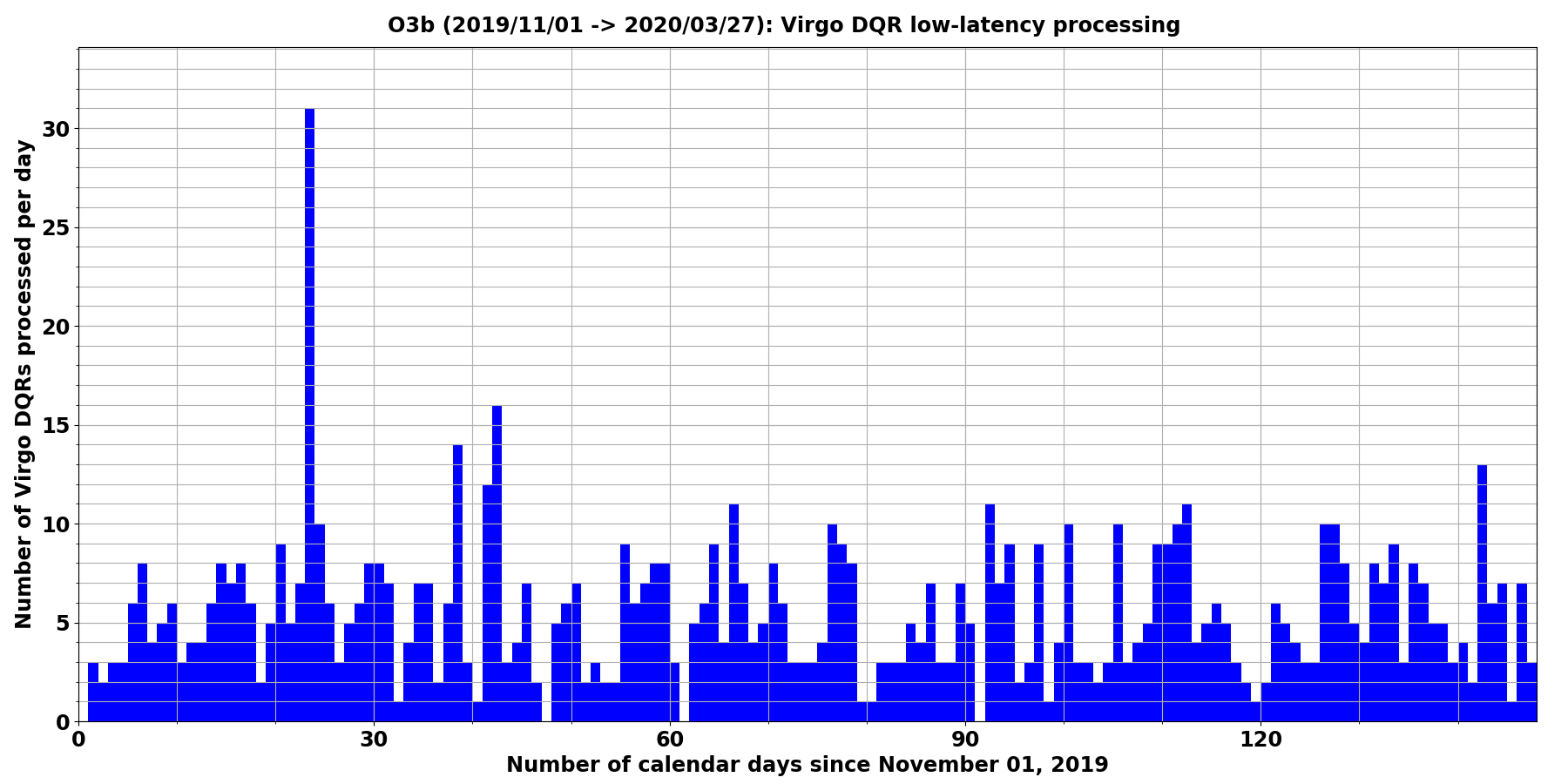}
  \caption{Number of Virgo \acp{dqr} automatically processed per day during the O3b run. The peak of 31 entries corresponds to November 24, 2019 when there was a transient problem with the Virgo $h(t)$ reconstruction: that generated several online triggers, finally including S191124be that passed the public alert threshold and was promptly retracted.}
  \label{fig:nDQRsPerDay_O3b}  
\end{figure}

During O3, 24 public alerts out of 80 have been retracted: 8 during O3a and 16 during O3b. Out of these retractions, only two were due to Virgo data:
\begin{itemize}
\item S191124be~\cite{S191124be} was due to a problem in the noise removal procedure included in the reconstruction of the $h(t)$ GW stream~\cite{VIRGO:2021umk}. Two such cleaning algorithms running in sequence started interfering, leading to a noise increase over time. An online pipeline started triggering on that excess noise, creating several non-astrophysical \ac{gw} candidates in rapid succession (Figure~\ref{fig:nDQRsPerDay_O3b}) until one of them had a false alarm rate lower than the public alert threshold. That led to the generation of an automated alert that was then quickly retracted.
\\ A similar problem should not happen again in future runs for three reasons: i) improved noise cleaning procedures are being developed within the Virgo $h(t)$ reconstruction; ii) an online monitoring dedicated to such noise removal interferences will be in place during O4; iii) a monitoring of the pipeline trigger rates in \ac{gracedb} will be running as well during future data taking periods, in order to spot quickly any misbehavior, like an excess trigger rate (thee case of S191124be) or the opposite: a too long data-taking time period without any trigger, even of low significance. 
\item S200303ba~\cite{S200303ba} was a single-pipeline trigger with most of its \ac{snr} concentrated in Virgo. At that time, Virgo data were quite noisy due to bad weather. An \texttt{Omicron}-scan around the trigger time (Figure~\ref{fig:S200303baOmicron}) showed evidence of scattering light noise at low frequency. The unusually-long delay to send out the retraction circular (about 80~min) was partly due to an issue with the Gamma-ray Coordinates Network broker connection.
\begin{figure}[htb!]
  \center
  \includegraphics[width=0.8\textwidth]{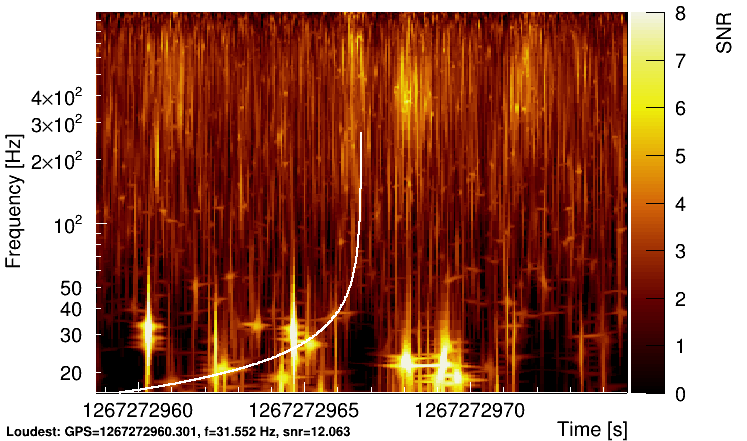}
  \caption{\texttt{Omicron} spectrogram around the time of the S200303ba trigger. The search template time-frequency track (solid line) overlaps with low-frequency scattering-light glitches (yellow color) caused by bad weather.}
  \label{fig:S200303baOmicron}  
\end{figure}

\end{itemize}

\subsubsection{Virgo contribution to O3 public alerts}

Out of the 56 non-retracted O3 public alerts, 42 involved the Virgo detector.
For 10 out of the 14 LIGO-only alerts, Virgo was not controlled in its nominal configuration at the GPS time of the trigger.
This fraction is consistent with the average duty cycle of Virgo during O3 (see Section~\ref{subsection:O3_perf}).
For the four remaining alerts, described in detail next, Virgo was fully controlled at the time of the trigger and had a \ac{bns} range consistent with its typical performance at that moment.

S190720a occurred during a $\sim 1$ min segment between lock acquisition and beginning of nominal observing mode, so Virgo data were not used for low-latency analyses.
Offline analyses later confirmed S190720a as a significant detection and were able to use the low-noise Virgo data, finding a non-negligible amount of signal power in them.
S190720a was published as GW190720\_000836 in GWTC-2~\cite{Abbott:2020niy}.

S190910d occurred during nominal observing mode in Virgo.
It was a marginal candidate, only reported by a subset of the low-latency searches.
These searches did not find a significant amount of signal power in Virgo data, and did not report Virgo as being used for the candidate.
S190910d was not confirmed by offline analyses.

S190923y occurred while Virgo was undergoing commissioning activity.
It was not confirmed by offline analyses.

S200225q occurred while Virgo was undergoing a calibration run.
Offline analyses confirmed S200225q as a significant detection and were able to include the low-noise Virgo data, although no significant signal power was found there.
S200225q was published as GW200225\_060421 in GWTC-3~\cite{GWTC3}.

\section{Global data quality studies}
\label{section:dq_studies}
\markboth{\thesection. \Sectionname}{}
This final section presents examples of global data quality studies made during or after the O3 run: noise transients, spectral analyses, 
classification of auxiliary channels based on their potential sensitivity to \ac{gw} signals
and offline data quality studies leading to the final Virgo O3 dataset.

\subsection{Glitches and pipeline triggers}

\subsubsection{Glitch rates during the O3 run}

\begin{figure}
    \begin{center}
      \includegraphics[width=0.98\columnwidth]{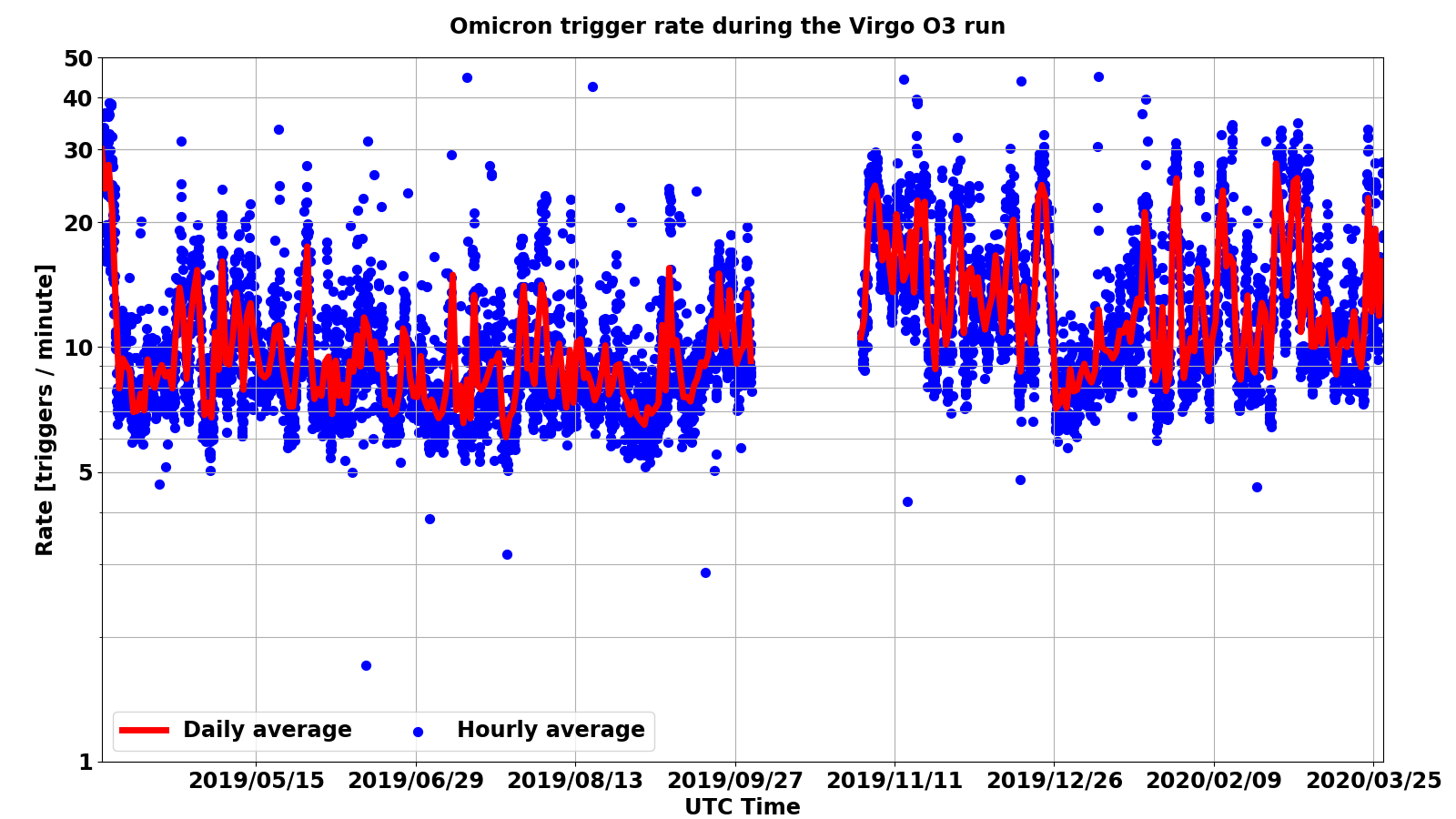}
    \end{center}
    \caption{Virgo glitch rate, using \texttt{Omicron} triggers, for the final O3 dataset (Science segments that have not been CAT1-vetoed). The blue dots are averages over one hour while the red curve shows the corresponding weekly moving average. The gap in between O3a and O3b corresponds to the 1-month commissioning break.}
    \label{fig:omicron_trigger_rates:global}
\end{figure}

\begin{figure}
    \begin{center}
      \includegraphics[width=0.98\columnwidth]{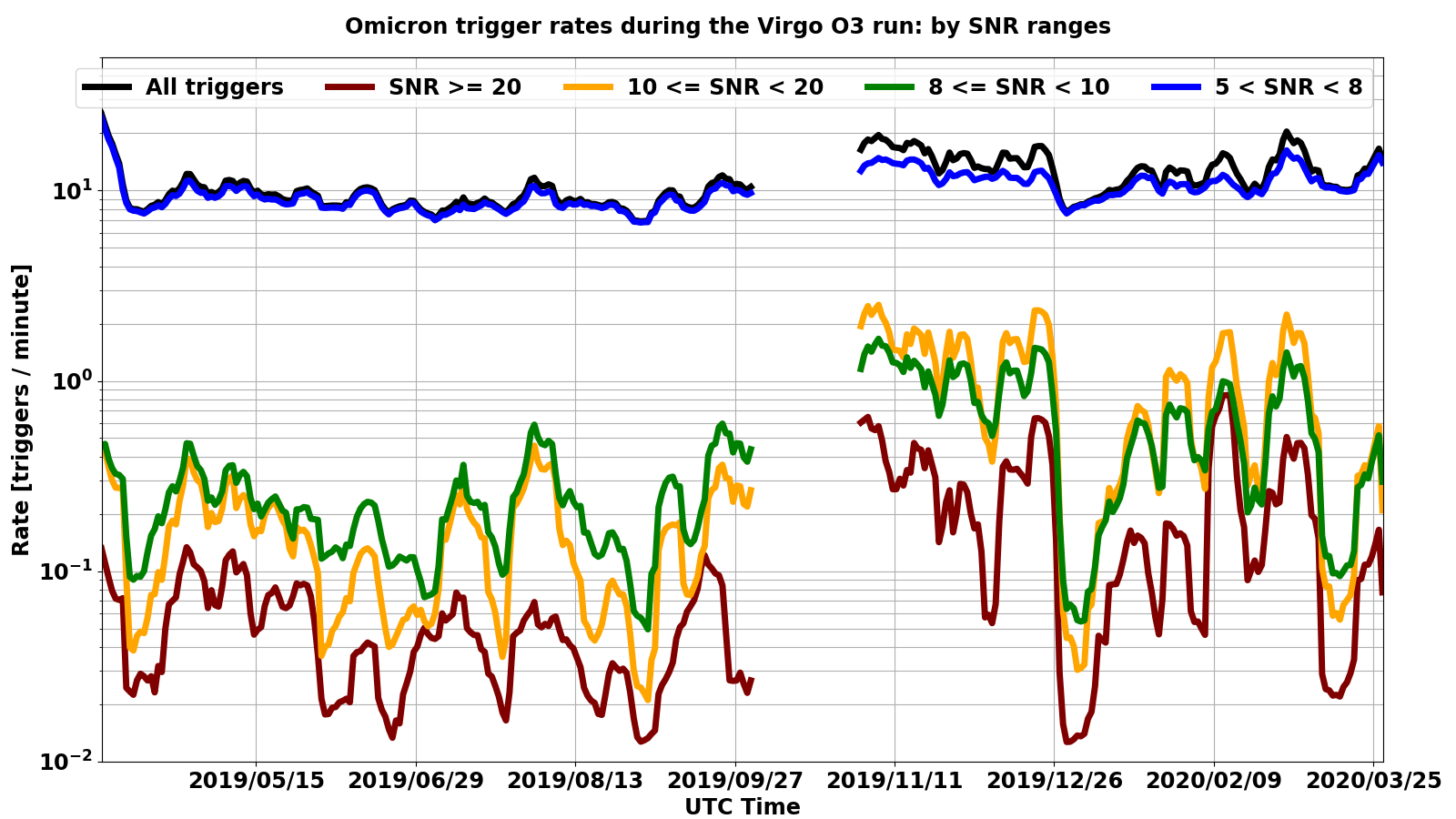}
      \includegraphics[width=0.98\columnwidth]{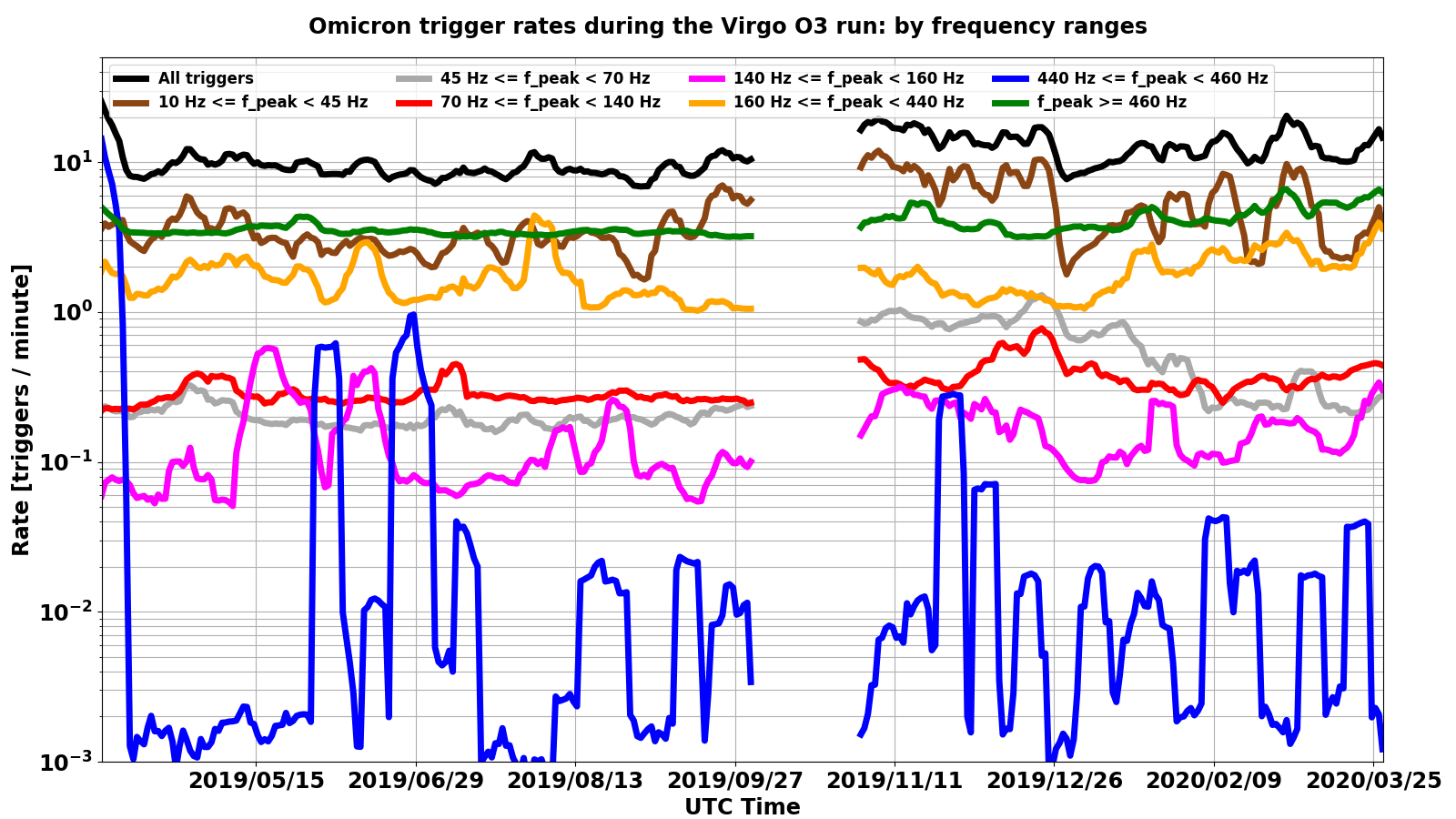}
    \end{center}
    \caption{Glitch rates (weekly moving average) using \texttt{Omicron} triggers during the O3 run for Virgo--- the gap in between O3a and O3b corresponds to the 1-month commissioning break. The top plot breaks the glitch rate into \ac{snr} ranges, while the bottom one categorizes it in terms of frequency ranges for the glitch peak frequency. The choice of the frequency bands has been mainly driven by the need to isolate some frequencies: 50~Hz (mains fundamental power in Europe), 150~Hz (second harmonics of the mains) and the range around 450~Hz (another harmonics of the mains, plus the range in which the frequencies of the suspension wire violin modes are located).}
    \label{fig:omicron_trigger_rates:details}
\end{figure}

During data taking, \texttt{Omicron} runs online on a few hundred channels, including the \ac{gw} strain $h(t)$, and monitors glitches in real time: these triggers are stored on disk with a few minutes latency. Figure~\ref{fig:omicron_trigger_rates:global} displays the evolution of the glitch rate during the O3 run. Figure~\ref{fig:omicron_trigger_rates:details} provides more details by breaking the global \texttt{Omicron} glitch rate into \acp{snr} (top plot) and peak frequencies (bottom plot). In these plots, the glitch rates have been smoothed by computing their weekly moving average to ease the reading.

The large majority of glitches identified by \texttt{Omicron} have a moderate \ac{snr}: between 5 (the minimum value from which the \texttt{Omicron} trigger is kept) and 8. The highest trigger rate at the very beginning of O3a corresponding to glitches with a peak frequency between 440 and 460~Hz is an artefact due to a mis-configuration of the \texttt{Omicron} online server that was quickly fixed. The significant increase of the trigger rate in O3b with respect to O3a is mainly due to the bad weather conditions during the fall and winter seasons (see \cite{o3virgoenv} for more details). The weather was actually very quiet in January 2020 and the associated drop in glitch rate is quite strong.

\subsubsection{Offline searches}

Non-stationary instrumental noise can potentially impact searches for transient
\ac{gw}, which must include methods to robustly separate
astrophysical candidates from noise fluctuations. Despite the power of such
methods, inspecting the candidates produced by a search remains a sensitive way
to identify problematic operating conditions of \ac{gw} detectors.

In this section, we focus on candidates produced from Virgo O3 archival data
by one of the pipelines used by the LIGO and Virgo collaborations
to detect compact binary mergers, namely \texttt{PyCBC}~\cite{PyCBCOfflineO3}.
This analysis performs a broad-space search for compact binary mergers involving
neutron stars, black holes, or a mix of both.
It uses a bank of model waveforms and matched filtering to generate candidates
from LIGO and Virgo data.
Each single-detector candidate is ranked by a combination of its matched-filter
\ac{snr} and various statistics designed to reject candidates produced by
non-stationary noise.

\begin{figure}
    \begin{center}
        \includegraphics[width=0.48\columnwidth]{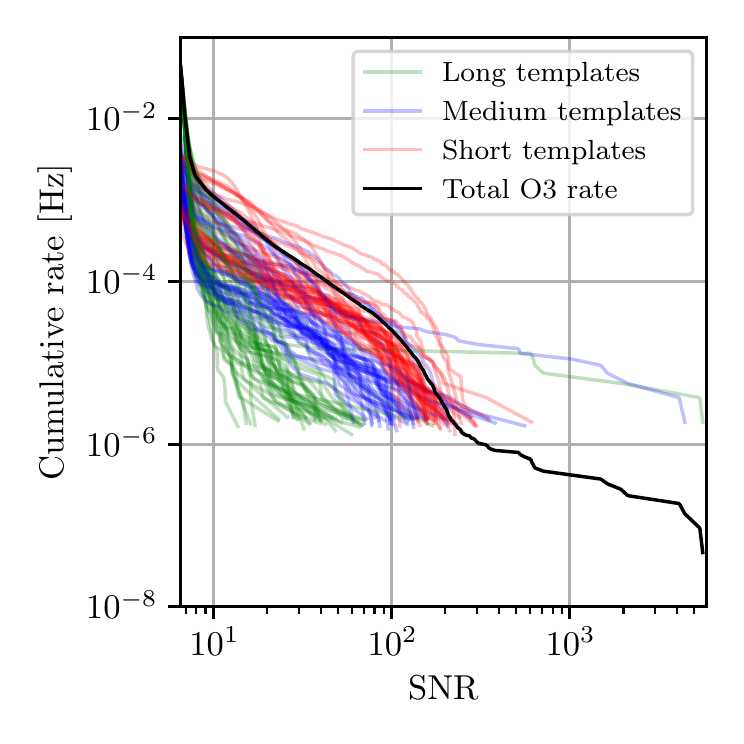}
        \includegraphics[width=0.48\columnwidth]{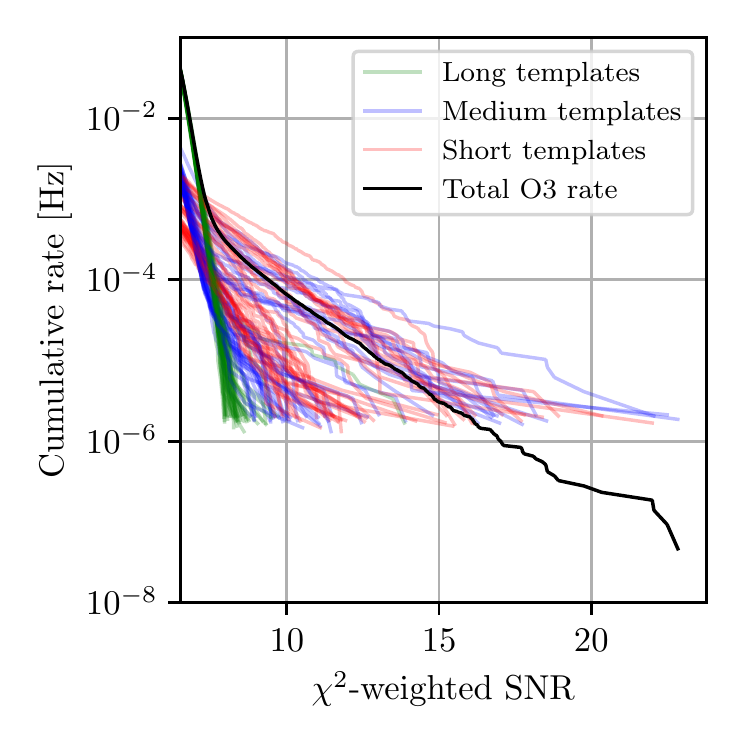}
    \end{center}
    \caption{Rate of compact binary merger candidates produced from Virgo
             O3 data using a broad-space search based on \texttt{PyCBC}.
             Left: rate as a function of the matched-filter \ac{snr}.
             Right: rate as a function of the reweighted \ac{snr},
             which combines the \ac{snr} and two $\chi^2$ discriminators.
             Red, blue and green curves correspond respectively to templates
             lasting less than $0.6$ s, $0.6$--$4$ s, and more than $4$ s.
             Curves of the same color represent different chunks of Virgo data,
             each lasting $\sim 5$ days.
             The black curves show the rate over the entire run and search space.
             They extend to lower rates than the individual chunks due to the
             much longer duration of the entire run.}
    \label{fig:pycbc_trigger_rate}
\end{figure}

Figure~\ref{fig:pycbc_trigger_rate} shows the rate of candidate events recorded
by \texttt{PyCBC} from Virgo data.
The horizontal axis shows either the matched-filter \ac{snr} of the candidate
(left plot) or a ranking statistic which combines the \ac{snr} of the candidate
and two $\chi^2$ signal-based discriminators~\cite{Allen:2004gu, Nitz:2017lco}
(right plot).
The vertical axis shows the rate of candidates that are ranked higher than the
value in the horizontal axis
\footnote{At this early stage of the \texttt{PyCBC} analysis, candidates with
merger times within fractions of a second from each other can be highly
correlated, because a given transient in the data typically ``rings off''
several templates with high overlaps between each other.
The estimated rate of candidates is biased if this correlation is not accounted for.
We do so by means of a clustering procedure: a given candidate is ignored if a
higher-ranked one exists within a time window of $\pm 5$ s.}.
If the Virgo noise had been Gaussian and stationary throughout O3, we would expect the
rate to decrease exponentially for larger and larger values of the ranking, and
be independent on the template parameters and particular chunks of data.
Instead, the rate-vs-\ac{snr} curves show a more complicated behavior, with a large
variation across the search space and particular data chunks. We observe a
non-negligible rate at \acp{snr} as high as 100, while astrophysical signals are
typically expected to have \acp{snr} between ${\sim}$5 and ${\sim}$10.
After the application of the $\chi^2$ discriminators, the behavior changes
drastically, and the exponential behavior of the rate is recovered, at least as
long as we restrict to a subset of the search space.
We still observe a large variation of the exponential slope and amplitude across
the search space and data chunks, except for the longest templates (green curves
of Figure~\ref{fig:pycbc_trigger_rate}), which are more robust to instrumental
artifacts due to their particular time-frequency signature.
However, the same variation is also seen with candidates from the LIGO detectors,
and it is taken into account by the analysis when ranking the multidetector
candidates~\cite{Nitz:2017svb}.

A detailed inspection of the candidates in the tails of these plots shows that
the highest \acp{snr} can be attributed to a single segment of  ${\sim}$15~min of data on
November 11, 2019. These data contain narrowband, loud and rapidly-varying excesses of
power (coming from transient problems with the noise subtraction algorithms used to 
reconstruct the \ac{gw} strain channel $h(t)$)
which temporarily affected the data conditioning algorithm used by \texttt{PyCBC}.
Most of these high-\ac{snr} triggers were removed by the $\chi^2$ discriminators,
effectively vetoing the entire problematic segment.
On the other hand, most top candidates by $\chi^2$-weighted \ac{snr} are clearly
associated with scattered-light glitches.
We conclude that the $\chi^2$ discriminators used by \texttt{PyCBC}, which were
designed for and tuned on LIGO data, are also reasonably effective in Virgo, and
should be further developed to more effectively reduce the impact of scattered light.

\subsection{Channel safety: channel (in)sensitivity to gravitational waves}

Many Virgo data quality analyses aim at ensuring that \ac{gw} candidates are of astrophysical origin and not caused by terrestrial noise. Typically, searches for correlations between auxiliary channels (monitoring the environment, the detector status, the accuracy of its control, etc.) and the $h(t)$ strain channel are run to produce vetoes, that reject times when such correlations are identified. This strategy can lead to a loss of interesting signals if any of the auxiliary channels is sensitive to \acp{gw}, which means that it picks up disturbances induced in the detector by these. Hence, a good knowledge of the couplings of auxiliary channels to $h(t)$ is essential. To gather such information, a statistical analysis of all auxiliary channels is performed, using the approach proposed in~\cite{Essick2021}.

This method relies on hardware injections that mimick the effects of \acp{gw} on the detector, by moving in a deterministic way one of its test masses. They are used to workaround the fact that the transfer functions between $h(t)$ and most auxiliary channels are not well-known, nor understood. The injected signals are 0.6~s long sinusoidal Gaussian functions of various frequencies (between 19~Hz and 811~Hz) and amplitudes (SNR between ${\sim}$20 and ${\sim}$500). The frequencies injected are chosen to scan the entire detection band while avoiding any known resonant frequency (like violin modes). Each waveform is injected three times, spaced by 15~s.

This {\em safety} analysis assumes that glitches in a given auxiliary channel are distributed according to a stationary Poisson process, whose rate and $p$-value time series are measured using stretches of data during which no hardware injection is performed. These $p$-value time series are used to define a classification threshold. Then, a null test is applied to see whether the $p$-value distribution changes significantly in the presence of hardware injections. Auxiliary channels that exhibit anomalously small $p$-values (i.e. lower than the defined threshold) are classified as {\em unsafe}, meaning that they are likely to mirror excess power coming from the strain channel. The other channels, called {\em safe} are the only ones used to produce vetoes.

Virgo DetChar hardware injections were organized at short notice, in the few days between the anticipated end of the O3b run (because of the pandemic) and the moment when the detector was switched off. Among the ${\sim}$2500 auxiliary channels analysed, 69 are found to be unsafe. 
The safety analysis of these data allowed to validate the existing sets of safe and unsafe channels determined by a previous study. These results matched as well the a priori safe status one could infer based on the definition of the auxiliary channels, i.e. which measurements they perform and how they do them.

\label{subsection:safety}

\subsection{Spectral noise}

The term spectral noise, introduced in Section~\ref{section:data_detchar}, identifies the class of detector disturbances appearing as an persistent excess in the noise power spectrum estimation of the data.

Spectral noise has a negative impact especially on searches for persistent \acp{gw}, which aim at detecting astrophysical or cosmological signals mainly through the identification of their spectral features.
Two typical signal categories of persistent waves are \ac{cw}~\cite{cw_review_keith} and a \ac{sgwb}~\cite{stoch_review_nelson}. The signals are very weak with respect to the already detected coalescing binary emission. Due to their persistent nature, they can be looked for in the frequency domain where the accumulated power over long observation times can show up at a detectable level, after applying effective signal processing techniques.
Moreover, some spectral features of the signals can help in discriminating them from detector noise. On the other hand, spectral noise can mask signals, or produce false candidates, in both cases reducing the search sensitivity.

Searches for persistent signals are typically run off-line, once long stretches of data have been collected.
An early identification of spectral disturbances and of their instrumental source would allow to remove, or at least reduce, the source of noise, thus improving the quality of the data. 

Different actions can be accomplished at the detector characterization level in support of data analysis.
A first action is to identify, and possibly remove, the instrumental source of spectral noises as soon as possible during a data taking period. This is a non trivial task that typically requires a significant amount of work to nail down which detector component is responsible for a given disturbance and to eliminate the noise source, which may imply to replace the noisy component (for instance a cooling fan, an electric motor, etc.)~\cite{EnvHuntVirgoO3}, to shut it down (if not needed) or to modify it properly.
This could consist, for instance, in shifting the frequency of a calibration line which non-linearly couples to another noise source, in order to move the noise line frequency into a less relevant band for the GW search~\cite{Aasi:2012wd,cw_vsr4_nb}.

A second action is the use of additional techniques to
 differentiate between possible signals and other spectral features. These methods strongly depend on the analysis and 
on the type of \ac{gw} signals searched
An example of such techniques
relies on the Doppler effect. An astrophysical \ac{cw} signal is expected to be modulated in frequency by the Doppler effect, due to the Earth rotation, which induces a shift $\Delta f (t)\simeq f_0\frac{\vec{v}(t)\cdot \hat{n}}{c}$, where $f_0$ is the source frequency, $\vec{v}$ the detector velocity, $\hat{n}$ the unit vector identifying the sky direction and $c$ the speed of light. 
\ac{cw} searches correct this Doppler effect, thus any monochromatic line present in the h(t) signal is spread
by a maximum amount of $\Delta f_{max} \simeq 10^{-4}f_0\cdot cos{\beta}$--- where $\beta$ is the ecliptic declination. This shift corresponds to up of hundreds or even thousands of frequency bins for typical \ac{cw} searches. 

Potential candidates found in the analysis lead to follow-up 
investigations to identify a possible instrumental source. This follow-up is also based on a combination of DetChar activity, to spot the source of the disturbances, and application of CW or SGWB algorithms to build confidence in the astrophysical nature of the candidate, see e.g. \cite{2022arXiv220100697T}.

Although spectral noises cannot always be removed, it is still useful to characterize them by constructing a list of noisy lines.
This list can be used to exclude those disturbing frequency bands from the analysis, or to veto candidates with frequency 
too close to those of these noisy lines

The identification of lines is typically done by automated pipelines (see Sections.~\ref{sec:spectro}, \ref{sec:noemi} and \ref{sec:bruco}), based on   
\begin{enumerate}
\item user-defined thresholds set on data power spectrum or on line {\it persistency}, defined as the fraction of \ac{fft}, compared to the total number covering the full observation time, in which the ``normalized'' power content of a given frequency bin was above such a threshold (typically set to six times the average value);
\item by highlighting coincidences or significant coherence among different channels;
\item by highlighting a pattern in time-frequency maps of the data. 
\end{enumerate}
Candidates found in \ac{gw} searches are subject to verification steps, in which the identification of possible noise counterparts is done by processing the data in the relevant frequency band and period of time and/or running manually one or more of the previously mentioned line identification pipelines, described in Sections.~\ref{sec:spectro}, \ref{sec:noemi}, \ref{sec:bruco}.  
In the following we report and discuss a few examples of lines identified in Virgo O3 data. Readers can refer to the LIGO-Virgo \ac{gwosc}~\cite{GWOSC_O3_lines} for the full official list of lines.

\subsubsection{Combs}

Combs are families of lines separated by a constant frequency interval. 
Typically, noise combs are electromagnetic disturbances 
generated by digital devices (e.g. microprocessors, programmable communication devices like logical controllers, ethernet cables, wireless repeaters) that leak into the strain signal. 
Comb lines can have 
an impact on searches for persistent \acp{gw} due to their large number and usually high strength. This makes the identification of combs an important task. There are several combs present in Virgo O3 data, which we describe in the following.

A 1~Hz spaced comb with 0~Hz offset was already present during previous runs. 
A new 1~Hz comb discovered during O3 has a 0.333~Hz offset with respect to integer frequencies. This comb was discovered 
following investigations of a line at 22.333~Hz that falls 
within a region of interest for the Vela pulsar \ac{cw} search.
The instrumental origin of the comb has been confirmed by finding lines at the same frequency in the magnetometers deployed at EGO.

\begin{figure}
    \begin{center}
        \includegraphics[width=\columnwidth]{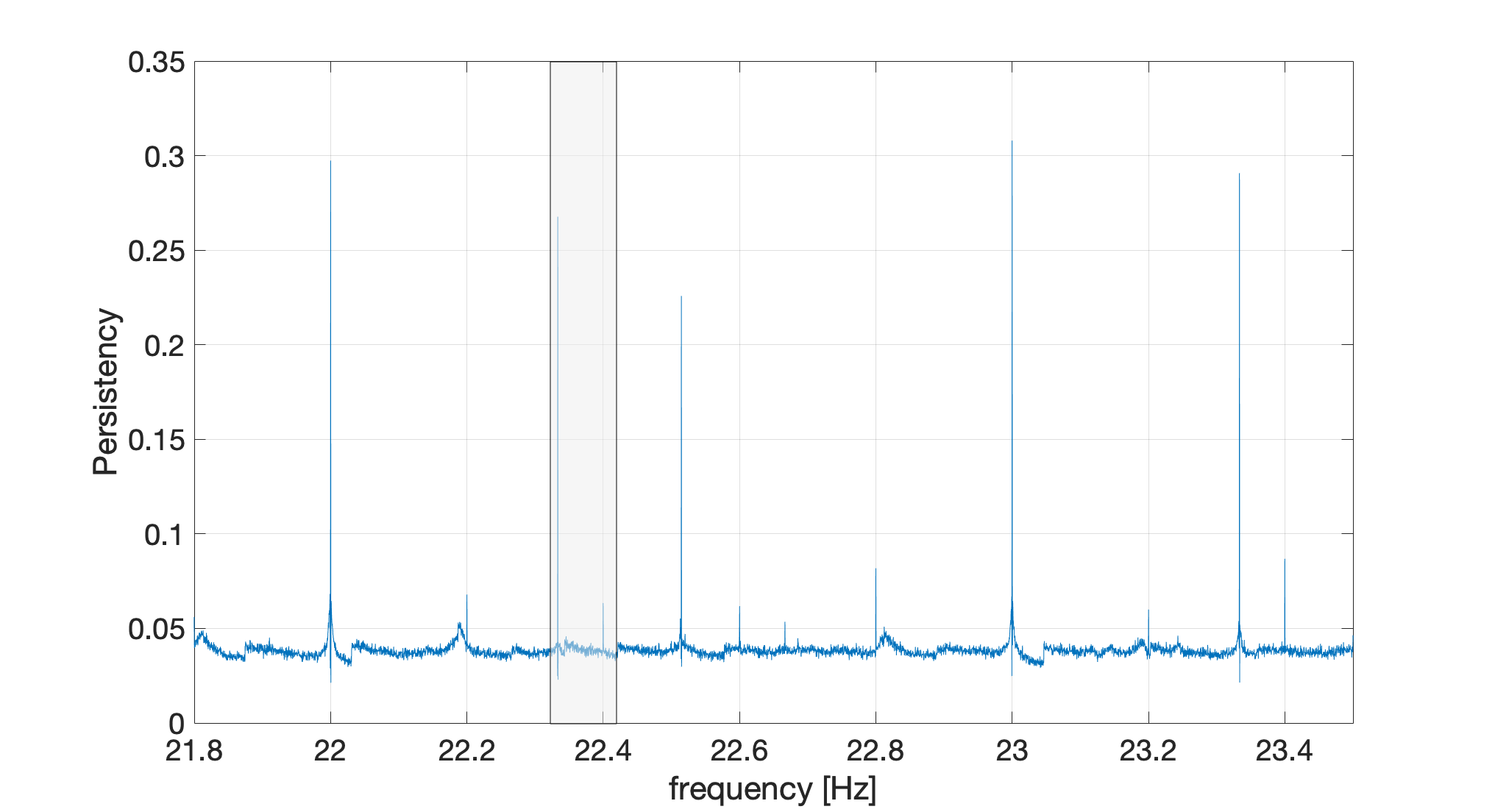}
    \end{center}
    \caption{Plot of line {\it persistency} over the frequency band 20-30~Hz. The grey box identifies the frequency range covered by a narrow-band search of \ac{cw} signals from the Vela pulsar in O3a data. The line at 23.333~Hz, which contributed to produce a candidate in the search is clearly visible. In fact, several other lines belonging to the 1~Hz {\it comb}, both at integer frequencies and shifted by 0.333~Hz, and to a weaker 0.2~Hz comb, are present.}
    \label{fig:line23}
\end{figure}

Figure~\ref{fig:line23} shows the line {\it persistency} computed over the frequency range 21.8-23.5~Hz on O3 Virgo data. Both 1~Hz combs
are clearly visible. Furthermore, there is a comb with 0.2~Hz spacing, whose origin is unknown. The grey area indicates the frequency region explored by a narrow-band \ac{cw} search targeting the Vela pulsar. The strong line at 22.333~Hz produced an outlier in the search, which was discarded after its instrumental origin was identified.

Finally, two more combs which have been identified by DetChar studies, have both $\sim$9.99~Hz spacing, one with 0~Hz offset and the other with 0.5~Hz offset.

\subsubsection{Wandering line around 83~Hz -- 84~Hz}
\label{sec:wanderingline83}

A wandering line is a peculiar kind of spectral noise where the frequency of a spectral line changes with time, with no apparent reason.
This is also called a \emph{drifting line} once the mechanism driving the frequency change is at least partially identified, making its variations not entirely random anymore.

An example that triggered lots of DetChar investigations during O3 is the line, normally located between 83 and 84~Hz, as shown in Figure~\ref{fig:o3_spectro}, that reached about 110~Hz at the maximum of its excursion and had variations of a few Hertz over about one hour~\cite{DiRenzo2019:elog,Fiori2018:elog}.
Its origin dates back to the Virgo commissioning run 10 (C10) of August 2018~\cite{DiRenzo:2020}, and possibly even earlier, in the preparatory phase preceding O2~\cite{Mantovani2017:elog}.
Neither of the mechanism that make the line to depart from its typical frequency of about 83~Hz or what produces its variations with time have ever been understood, although several data analysis techniques have been applied and newer ones developed 
for \emph{line tracking}~\cite{DiRenzo:2020}.
An analysis with Bruco (see~\ref{sec:bruco}) revealed no witness channel coherent with $h(t)$ around that line.
Moreover, we tracked the frequency evolution of this line, and we correlated the corresponding time series with the auxiliary channels monitoring Virgo~\cite{DiRenzo2019:elog}.
This technique has proven successful in the past, in the case of drifting lines driven by the temperature of some optical components~\cite{Fiori2017:elog}, but has produced no convincing correlation in the case of this line, whose origin has remained unknown. 

\subsubsection{Spectral noise bump around 55~Hz}

Figure~\ref{fig:spectral_noise}a) shows the power spectrum of the Virgo \ac{gw} strain channel $h(t)$ computed at two different dates, February 26 and March 2, 2019 (before the start of the O3 run), showing that a wide bump around 55~Hz had been cured in the meantime. Indeed, a detailed study had shown that this disturbance was present most of the time and was observed also in the \ac{prcl} channel. This allowed to remove most of this noise excess when producing the reconstructed strain $h(t)$, by accurately subtracting the remaining \ac{prcl} contribution~\cite{VIRGO:2021umk}. Note that this 55~Hz bump affected the frequencies around 55.6~Hz, where the \ac{cw} signal possibly emitted by pulsar PSR J1913+1011 is expected. Furthermore, this bump was located within the most sensitive region of the Virgo spectrum for an (isotropic) stochastic background search.

\subsubsection{Spectral noise around the 50~Hz power line frequency}

The GW strain signal in the frequency region between 45~Hz and 55~Hz was significantly affected by ambient electromagnetic fields originating from the interferometer infrastructure. This noise was studied and mitigated in subsequent steps during the run~\cite{EnvHuntVirgoO3}.

\begin{figure}
    \begin{center}
        \includegraphics[width=\columnwidth]{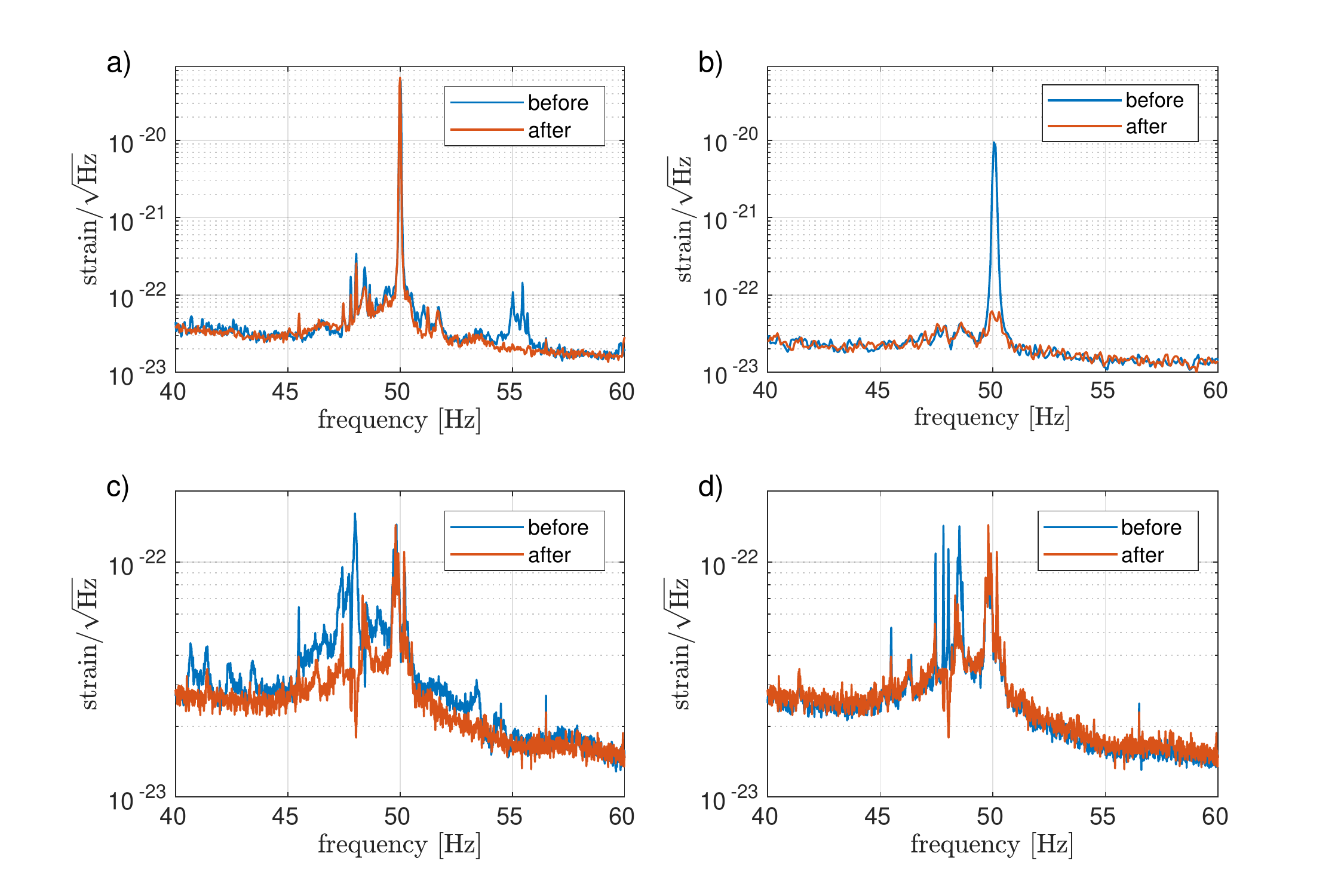}
   \end{center}
    \caption{Four steps of the reduction process of the strain spectral noise between 40~Hz and 60~Hz during O3. 
In all sub-figures the blue curve is before the mitigation, and the red curve is after the mitigation. a) Cancellation of the spectral noise structure around 55~Hz which was common to the \ac{prcl} signal.
b) Subtraction of the 50~Hz line associated to the power grid.
c) Mitigation of a wide-band noise associated to the motor driver crate of the WE suspension.
d) Suppression of a few noise lines associated to mechanical modes of the test masses payload.
}
    \label{fig:spectral_noise}
\end{figure}

The intense 50~Hz line, corresponding to the frequency of the electricity mains, was mitigated and substantially eliminated from $h(t)$ (see Figure~\ref{fig:spectral_noise}b), by implementing a feed-forward noise cancellation scheme using as sensor a voltage monitor of the detector uninterruptible power supply system~\cite{EnvHuntVirgoO3}.
This operation did not reduce the 50~Hz harmonics also present in the $h(t)$ spectrum
(see Figure~\ref{fig:O3_sensitivities}) because they are not due to a non-linear response of the interferometer. 
They are present in the global environmental disturbances and enter the GW strain channel through different 
coupling paths.

Sidebands of the mains frequency, at approximately 49.5~Hz and 50.5~Hz, were generated by the pulse width modulation of the electric heater controller of the \ac{imc} building. The noise was mitigated by decoupling the electric ground of the building from the central experimental area with an isolation transformer.

Figure~\ref{fig:spectral_noise}c) illustrates a wide-band noise affecting the same region. 
The origin of this noise was eventually found to be a noisy static voltage accidentally 
applied to the signal wires of the motors used for positioning and balancing the WE mirror suspension, then coupling capacitively to the mirror coil actuator wires. 
The noise was mitigated by un-plugging the drivers of the motors, which are not used in science mode.

Finally, Figure~\ref{fig:spectral_noise}d) illustrates a family of lines between 47~Hz and 49~Hz which
have been identified as vertical mechanical modes of the last stage of the test mass suspension
system. These modes are excited by ambient magnetic fields coupling to the magnetic actuators along the suspension chain. This noise was suppressed by an active mechanical damping of the modes.

\subsection{Offline data quality}

\subsubsection{Offline studies and checks}
\label{sec:dq_offline}

While Section~\ref{sec:onlinedq:cat1} describes the {\em online} CAT1 vetoes, we focus here on the final set of offline CAT1 vetoes. They supersede online vetoes and have been used by all analyses processing the final O3 Virgo dataset. These include analyses using the O3 LIGO-Virgo public dataset: that is why the \ac{gwosc} website~\cite{RICHABBOTT2021100658} includes detailed public information about these vetoes~\cite{GWOSC_O3_vetoes}.

Like the online CAT1 vetoes, all these veto flag segments of bad data that are unusable. They are defined with a 1~s granularity and the figure-of-merit used to quantify their impact is their dead time, that is the fraction of Science time that is removed by applying them individually. Yet, the vetoes are not independent and they may overlap. Therefore, they are meant to be applied globally on the dataset, by taking the logical OR of all of them.

The offline vetoes defined during the O3 run can be classified into three main categories.

\begin{itemize}
\item The duplication---after crosscheck and potential additions or fixes---of online CAT1 vetoes: this includes the saturations of dark fringe photodiodes or mirror suspensions, and the monitoring of the reconstructed \ac{gw} strain $h(t)$.
\item The upgrade of existing online vetoes: the excess rate of glitches is monitored offline using $h(t)$ whereas only the \ac{darm} channel could be used online due to latency constraints.
\item The addition of new vetoes, based on information that was not available in low latency, or that was not known at the time online flags were generated. These categories are described below.
\begin{itemize}
\item Checks of the consistency and of the completeness of the files storing the $h(t)$ GW stream: these vetoes flag segments in which $h(t)$ is missing or contains missing samples.
\item The $h(t)$ stream is reconstructed by blocks of eight consecutive seconds of data. Therefore, a control loss can possibly impact up to the eight seconds of data that predate it. As the exact time of a control loss is not easy to define, the last ten seconds preceding each recorded control loss have been removed.
\item The Science dataset has been scanned accurately to identify segments during which the detector was not taking good quality data, contrary to what its status was indicating. These segments were removed from the final dataset.
\item Finally, a workaround was applied to the detector control system during some weeks in O3b in order to mitigate transient data losses due the failure of an hardware component. That patch allowed to maintain the working point of the instrument, thus sparing a ${\sim}$20~min control acquisition procedure each time it prevented a global control loss. Yet, the application of that workaround could degrade the quality of the data. Thus, the impacted segments were removed from the final dataset, with some safety margin on both ends (the last 10 seconds before having the control patch be applied automatically, and the first 110 seconds following the end of the transition back to the nominal control system).
\end{itemize}
\end{itemize}

Table~\ref{table:offline} summarizes the impact of CAT1 vetoes on the final O3 Science dataset: overall, only 0.2\% of the Science data have had to be removed due to various problems.

\begin{table}[htbp!]
\caption{\label{table:offline}Virgo O3 offline Science dataset and CAT1 vetoes.}
\centering
    \begin{tabular}{c|ccc}
        \toprule
            & O3a              & O3b               & O3a + O3b\\
        \hline
        Science dataset & 12,057,731~s & 9,611,843~s & 21,669,574~s \\
        Logical OR & 18,802~s & 20,636~s & 39,438~s \\
        of all offline CAT1 vetoes & (0.16\%) & (0.22\%) & (0.18\%) \\
        \bottomrule
    \end{tabular}
\end{table}

Conversely, a few minutes of good quality data that had not been included in the online Science dataset for various and clearly understood reasons (software issue, human error, etc.) were added to the offline, final, dataset.

\subsubsection{Event validation}\label{sec:offlinedq:validation}
To assess whether the detection alerts produced by transient searches~\cite{PyCBCLiveO3,PyCBCLiveO2,Adams:2015ulm,Klimenko:2016} should be considered as ``candidate events'', a procedure of \emph{validation} is implemented after each generated trigger~\cite{GWTC3,LIGO:2021ppb}. 
This task has the role to verify if data quality issues, such as 
instrumental artifacts, environmental disturbances, etc., can impact the analysis results and decrease the confidence of a detection,
or even foster a rejection~\cite{2019ApJ_low-lat}.

The validation of the online triggers found by \ac{gw} transient searches includes two separate stages.
A prompt evaluation is typically completed within few tens of minutes after an event trigger has been generated, as represented by the data flow in Figure~\ref{fig:DataflowDetChar}.
Its goal is to determine a preliminary detection confidence and sky localization, in order to deliver public alerts to the astronomy community and support for multi-messenger follow up observations~\cite{2019ApJ_low-lat}, as described in Section~\ref{section:public_alerts}, or to vet that trigger if evidence of severe contamination from non-astrophysical artifacts is present.
A team of DetChar shifters is in charge of this task as part of the rapid-response team (Section~\ref{section:oncall_RRT}). 
The decision about the event is primarily based on the quick results provided by the \ac{dqr} within a few minutes from the trigger.
This decision takes into account the evaluation of the operational status of the detector and its subsystems, the environmental conditions, as well as preliminary checks on the strain data.
In particular, the shifters are asked to verify the presence of excess noise, namely glitches, around the time of the trigger and the validity of the hypotheses of stationarity and Gaussianity of the data, as discussed in Sections~\ref{sec:tools:Bristol} and~\ref{sec:tools:glitch}.
Moreover, it is examined the possible presence of correlations between the strain data and the auxiliary sensors, which may advise a non-astrophysical origin of the trigger.  

With higher latency, a second stage of validation is performed by a \emph{validation team} to finally check candidate events before publications, including those found by offline analyses~\cite{Aubin:2020goo,PyCBCOfflineO3}.
Besides of (double-)checking the astrophysical origin of the event trigger, the main purpose of this process is to carefully assess whether the parameter estimation of the source properties can be affected by noise artifacts.   
This procedure takes advantage of dedicated reruns of the \ac{dqr}, as well as from additional tools and metrics, including, for example, signal consistency checks~\cite{LIGO:2021ppb,Mozzon:2020gwa}.

For those events where non-stationary noise, such as glitches, are found in the vicinity of the putative \ac{gw} signal, or even overlapping with it, a procedure of noise \emph{mitigation} is implemented~\cite{Davis:2022dnd,Cornish:2014kda}.
During O3b, such process has involved 12 events, including one with Virgo data, GW191105e~\cite{GWTC3,GCN26182}, where the process of mitigation and validation of the data quality has improved the parameter estimation results and credibility. Various O3a events have undertaken a preliminary version of this procedure~\cite{Abbott:2020niy}.

\section{Preparation of the O4 run}
\label{section:outlook}
\markboth{\thesection. \Sectionname}{}
The LIGO-Virgo O3 run has lead to the discovery of dozens of new \ac{gw} signals from compact binary mergers, which have boosted our knowledge of these populations in our local Universe and allowed further, more stringent, tests of general relativity. The O3 run has also been the first long data-taking period for the \ac{adv} detector. Thus, it represents a full-scale, extended and non-stop stress test of the organization and work methods of the Virgo DetChar group. The experience accumulated during these 11 months will form the base of the DetChar activities, both to prepare and operate for O4 and the following runs.

Although the Virgo DetChar group has fulfilled all its main requirements during the O3 run, work has been going on since then to improve its performance and extend its activities. In particular, the anticipated differences between the O3 and O4 runs lead to new challenges that the group should tackle. The \ac{adv} detector will have evolved significantly, with the completion of the Phase I of the \ac{adv}+ project. The main changes on the instrument side are the addition of the signal-recycling mirror in between the beam splitter and the output port of the Virgo interferometer, a higher input laser power and the implementation of frequency-dependent squeezing. This new configuration will require dedicated instrument characterization activities, while many new data quality features will have to be discovered, understood and later mitigated or solved. On the data analysis side, progress in terms of sensitivity while keeping the network duty cycle high will lead to more \ac{gw} detections. On the one hand, more work will be required to validate this excess of signal candidates compared to O3; on the other hand, the triggers passing a given false alarm rate threshold will remain dominated by noises, meaning that the bulk of computing resources used by the Virgo DetChar group will not change significantly.

Gathering experience from the past and predictions for the future, a few top priorities have emerged for the DetChar group. A first and obvious one is to broaden the scope of the DetChar monitoring, to make sure that no relevant area remains uncovered, from raw data to the final analyses. Then, the latency of the various DetChar products should be decreased when it is relevant and possible: either by making the corresponding software framework more efficient, or by processing new data more regularly. Finally, some emphasis should be put on increasing the automation of the DetChar analyses and the reporting of their results. In that respect, the \ac{dqr} is a good example of the realization of these plans. Parallel to common LIGO-Virgo-KAGRA developments on the framework architecture to make \acp{dqr} more uniform among the three collaborations and to improve its performance, additional data quality checks will be implemented. They will provide combined results that should give a partial digest of the global vetting of a given \ac{gw} signal candidate. 

The increase of the information available and the help to identify quickly its most relevant points should allow maintaining, if not improving, the high and steady level of Virgo performances observed during O3.

\section*{Acknowledgements}
{\bf September 2021 version -- \url{https://tds.virgo-gw.eu/ql/?c=17224}}

The authors gratefully acknowledge the Italian Istituto Nazionale di Fisica
Nucleare (INFN), the French Centre National de la Recherche Scientifique (CNRS)
and the Netherlands Organization for Scientific Research (NWO), for the construction
and operation of the Virgo detector and the creation and support of
the EGO consortium. The authors also gratefully acknowledge research support
from these agencies as well as by the Spanish Agencia Estatal de Investigaci\'on,
the Consellera d'Innovaci\'o, Universitats, Ci\`encia i Societat Digital de la
Generalitat Valenciana and the CERCA Programme Generalitat de Catalunya,
Spain, the National Science Centre of Poland and the European Union--- European
Regional Development Fund; Foundation for Polish Science (FNP), the
Hungarian Scientific Research Fund (OTKA), the French Lyon Institute of Origins
(LIO), the Belgian Fonds de la Recherche Scientifique (FRS-FNRS), Actions
de Recherche Concert\'ees (ARC) and Fonds Wetenschappelijk Onderzoek---
 Vlaanderen (FWO), Belgium, the European Commission. The authors gratefully
acknowledge the support of the NSF, STFC, INFN, CNRS and Nikhef for
provision of computational resources.

{\it We would like to thank all of the essential workers who put their health at risk
during the COVID-19 pandemic, without whom we would not have been able to
complete this work.}

{\it The authors would also like to thank Samuel Salvador for his extensive and careful proofreading of the manuscript.}

\markboth{\Sectionname}{}

\addcontentsline{toc}{section}{List of Abbreviations}
\printacronyms[name=List of Abbreviations]

\appendix
\addtocontents{toc}{\addtolength{\cftsecnumwidth}{5em}}
\addtocontents{toc}{\addtolength{\cftsubsecnumwidth}{5em}}

\section{Additional tool information}
\label{section:appendix}
\markboth{\uppercase{\thesection.} \Sectionname}{}
In this Appendix, we describe in more detail some of the data analysis techniques presented in Section~\ref{section:tools}, and their implementations into tools.
\ref{appendix:BRiSTOL} and \ref{appendix:Rayleigh} provide additional information on the statistics adopted to test the hypotheses of stationarity and Gaussianity with the \ac{bristol} and \texttt{rayleighSpectro} tools, and their interplay.
\ref{section:appendix.bruco} and~\ref{subsec:monet} are devoted to the tools for spectral noise investigations \texttt{BruCo} and \texttt{MONET}.
The former uses the coherence as a figure of merit to study the linear transfer of power between a set of auxiliary channels and a main channel, usually the GW strain $h(t)$.
The latter investigates non-linear couplings between channels, by studying the coherence of the main channel (usually $h(t)$) with a synthetic one, created by modulating a carrier signal (either an existing DAQ channel or a sinusoid with a fixed frequency) with the low-frequency part of an auxiliary channel.

\subsection{BRiSTOL - a Band-limited RMS Stationarity Test Tool}\label{appendix:BRiSTOL}

This statistical test aims at verifying the hypothesis of \emph{wide} (or \emph{weak}) \emph{sense} stationarity of the data,
i.e.~that the covariance function is left unchanged by shifts in time.
We test this by verifying that subsequent \ac{psd} estimates, in predefined frequency bands, are compatible with the same probability distribution. 
The corresponding test statistics are based on a set of \ac{brms} time series, estimated on an equal number of bands:
\begin{equation}\label{eq:defBRMS}
\mathit{BRMS}_t(b) = \sqrt{\int_{f\in b} \hat{S}_t(f)\,df}, \qquad \mathrm{for} \quad b\in\left\{[f_1 ^{\max}, f_1 ^{\min}],\ldots,[f_K ^{\max}, f_K ^{\min}] \right\}
\end{equation}
for data $x_{t_n}$ recorded at Nyquist rate $f_S$,  $t_n = t+n/f_S$, where $\hat{S}_t(f)$ is a \ac{psd} estimate referred to time $t$, and obtained with the \emph{periodogram} method~\cite{schuster1898}:
\begin{equation}\label{eq:periodogram}
    \hat{S}_t(f_k)=\frac{1}{N}\bigg|\sum_{n=0} ^{N-1} x_{t_n} e^{-2\pi i\, nf_k/f_S}\bigg|^2,\qquad f_k=\frac{k\,f_S}{N}, \quad\text{for}\quad k=0,\ldots,N-1
\end{equation}

Two modifications have been implemented to make \eqref{eq:defBRMS} more suitable for the study of transient noise, in particular to highlight slow non-stationarity.
Firstly, {spectral lines} (refer to Section~\ref{sec:noemi} for more details) have been removed from the integral.
These are narrow features in the \ac{psd} of the data, originating from resonances in various parts of the interferometer and their harmonics.
Their intensities can be orders of magnitude larger than the neighboring noise floor.
Hence, if a line is present in a band where we are about to compute the \ac{brms}, it is likely to dominate the final estimate, and also the corresponding fluctuations, preventing us from probing the features of the underlying noise floor.
To remove these lines, we identify them with an algorithm similar to the one developed for the \ac{noemi} pipeline~\cite{Accadia_2012}, and based on the \emph{prominence} of their \ac{psd}~\cite{acernese2005simple}.

Second, glitches are also typically removed from the \ac{brms} time series.
These fast transients can manifest at a rate of about 10 per minute, as shown in Figure~\ref{fig:omicron_trigger_rates:global}, which means that every data segment longer than a few seconds is likely to contain one of them.
To focus on slower noise transients, which typically are not targeted by tools specifically devoted to excess power identification, such as those presented in Section~\ref{sec:onlinedq}, we must exclude the data segments affected by glitches from the stationarity test.
This is done by means of an algorithm based on a rolling \emph{median absolute deviation}, defined as the median absolute difference from the median, to identify outliers in the \ac{brms} data.

Then, the time series corresponding to the resulting modified \ac{brms} are divided into ``chunks'' where estimate their empirical distribution function.
The stationarity hypothesis is tested by means of a two-sample Kolmogorov--Smirnov test~\cite{kolmogoroff1941confidence} for each pair of consecutive chunks, whose $p$-values are compared to a test significance $\alpha$ (to be decided in advance), and the (\emph{null}) hypothesis of stationarity rejected when the latter is exceeded.

There are two advantages in using the \ac{brms}.
First, averaging over the frequencies of each band has a similar variance reduction effect than the means in Welch's \ac{psd} estimation method~\cite{welch1967}. This in turn allows a finer time resolution while maintaining a moderate variance for our test statistics, that is, the empirical distribution of the \ac{brms}.
Second, the various non-stationarities typically manifest in specific frequency bands, closely related to the noise source that generated them. For example, the main harmonic of scattered light is usually visible below 30~Hz; non-linear and non-stationary couplings of the angular controls with the 150~Hz harmonic line are characteristic of a tight region around it, etc.
So, without losing much of resolution, we can perform the noise characterization directly on these bands instead of on each frequency bin comprising the spectrum of the signal.

\subsection{rayleighSpectro - Gaussianity test}\label{appendix:Rayleigh}
Similarly to what was discussed for the stationarity hypothesis, Gaussianity is likewise important to be tested separately in the different regions of the spectrum where noise sources can show up. 
\texttt{rayleighSpectro} does this by means of a consistency check on the \ac{psd} estimated from the data with what is expected for stationary Gaussian noise.
Indeed, if the data is compatible with the hypothesis od Gaussianity, the periodogram estimator in Equation~\eqref{eq:periodogram}
is asymptotically (with $N$) described by an \emph{exponential distribution} of parameter $S(f_k)^{-1}$~\cite{KOKOSZKA200049}, where $S(f_k)$ is the process \ac{psd}.
The corresponding \ac{asd} estimator, obtained as the square root of Equation~\eqref{eq:periodogram}, is described by a Rayleigh distribution with parameter $\sqrt{S(f_k)/2}$.
The scaling property of this distribution can be used to construct consistency tests.
For example, the standard deviation of the \ac{asd} estimates obtained on non-overlapping segments provides an estimator of the standard deviation of this variable, which equals $\sqrt{(4-\pi)\,S(f_k)}/2$.
Similarly, the mean of these estimates provides an estimator of the mean: $\sqrt{\pi S(f_k)}/2$.
The ratio of these two quantities gives a statistic that, at each frequency $f_k$, is \emph{asymptotically} equal to a constant whose numerical value is given in Equation~\eqref{eq:Rayleigh_test_statistic},
in the null hypothesis that the data is described by a stationary and Gaussian distribution.
The actual value of the previous quantity for a finite number of averages and the corresponding critical values
for performing statistical tests have been computed in~\cite{direnzo2020ragout,verkindt_spectro}.

By dividing the data into chunks of duration $\Delta t$, one can obtain a time--frequency map,
similar to a spectrogram, showing with time resolution $\Delta t$ the frequencies and times where the data significantly depart
from the expected value of Equation~\eqref{eq:Rayleigh_test_statistic}.
Smaller values of this statistic are associated with data having smaller fluctuations than those expected for a Gaussian process; {spectral lines} usually behave in this way.
Larger values are instead typical of non-stationary noises, such as glitches, that produce a larger variance of the \ac{asd} estimates.

The interplay between this tool and \ac{bristol} for the assessment of the stationarity and Gaussianity of the data is the following.
The latter assesses where the data is compatible with the hypothesis of wide sense stationarity, that is, the second order moments (i.e.~the covariance or the RMS) are left unchanged by shifts in time.
This corresponds also to \emph{strong sense stationarity} if the data is also Gaussian, that is, completely characterized by its mean and covariance functions, as tested by \texttt{rayleighSpectro}.
Conversely, deviations from these assumptions can be tested independently.

\subsection{BruCo}
\label{section:appendix.bruco}

The \ac{bruco} python code (version 2017-01-23)~\cite{Vajente:2008bka}, 
is publicly available from the git repository~\cite{BRUCO_git}, which also provides 
a description of the argument list. An instance 
of the repository is kept with Virgo-specific data
access features.

\ac{bruco} computes the magnitude-squared coherence between a main channel 
(typically, but not necessarily, the detector strain channel $h(t)$)
and all auxiliary channels that, at the time of interest, are recorded by the \ac{daq} system.
Optionally, a set of redundant channels which are known a priori to be correlated with the main one, 
can be excluded. In Virgo, during the O3 run, there were approximately 3,000 non-redundant channels with a 
sampling frequency $\geq$1~kHz.
To deal with the high computational load required by this analysis, \ac{bruco} implements the option of multi-core 
parallel processing in up to $10$ threads.

In the \ac{bruco} implementation adopted for Virgo during O3,
a continuous Science data segment
of length $T=800$~s is selected for the main channel $h(t)$ and, in turn, for each auxiliary channel $n(t)$.
Each data segment is resampled to a targeted output frequency of 2~kHz with the Fourier 
resampler {\it scipy.signal.resample}, 
divided into $N_{ave}=100$ sub-segments (8~s long) and 
the averaged magnitude-squared coherence is computed, as: 

\begin{equation}
C_{h,n}(f_i) = \frac{\mid < FFT_n(f_i)^{*} FFT_h(f_i) >\mid ^{2}}{< \mid FFT_h(f_i)\mid ^{2}>< \mid FFT_n(f_i)\mid ^{2}>}
\label{eq:coherence}
\end{equation}

where $FFT$ denotes the windowed fast Fourier transform, "$<>$" denotes the averaging operation,
$f_{i}$ is the $i^{\textrm{th}}$ frequency bin, and "$^{*}$" the complex conjugate operation.
With these parameters, the frequency resolution is $df = N_{ave} / T = 0.125$~Hz.
Coherence is examined up to 1~kHz, 
and its value is deemed significant if it exceeds a threshold set to 0.03, a value
corresponding to the 95\% confidence level of the distribution of averaged coherence between 
random data~\cite{Piersol}, given the selected parameters.

\begin{figure}[htb!]
    \includegraphics[width=0.98\textwidth]{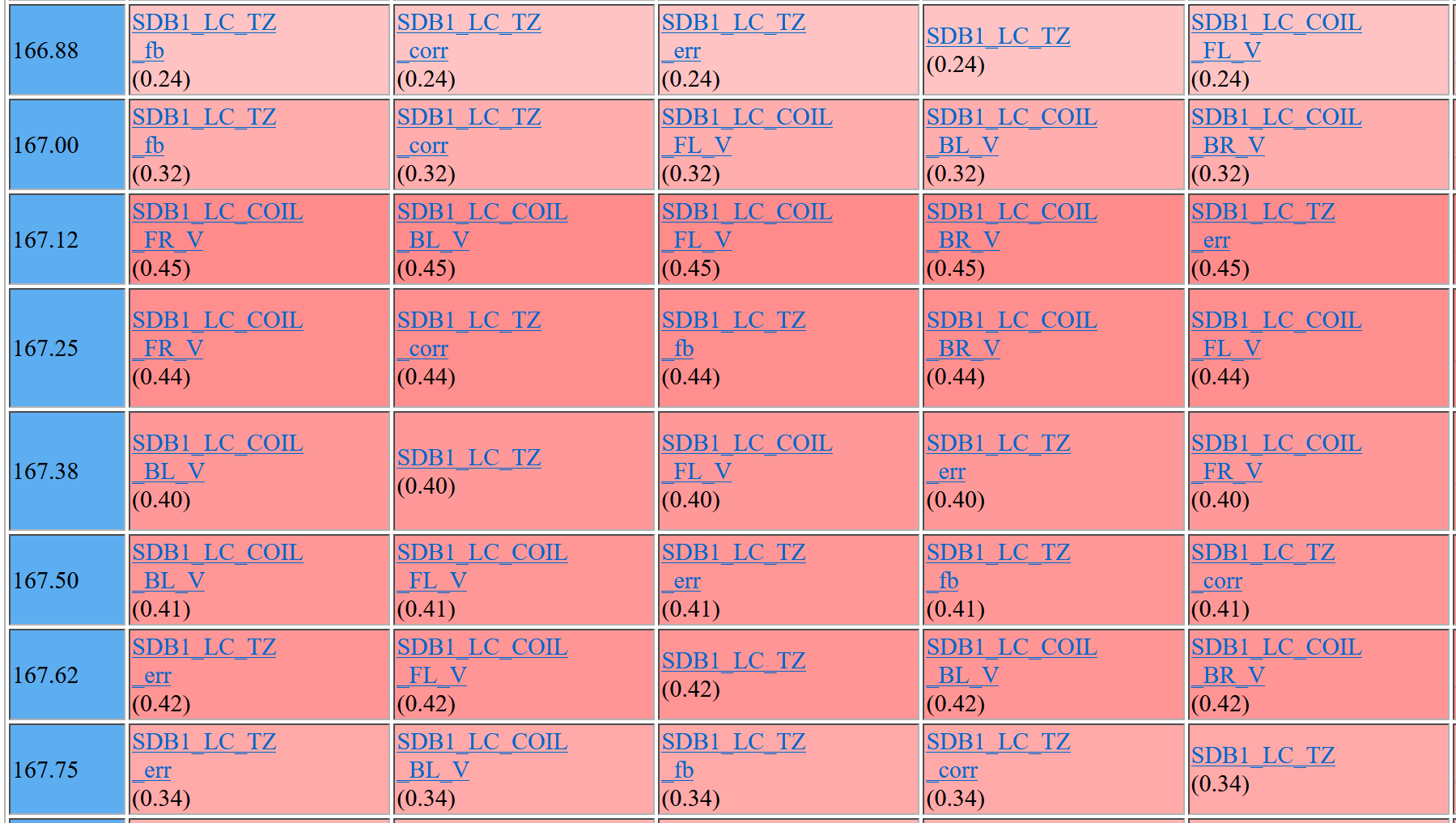}
    \caption{An example of \ac{bruco} result webpage displaying, for each frequency bin (leftmost column), the most coherent channels sorted by decreasing coherence value, also represented by the red shade intensity. 
These data are from November 11, 2019.
The large coherence detected at frequencies $155-170$~Hz triggered some further investigations of the noise~\cite{Cieslar:elog}.}
    \label{fig:brucopage}
\end{figure}

\begin{figure}[htb!]
    \includegraphics[width=0.9\textwidth]{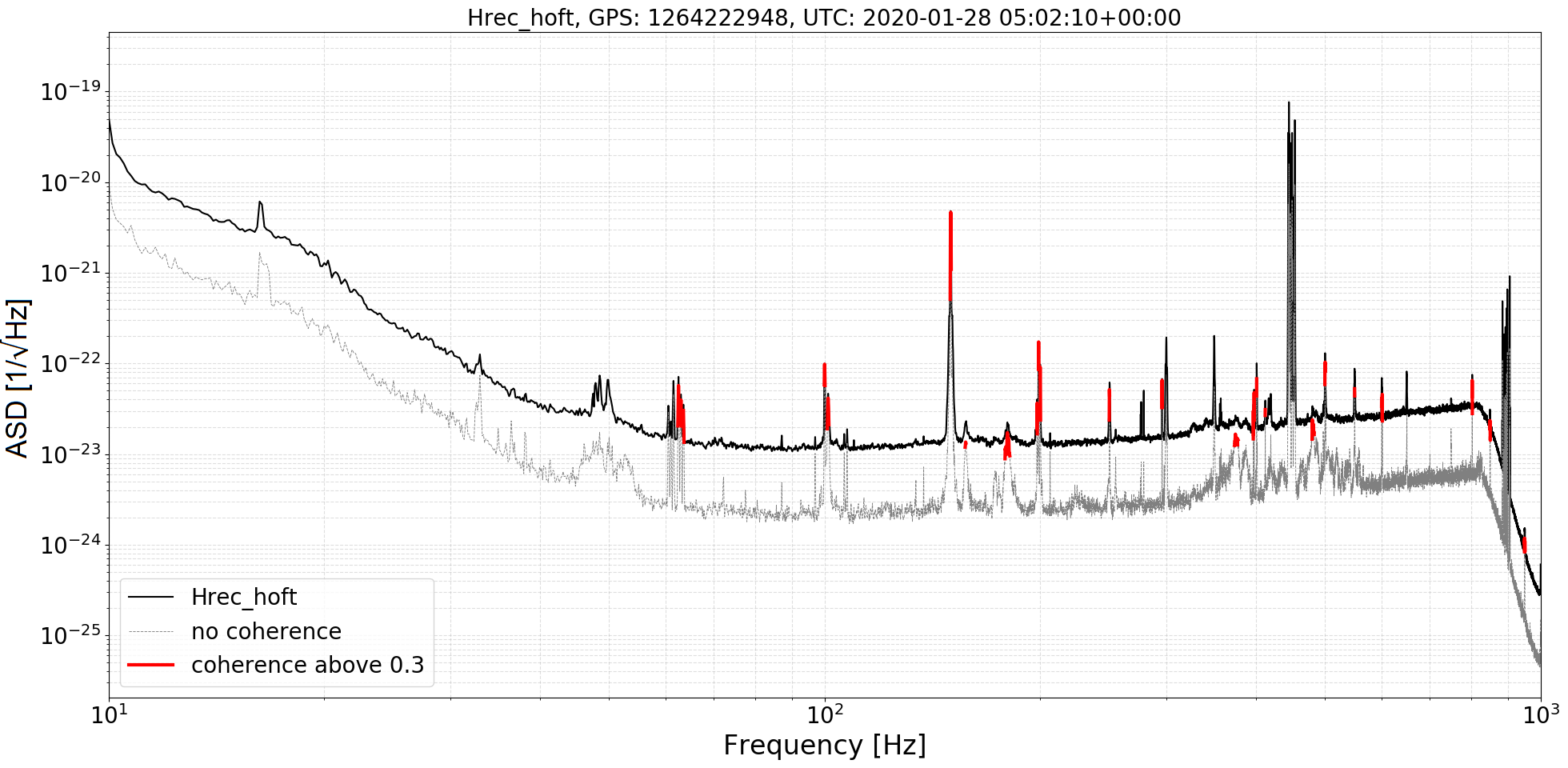}
    \caption{A \ac{bruco} \ac{vim} daily combined projection of the GW strain channel $h(t)$, showing coherences greater than 0.3 for the same GPS time 1264319383 (2020/01/29 at 07:49:25 UTC) as the previous plot.}
    \label{fig:bruco_coheprojection}
\end{figure}

Daily \ac{bruco} results are html-formatted and made accessible in a dedicated \ac{vim} web page (see 
Section~\ref{sec:tools:VIM}). The \ac{bruco} \ac{vim} summary page allows to quickly spot noise paths 
contributing to the \ac{gw} strain $h(t)$ in specific frequency bands. 
A summary table is generated: each row corresponds to a given frequency bin and contains the list of the most coherent channels in descending order.
The cell background is color coded in shades of red 
from full red (maximum coherence: 1) to white (no coherence) as shown in Figure~\ref{fig:brucopage}.
For each auxiliary channel, a plot (see Figure~\ref{fig:bruco_1}) of the {\it projected coherence} 
quantity, $ h_{n}(f)=<FFT_{h}(f)>\sqrt{C_{h,n}(f)}$, is produced and linked to the table. 
In the hypothesis of linear coupling, this quantity estimates the contribution to the strain channel of the noise 
witnessed by the $n^{\textrm{th}}$ auxiliary channel~\cite{Piersol}.
Additionally, the \ac{vim} daily summary page contains the list of the top ranked channels in the frequency 
bins with coherence greater than 0.3 (Table~\ref{tab:Bruco_cohshort}), and a plot of the combined 
projected coherence greater than 0.5 (Figure~\ref{fig:bruco_coheprojection}).

\begin{table}[htb!]
    \caption{\label{tab:Bruco_cohshort}An excerpt (15 top lines) from top ranked channels' summary with coherence greater than 0.3 for GPS time 1264319383 (2020/01/29 at 07:49:25 UTC).
}
    \begin{tabular}{r | r | l}
        \toprule
        Frequency [Hz] & Coherence & Type of channel \\
        \hline
         9.375 & 0.348 & West arm transmitted power measurement \\
         \multirow{2}{*}{9.500} & \multirow{2}{*}{0.371} & Angular correction of the last stage \\ 
          & & of \ac{wi} payload suspension \\
        \multirow{2}{*}{13.375} & \multirow{2}{*}{0.308} & Sensing signal measured at the dark fringe port, \\ 
         & & used to control optical cavity alignment \\
        13.625 & 0.349 & Same sensing signal \\
        13.750 & 0.347 & Same sensing signal \\
        14.125 & 0.311 & Same sensing signal \\
        14.250 & 0.310 & Same sensing signal \\
        14.375 & 0.335 & Same sensing signal \\
        14.625 & 0.309 & Same sensing signal \\
        15.375 & 0.323 & Same sensing signal \\
        \multirow{2}{*}{16.250} & \multirow{2}{*}{0.589} & Longitudinal correction of the last stage \\ 
         & & of \ac{bs} payload suspension \\
        16.375 & 0.532 & Same longitudinal correction \\
        16.750 & 0.303 & Same sensing signal \\
        \multirow{2}{*}{60.375} & \multirow{2}{*}{0.403} & Calibration signal applied \\ 
                                                         & & to the \ac{we} test mass actuators \\
        \bottomrule
    \end{tabular}    
\end{table}

\subsection{MONET}\label{subsec:monet}
\ac{monet} computes the magnitude-squared coherence between a main channel (typically the DARM channel) and a list of other signals, that are constructed multiplying in the time domain a carrier signal and a set of modulating auxiliary channels to  which a low-pass filter (with a typical cutoff frequency of a few Hz) is applied; the carrier signal can be a real channel or a simulated signal (e.g. a sinusoidal signal).
Similarly to what is done with \ac{bruco}, also with \ac{monet} continuous science data segments of a specific time length $T$ are selected for all the channels to be investigated; then each data segment is resampled to a targeted output frequency (\mbox{$f^{\rm out}$}) and, finally, the magnitude-squared coherence is computed with Equation~\ref{eq:coherence}. For the analysis of O3 Virgo data we typically used a cutoff frequency of 5 Hz, $T=1200$ s and $f^{\rm out}$=1 kHz.

\ac{monet} can be executed on demand, investigating dozens of auxiliary channels and spectral lines every single run. The outputs are organized in a directory structure:
\begin{itemize}
\item[$\bullet$] A main directory, whose name indicates the main channel, the initial gps time and the time length of the segment of data to be analysed.
\item[$\bullet$] A secondary directory, named after the carrier signal used. 
\item[$\bullet$] Several sub-directories, one for each modulator channel.
\end{itemize}

\begin{figure}[h!]
\begin{center}
\includegraphics[width=0.9\textwidth]{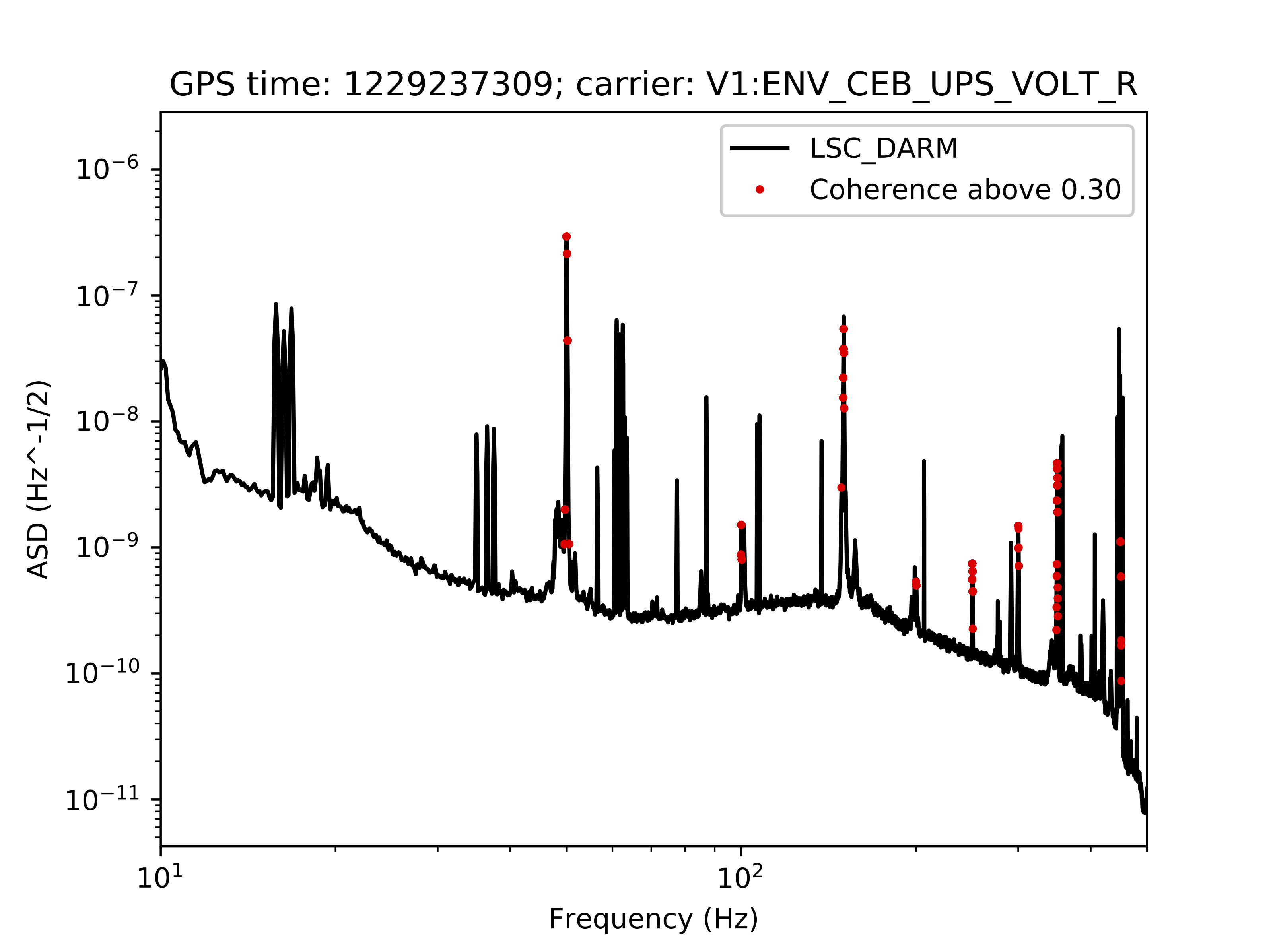}
\caption{\ac{asd} of the \ac{darm} channel (in black), together with red points that mark the frequencies at which the coherence is above the threshold, fixed in this case at 0.3; the initial GPS time of the analysed data and the chosen carrier signal are indicated on the top of the figure.} \label{fig:ASD-monet}
\end{center}
\end{figure}

The secondary directory contains an ascii table and a summary figure. The ascii table contains three columns: the frequency bins, the computed above-theshold coherence values for each bin (ordered from the lowest to the highest) and the corresponding modulator channels name. Figure~\ref{fig:ASD-monet} shows the MONET summary plot for the DARM channel during O3. Superimposed to the main channel's \ac{asd} there are red points marking the frequencies at which coherence above threshold is found with at least one auxiliary channel.

\begin{figure}[h!]
\begin{center}
\includegraphics[width=0.9\textwidth]{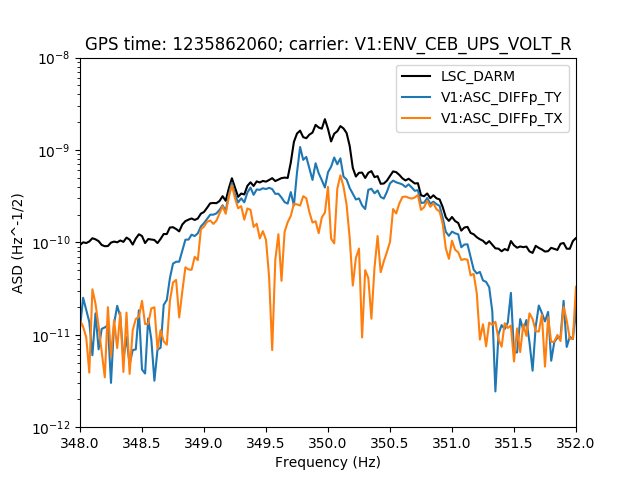}
\caption{\ac{asd} of the main channel (DARM) around the 350~Hz spectral line (in black), together with the noise projection, based on the coherence values obtained with the modulator channels ASC$\_$Diffp$\_$TY and ASC$\_$Diffp$\_$TX, in blue and orange respectively (these two channels are used to control the angular movement of the mirrors in the detector arms, to guarantee the proper ricombination of the laser beams at the beam splitter); the initial gps time of the analysed data and the chosen carrier signal are indicated on the top of the figure.}\label{fig:ASDzoomed-monet}

\end{center}
\end{figure}

In each sub-directory, a table and a plot are generated, in which the coherence values associated with the specific modulator channel for each frequency bin are reported; several other plots are also produced, in which the \ac{asd} of the main channel is reported, together with the noise projection based on the coherence values, around the specific spectral lines to be investigated (see Figure~\ref{fig:ASDzoomed-monet}).

\clearpage

\section*{References}
\markboth{REFERENCES}{}
\bibliographystyle{iopart-num}
\bibliography{references}

\end{document}